\documentclass[doublecol]{epl2}

\usepackage{amsmath}
\usepackage{amsfonts}
\usepackage{amssymb}
\usepackage{graphicx}
\usepackage{bm}
\usepackage{color}
\usepackage[hypertexnames=false]{hyperref}
\usepackage{ulem}
\usepackage{wasysym}
\usepackage{bm}

\title{Statistics of Conductances and Subleading Corrections to Scaling
near the Integer Quantum Hall Plateau Transition}
\shorttitle{Subleading Corrections to Scaling near the Integer Quantum Hall Transition}

\author{H. Obuse\inst{1,2}
\and S. Bera\inst{3}
\and A. W. W. Ludwig\inst{4}
\and I. A. Gruzberg\inst{5}
\and F. Evers\inst{1,6,7}}
\shortauthor{H. Obuse \etal}

\institute{
  \inst{1} Institut f\"ur Nanotechnologie, Karlsruhe Institute of
Technology (KIT), 76021 Karlsruhe, Germany\\
  \inst{2} Department of Applied Physics, Hokkaido University, Sapporo
060-8628, Japan\\
  \inst{3}Institut N\'eel, CNRS and Universit\'e Joseph Fourier, 38042
  Grenoble, France\\
  \inst{4}Department of Physics, University of California,  Santa
  Barbara, CA 93106, USA\\
\inst{5}
Department of Physics, The Ohio State University, 191 W. Woodruff Ave., Columbus, OH 43210, USA\\
\inst{6}Institut f\"ur Theorie der Kondensierten Materie, Karlsruhe
Institite of Technology (KIT), 76128 Karlsruhe, Germany\\
\inst{7}DFG-Center for Functional Nanostructures, Karlsruhe
Institite of Technology (KIT), 76131 Karlsruhe, Germary
}
\abstract{
We study the critical behavior near the integer quantum Hall  plateau transition by focusing on the multifractal (MF) exponents $X_q$ describing the scaling of the disorder-average moments of the point contact conductance $T$ between two points of the sample, within the Chalker-Coddington network model. Past analytical work has related the exponents $X_q$ to the MF exponents $\Delta_q$ of the local density of states (LDOS). To verify this relation, we numerically determine the exponents $X_q$ with high accuracy. We thereby provide, at the same time, independent numerical results for the MF exponents $\Delta_q$ for the LDOS. The presence of subleading corrections to scaling makes such determination  directly from scaling of the moments of $T$ virtually impossible. We overcome this difficulty by using two recent advances. First, we construct pure scaling operators for the moments of $T$ which have precisely the same leading scaling behavior, but no subleading contributions. Secondly, we take into account corrections to scaling from irrelevant (in the renormalization group sense) scaling fields by employing a numerical technique (``stability map'') recently developed by us. We thereby numerically confirm the relation between the two sets of exponents, $X_q$ (point contact conductances) and $\Delta_q$ (LDOS), and also determine the leading irrelevant (corrections to scaling) exponent $y$ as well as other subleading exponents. Our results suggest a way to access multifractality in an experimental setting.
}
\begin{document}

\maketitle

The integer quantum Hall (IQH) effect has been an exciting area of research in condensed matter physics for three decades \cite{vonKlitzing, QHeffects}. Recently, renewed interest put the IQH plateau transition, an Anderson
(de-)localization transition driven by disorder \cite{Evers08}, in the focus of intense experimental \cite{Amsterdam-group, Tsui-group, Amado10, Saeed11, Huang12, Shen12} and theoretical research
\cite{Zirnbauer99, Pruisken, Obuse08, Evers08b, Slevin09, Burmistrov10, Amado11, Bettelheim12, stabilitymap}.
Of particular interest is the scaling of transport and other properties at the transition, including the scaling of moments of the local density of states (LDOS) $\rho(\bf r)$.

So far, there is no widely accepted analytical theory of the critical phenomena near the transition \footnote{But see Ref.\cite{Bettelheim12} for recent developments.}.
Meanwhile, numerical simulations have revealed rich critical behavior of the LDOS that exhibit multifractality in the form of an infinite set of exponents (see \cite{Evers08, Mirlin10} for a recent review) that describe the scaling of the moments $\langle \rho^q({\bf r})\rangle \sim L^{-\Delta_q}$ with the system size $L$ \cite{DuplantierLudwig91, Subramaniam06, Obuse07}. Unfortunately, the probability distribution for the LDOS, and hence $\Delta_q$, are difficult to access experimentally, but there are notable attempts in this direction \cite{Morgenstern03, Richardella10}. Transport measurements are typically easier to perform, and this makes understanding relations between multifractality of the critical LDOS and conductances quite important.

In Refs. \cite{Janssen99} and \cite{Klesse01} the authors have introduced the notion of the point contact conductance (PCC) $T({\bf r}_1,{\bf r}_2)$ between two points ${\bf r}_1$ and ${\bf r}_2$ of the sample, and considered the scaling of its moments at criticality (we will use the short-hand notations $T \equiv T({\bf r}_1,{\bf r}_2)$ and $r \equiv |{\bf r}_1 - {\bf r}_2|$ in the following), described by
\begin{align}
\label{e1}
\langle T^q \rangle &\sim r^{-X_q}.
\end{align}
The following relation between the transport exponents $X_q$ and the LDOS exponents $\Delta_q$ was derived \cite{Janssen99,Klesse01,Evers01}:
\begin{align}
\label{e2}
X_q = \begin{cases} 2\Delta_q, &  \text{for } q < 1/2, \\ 2\Delta_{1/2}, & \text{for } q \ge 1/2.
\end{cases}
\end{align}
While this relation was derived using the Chalker-Coddington (CC) network model \cite{chalker88}
for the IQH transition, it connects two sets of universal critical exponents, and as such
is expected to be generally valid, independent of the microscopic model. In fact, it is
believed that similar relations hold for Anderson transitions in all symmetry
classes \cite{Evers08, Gruzberg12}.

In this Letter, we  subject eq. (\ref{e2}) to a sensitive test by determining
numerically the transport exponents $X_q$ and then comparing them with earlier
numerics for $\Delta_q$ obtained
from the scaling behavior of the LDOS (or wavefunction) moments
\cite{Obuse08, Evers08b}. For the high-precision comparison that we are
aiming at, a careful treatment of subleading power-laws is required. As
recent work \cite{stabilitymap, Slevin09, Amado11} demonstrates,
the IQH transition poses a particular challenge
in this respect.
It is well known\cite{Evers08} that
corrections to scaling decay near the IQH fixed point with
an irrelevant exponent, $|y|$, that is certainly smaller than one, probably  smaller
than 0.5. This unusually small value requires us to keep
several terms in power series expansions off scaling functions, see below. Therefore,
the IQH fixed point is much more difficult to access than the
three-dimensional Anderson transition
where much larger values ($|y|\approx 1.5$ \cite{rodriguez10}
and $|y| \approx 3.3$ \cite{slevin13})
were reported in the orthogonal class.
Even though subleading terms do not influence the true asymptotic
scaling behavior they are still important in practice,
in particular because they determine the size of the critical region.

We remind the reader that two types of corrections to scaling generally exist.
First, a particular physical observable may be a combination of several pure scaling
operators. It is know, for example, that $T^q$ is not a pure scaling operator \cite{Janssen99},
while $\rho^q(\bf r)$ is \cite{Gruzberg11, Gruzberg12}. Thus, eq.~(\ref{e1}) should be
understood as characterizing the leading long-distance behavior of the observable $\langle T^q\rangle$,
subject to subleading corrections from the admixture of other pure scaling operators, characterized by
certain subleading scaling dimensions $\gamma_q$. Secondly, even correlation functions of pure scaling
operators exhibit what is called {\it irrelevant} corrections to scaling due to the fact that they
are calculated using a critical Hamiltonian different from the fixed point Hamiltonian.
Deviations from the fixed point are controlled by irrelevant exponents
\footnote{In $D$ spatial dimensions, $y_i= (D-x_i)$  is the scaling
dimension of the coupling constant of an operator of scaling dimension $x_i$ added to the Hamiltonian.}
$y_i  < 0$
which are the same for all scaling operators. The second goal of the paper is to determine the
leading irrelevant (i.e. least irrelevant) exponent $y$ (which should be independent of $q$)
as well as the  scaling dimensions $\gamma_q$ of the above mentioned subleading
operators.

Keeping these goals in mind, we write the scaling function for moments of the PCC as
\begin{align}
\label{e3}
\langle T^q \rangle &= \textstyle c_q r^{-X_q}\Big(1+\sum_{n=1}^{N_p} a^{(n)}_q r^{ny} \Big) \nonumber \\
&\quad \textstyle + d_q r^{-\gamma_q}\Big(1+ \sum_{n=1}^{N_s} b_q^{(n)}r^{ny} \Big).
\end{align}
Here in addition to the leading scaling operator characterized by
the scaling dimension $X_q$ we have retained only one subleading
operator characterized by the scaling dimension $\gamma_q > X_q$,
and in both contributions we have retained only one irrelevant, i.e.
the leading corrections to scaling exponent $y$. We have also truncated
the expansions of the scaling function in powers of $r^y$ at orders $N_p$ and $N_s$
for the two scaling operators, for the purpose of the numerical analysis
below 
\footnote{More generally, $\langle T^q \rangle$ should include several irrelevant exponents, $y_1, y_2, \cdots$,
and contributions from more subleading scaling operators characterized by exponents $\gamma_q^{(1)}, \gamma_q^{(2)}, \cdots$,
and these contributions should not be truncated.}.

Because of the presence of the subleading scaling dimension $\gamma_q$ in eq. (\ref{e3}),
it is difficult to get an accurate estimate of the leading scaling dimension $X_q$
from fitting numerical data for the moments $\langle T^q \rangle$ to eq. (\ref{e3}).
To circumvent this difficulty, we will consider $p_q(T) \equiv P_{-q}(2/T -1)$,
where $P_q(x)$ is the associated Legendre function of the first kind \cite{DLMF}.
As explained below, this
quantity has the following properties: (i) its leading scaling behavior is the
same as that of $\langle T^q \rangle$ for $q \le 1/2$, and (ii) it is a pure scaling operator.

Property (i) follows if we consider long distances $r$, where $T \ll 1$. Then
$x \equiv 2/T - 1 \gg 1$, and we can use the standard asymptotics 
\footnote{See sect. 14.8 (iii) in Ref. \cite{DLMF}. For $q = 1/2$ there
is an extra logarithmic correction: $P_{-1/2}(x) \sim x^{-1/2}\ln x \sim T^{1/2}\ln\dfrac{1}{T}$.}
\begin{align}
\label{e4}
P_{-q}(x) \sim \begin{cases}
x^{-q} \sim T^q, & q < 1/2, \\ x^{q - 1} \sim T^{1-q}, & q > 1/2.
\end{cases}
\end{align}

Property (ii) can be derived as follows. In Ref.\ \cite{Klesse01} the following formula was derived within the CC model:
\begin{align}
\label{e5}
2\pi \nu \Big\langle \rho_2 f\Big( \frac{\rho_2}{\rho_1} \Big) \Big\rangle &=
\bigg\langle \int_{-\pi}^{\pi} \frac{d \phi}{2 \pi} f\bigg( \frac{ |1 + e^{i \phi} \sqrt{1-T}|^2 } {T} \bigg)
\bigg\rangle.
\end{align}
Here $\rho_i \equiv \rho({\bf r}_i)$ is the LDOS at point ${\bf r}_i$,
$\nu$ is the mean level density, and $f(z)$ is an arbitrary function. If
we choose $f(z)=z^{-q}$ \cite{Evers01}, the integral on the right-hand
side of eq. (\ref{e5}) becomes 
\footnote{See eq. (14.12.7) in Ref. \cite{DLMF}.}
\begin{align}
\label{e6}
\frac{1}{\pi}\int_0^{\pi} \!\!\! \frac{d\phi}{[x + (x^2 - 1)^{1/2} \cos \phi]^{q}} = P_{-q}(x) = P_{q-1}(x),
\end{align}
where the last equality is a symmetry property of $P_q$. Thus, we obtain the following relation:
\begin{align}
\label{e7}
2\pi \nu \big\langle \rho^{q}({\bf r}_1) \rho^{1-q}({\bf r}_2) \big\rangle &= \langle p_q(T) \rangle.
\end{align}
The left-hand side of this relation is a correlation function of pure scaling operators \cite{Gruzberg11} with dimensions $\Delta_q = \Delta_{1-q}$ \cite{Mirlin06}, which demonstrates the point (ii) above.

The arguments above allow us to write
\begin{align}
\langle p_q(T) \rangle &= \bar{c}_q r^{-\bar{X}_q} \textstyle \Big(1+\sum_{n=1}^{N_p} \bar{a}_q^{(n)} r^{n \bar{y}}\Big),
\label{e8} \\
\bar{X}_q &= 2\Delta_q.
\label{e9}
\end{align}
In contrast to eq.\ (\ref{e2}), $\bar{X}_q$ should be equal to $2 \Delta_q$ for any $q$. Also, unlike in eq. (\ref{e3}),
there are no admixtures of subleading scaling operators in eq.\ (\ref{e8}), which makes fitting numerical
data to eq. (\ref{e8}) much better controlled. This allows us to extract reasonable numerical
estimates of exponents $\bar{X}_q$. However, the numerical values of $\bar{X}_q$ obtained in this way turn
out to be not precise enough for a high accuracy test of eq. (\ref{e9}). The limitations are set by
statistical noise in the raw data, which is of the order of $0.1\%$ of the relative standard error.
Also, the range of available distances ($r=3$--$59$ lattice constants) is not sufficient to separate
the different power-law contributions in eq. (\ref{e8}) from each other. Going one step further,
we solve this problem by employing conformal invariance.

By using a logarithmic function we conformally map the 2D plane to a cylinder with circumference $M$ \cite{Ludwig90,Janssen94,Dohmen96,Obuse10}. In this quasi-one dimensional (Q1D) geometry the distance between point contacts along the cylinder is denoted as $L$. For $L \gg M$, the PCC $T$ in the Q1D geometry should have the same scaling properties as the two-terminal conductance $g$ of the cylinder of length $L$. Therefore, we compute numerically moments of $g$ and fit them to the following scaling function\cite{Ludwig90,Obuse10},
to be contrasted with eq. (\ref{e3}):
\begin{eqnarray}
\left\langle g^q \right\rangle &=
c_q' \exp\left[-\pi \left(X_q' + \sum_{n=1}^{N_p} {a'}_q^{(n)} M^{ny'}\right) \dfrac{L}{M} \right] \nonumber \\
&+ d_q' \exp\left[-\pi \left(\gamma_q' + \sum_{n=1}^{N_s} {b'}_q^{(n)} M^{ny'}\right) \dfrac{L}{M} \right].
\label{e10}
\end{eqnarray}
(Primed exponents from the Q1D geometry and unprimed exponents from the 2D geometry
are the same, in principle. We distinguish them nevertheless, in order to emphasize that
the numerical estimates that are obtained in practice
for primed exponents are significantly more reliable.)
Notice, that in the Q1D geometry the irrelevant terms appear as
corrections to the {\it exponents} $X_q'$ and $\gamma_q'$
\cite{Ludwig90,Slevin-Ohtsuki99, Bondesan12}. (This is obtained by the conformal perturbation theory directly in quasi-1D geometry, see, e.g., Refs. \cite{Cardy1986, Henkel-book}.)
This fact, together with the relative exponential suppression of the subleading
term (due to $\gamma_q' > X_q'$) leads us to a way to reliably extract
numerical values of all exponents. This method is especially effective
in cases where the subleading ($\gamma_q$) and irrelevant ($y$)
exponents happen to be numerically close, so that their contributions
tend to mask each other.

We can combine the mapping to the Q1D geometry with the use of Legendre functions, and we will demonstrate below that this leads to the best accuracy. However, for two terminal conductances in Q1D geometry, eq. (\ref{e5}) is, strictly speaking, not applicable. Thus, we do not expect $p_q(g)$ to be a pure scaling operator.
Consequently,  we use the following scaling function for this quantity
\begin{eqnarray}
\langle p_q(g) \rangle &=
{\bar c}'_q \exp\left[-\pi \left({\bar X}_q' + \sum_{n=1}^{N_p} {\bar a}_q^{\prime (n)} M^{n\bar{y}'}\right) \dfrac{L}{M} \right] \nonumber \\
&+ {\bar d}'_q \exp\left[-\pi \left({\bar \gamma}_q' + \sum_{n=1}^{N_s} {\bar b}_q^{\prime (n)} M^{n\bar{y}'}\right) \dfrac{L}{M} \right].
\label{e11}
\end{eqnarray}
which allows for an admixture of a subleading scaling operator.

We obtain the leading, subleading, and irrelevant exponents by
numerically fitting raw data for conductances to the
scaling functions (\ref{e3}), (\ref{e8}),
(\ref{e10}), and (\ref{e11}).
The quality of fitting for each data set is evaluated
in a standard way by employing the chi-square
and goodness of fitting.
The fitting is difficult for two reasons.
First, as already mentioned, corrections to scaling decay slowly near
the IQH fixed point. In general, several subleading terms need
to be included in order to obtain a consistent result.
Consequently, up to ten fitting parameters need to be included.
With so many parameters, the cost function has several
local minima even though the statistical uncertainty in our raw data is
below $0.02 \%$ for conductances in Q1D. In order to find the optimal fit we have proposed
in an earlier work to employ a ``stability map'' that we use
also here \cite{stabilitymap}. The statistical errors that we use
result from the $\chi^2$-analysis at the optimal point.
Second, results of our fitting depend
on the choice of the window of
the distance  between contacts or the system size
(e.g. insets fig. \ref{f2}).
This reflects a residual dependence on higher order terms
in the power expansion eq. \ (\ref{e8}) that we disregard
in our fitting to limit the number of free parameters, and leads to the second type of numerical uncertainty of our estimates.
In order to account for this situation, we
introduce ``practical error bars''.
They constitute the sum of both mentioned uncertainties
and thus are expected to give a conservative 
upper bound of the true errors.

Only upon combining all four essential
steps---the use of the Legendre functions, the mapping to the
Q1D-geometry, the stability map, and the practical error bars---are we able to achieve the
numerical accuracy necessary to finally confirm the relation
(\ref{e9}).

\begin{figure}[t]
\centering
\includegraphics[width=0.9\linewidth]{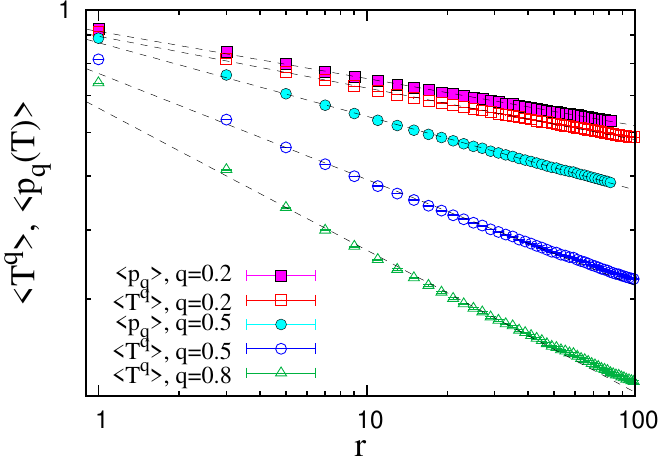}
\vspace{-3mm}
\caption{
The dependence of $\langle T^q \rangle$ and $\langle p_q(T) \rangle$  on $r$ (measured in lattice spacings) along a horizontal
line on a square of size $L=480$ with periodic boundary conditions at $q=0.2$, $0.5$, and $0.8$.
The dashed lines are guides to the eye. $\langle T^q \rangle$ is seen to deviate from the simple
power law (\ref{e1}) suggesting the presence of subleading corrections that require us to use eqs. (\ref{e3}) and (\ref{e8})
to fit numerical data for relatively short distances $r$. At longer distances corrections in powers of $r/L$ obscure
the $L\to\infty$ asymptotics. As $q$ increases, the difference in the slopes of $\langle T^q \rangle$ and $\langle p_q(T) \rangle$
also increases, leading to a difference in the exponents $X_q$ and $\bar{X}_q$. }
\label{f1}
\vspace{-3mm}
\end{figure}

Before presenting our numerical results, we make one more comment. Whenever moments of a random quantity exhibit multifractal scaling characterized by exponents $X_q$, the Legendre transform $F(a) = aq - X_q$,   $a = dX_q/dq$, is related to the probability distribution of this quantity \cite{Ludwig90,Janssen99}. Indeed, if we assume the following form for the distribution of a PCC $T$: $\mathrm{Prob}(T = r^{-a}) dT \sim r^{F(a)} da$ and {\it ignore correction terms}, then the moments of $T$ will scale for large $r$ as $\langle T^q \rangle \sim \int da \, r^{F(a) - aq} \sim r^{-X_q}$. Thus, for the probability distribution function of $\ln T$ we can write
\begin{align}
{\cal P}(\ln T,r) &= {\cal N} r^{F(a)}, & a &= -\ln T/\ln r,
\label{eq:lng_falpha}
\end{align}
where ${\cal N}$ is a normalization constant.
Analogously, the LDOS exponents $\Delta_q$ lead (in 2D) to the so-called singularity spectrum \cite{Evers08} $f(\alpha) = (\alpha - 2)q - \Delta_q + 2$, where $\alpha - 2 = d\Delta_q/dq$. The relation (\ref{e9}) leads to
\begin{align}
F(a) &= 2 [f(\alpha) - 2], & a &= 2(\alpha - 2),
\label{eq:F(a)_f(alpha)}
\end{align}
which can be used in eq. (\ref{eq:lng_falpha}). In the same way, eq.\ (\ref{e10}) for the two-terminal conductance in Q1D leads to the following probability distribution for $\ln g$:
\begin{align}
{\cal P}_{\text{Q1D}}(\ln g, L/M) = {\cal N}_{\text{Q1D}} e^{\pi F(a)
 \frac{L}{M}},\quad a = - \frac{M \ln g}{\pi L}.
\label{eq:lng_falpha_Q1D}
\end{align}
In practice, the normalization constant ${\cal N}_{\text{Q1D}}$ is
determined from the peak of ${\cal P}_\text{Q1D}(\ln g,L/M)$:
\begin{align}
{\cal N}_\text{Q1D}& = {\cal P}_\text{Q1D}(\ln g^{typ},L/M),
\label{eq:N_Q1D}
\end{align}
where $\ln g^\text{typ}$ is the typical value of $\ln g$, since
eq.\ (\ref{eq:F(a)_f(alpha)}) gives $F(a^\text{typ})\propto
f(\alpha_0)-2=0$, where $a^\text{typ} = -M\ln g^\text{typ}/\pi L$
is related to $\alpha_0$ from eq.\
(\ref{eq:F(a)_f(alpha)}), which gives a maximum of $f(\alpha)$, that is, $f(\alpha_0)=2$.

\section{Numerical analysis in 2D}
The PCC is computed numerically using the procedure described in Refs. \cite{Janssen99, Klesse01}. For a given sample of linear size $L$, this requires a solution of a linear system of equations of dimension $\sim L^2$ for each position of the two point contacts. In this work we investigate square systems with periodic boundary conditions in both directions, and place the two point contacts on the same horizontal row of links. Other boundary conditions will be treated elsewhere \cite{otherboundary}. In our simulations $L=480$ and we average over $10^5$ samples.

Figure \ref{f1} shows the moments $\langle T^q \rangle$ and $\langle p_q(T) \rangle$  for the torus geometry. When one contact moves along a row, $T(r)$ is periodic in $r$ with the period $L$. Hence, in this case the expansion (\ref{e3}) applies only at short distances where corrections of order $r/L$ can be neglected. By comparing $T(r)$ traces for different $L$-values we found that a sufficient condition is $r/L \apprle 0.1$ \cite{supplement}. Due to this limitation the window of $r$ values where we can hope to study the true asymptotic behavior is narrow, $r\apprle 41-57$ for the system sizes available to us. Under these conditions a reliable fit to eqs. (\ref{e3}) and (\ref{e8}) is very difficult and can be achieved only using the stability map \cite{supplement}.

\begin{figure}[t]
\centering
\includegraphics[width=0.9\linewidth]{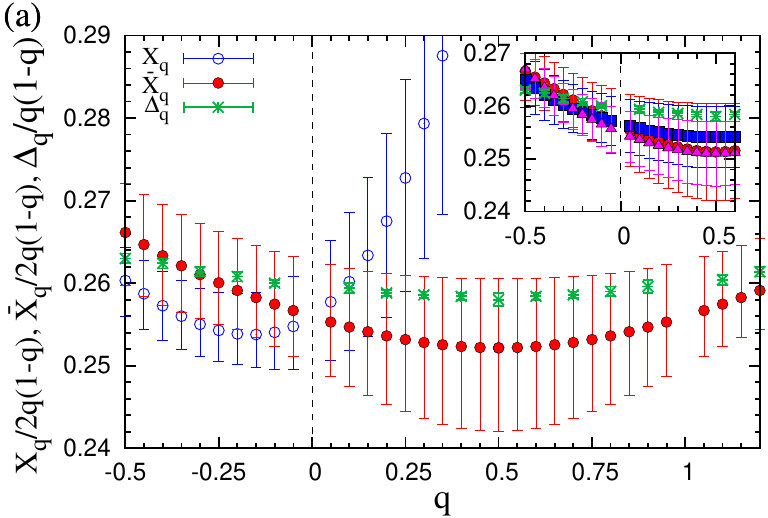}
\includegraphics[width=0.9\linewidth]{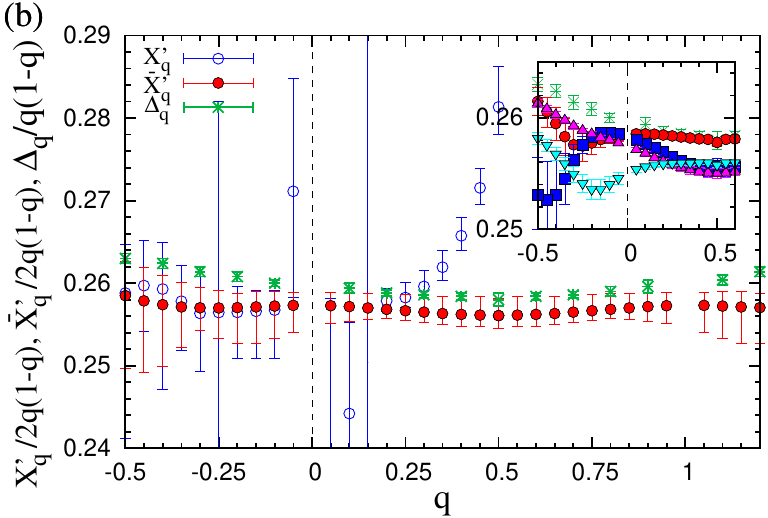}
\vspace{-3mm}
\caption{
(a) $X_q$ and $\bar{X}_q$ obtained from PCC in 2D  eq.\ (\ref{e3}) with
 $N_p=1$ and $d_q = 0$, see footnote\protect\footnotemark[6],
and eq.\ (\ref{e8}) with
 $N_p=1$. (b) $X'_q$ and $\bar{X}'_q$ obtained from the conductance $g$
 in Q1D and eqs.\ (\ref{e10}) and (\ref{e11}), both with $N_p=2$,
 $N_s=1$, see footnote\protect\footnotemark[7].
We also show the LDOS exponents
 $\Delta_q$ from Ref.\ \cite{Evers08b}. All exponents are divided by
 $2q(1-q)$, and, except for $\Delta_q$, are shown with the {\it
 practical} error bars, see the main text and \cite{supplement}. Note
 that the relative errors of the exponents $\bar{X}'_q$ for $0 < q < 1$
 are less than 2\%.
Inset in (a):
$\bar{X}_q$ calculated from data sets with different maximum
 distances $r_{\mathrm{max}}$; $r_{\mathrm{max}}{=}41$ ($\bullet$),
$49$ ($\blacksquare$), and
$57$ ($\blacktriangle$). Minimum distance
 is fixed: $r_\mathrm{min}{=}3$.
Inset in (b): $\bar{X}_q^\prime$ calculated from data sets with different maximum
 width $M_{\mathrm{max}}$;
$M_{\mathrm{max}}=192$ ($\bullet$),
$256$ ($\blacksquare$),
$384$ ($\blacktriangle$), and
$512$ ($\blacktriangledown$).
Minimum width $M_\mathrm{text}{=}32$.
Error bars represent standard (one-sigma) errors
from $\chi^2$-fitting analysis.
}
\label{f2}
\vspace{-5mm}
\end{figure}

Results of this extensive analysis are displayed in fig.\
\ref{f2}(a). It offers a comparison of the exponents $X_q$ and
$\bar{X}_q$ (both divided by $2q(1-q)$) as obtained from fits of
$\langle T^q\rangle$ and  $\langle p_q(T) \rangle$ to eqs. (\ref{e3})
and (\ref{e8}), as well as the LDOS exponents $\Delta_q$ (divided by
$q(1-q)$).
All leading exponents are shown with practical error bars. 
The fitting $\langle p_q(T) \rangle$ to eq.\ (\ref{e8}) is straightforward, and we obtain results that are
consistent with eq.\ (\ref{e9}), albeit with rather large error bars.
By contrast, subleading terms in $\langle T^q \rangle$ interfere strongly in the vicinity of $q=1/2$,
and a controlled fit to eq. (\ref{e3}) is not possible with our data
\cite{supplement}.

\begin{figure}[t]
\includegraphics[width=0.9\linewidth]{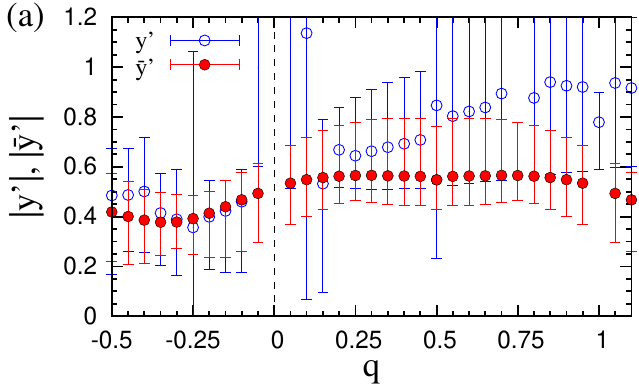}
\includegraphics[width=0.9\linewidth]{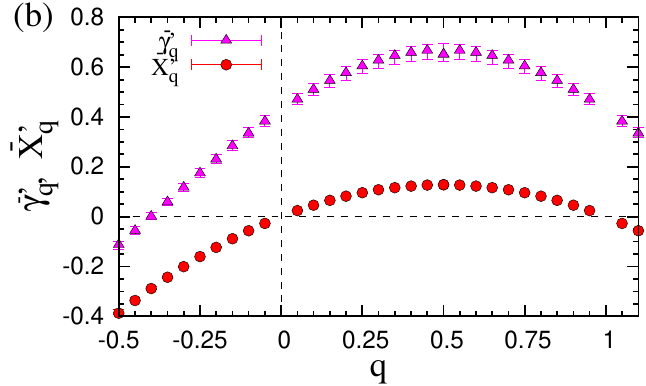}
\vspace{-3mm}
\caption{
(a) The irrelevant exponents $|y'|$ and $|\bar{y}'|$ obtained from fits
 of $\langle g^q \rangle$ and $\langle p_q(g) \rangle$ to
 eqs. (\ref{e10}) and (\ref{e11}). The large scatter in $|y'|$ reflects,
 once again, the difficulty to separate subleading exponents of the two
 kinds. The fact that $|\bar{y}'|$ is largely independent of $q$ in the
 range $0<q<1$ can be viewed as evidence for the quality of the
 fit. Moreover, the value $|\bar{y}'| \approx 0.6\pm 0.2$ is consistent
 with earlier estimates, $|y| \apprge 0.4$ \cite{stabilitymap}.
(b) Subleading exponent $\bar{\gamma}'_q$ from the scaling analysis of
 $\langle p_q(g) \rangle$ and comparison with $\bar{X}_q^\prime$
(Data given with practical error bars.)
}
\label{f4}
\vspace{-3mm}
\end{figure}

\section{Numerical analysis in Q1D} Practical error bars can be reduced by
an order of magnitude for exponents obtained in the long cylinder (Q1D) geometry, where the scaling of the moments $\langle T^q \rangle$ and $\langle g^q \rangle$ should be identical in the limit $L \gg M$. The conductance $g$ in Q1D is obtained by the transfer matrix method \cite{Kramer05}. The width $M$ is varied between 32 and 512. The range of $L$ used for the fitting is $2M$ to $10M$. The number of samples for each $M$ is $10^6$. The leading, subleading, and irrelevant exponents are all extracted from fits to eqs. (\ref{e10}) and (\ref{e11}) \cite{supplement}.

In fig. \ref{f2}(b) we show the exponents $X'_q$ and $\bar{X}'_q$
describing the scaling of $\langle g^q \rangle$ and $\langle p_q(g)
\rangle$.
As expected,
the error bars of $\bar{X}'_q$ are
dramatically smaller than those of $\bar{X}_q$ in 2D, which makes a
meaningful comparison with $\Delta_q$ possible. We thus obtain the first
important numerical result of this work: exponents $\bar{X}'_q/2$ and
$\Delta_q$ agree with accuracy better than 2\% in the range
$0<q<1$. This confirms the exponent relation (\ref{e9}). We interpret
the  small deviations visible outside the range $0<q<1$ as remnants of
higher order corrections in eq.\ (\ref{e8}) not used in the fitting. As
in 2D, errors in $X'_q$ are much larger, the fitting remains
uncontrolled near $q = 1/2$, and the validity of eq. (\ref{e2}) cannot
be established from our data.

Next, we show two plots that highlight an essential difference between the two types of subleading corrections.
fig. \ref{f4}(a) shows results for $|y'|$ and $|\bar{y}'|$ from fitting $\langle g^q\rangle$ and $\langle p_q(g) \rangle$ to eqs. (\ref{e10}) and (\ref{e11}). We see that $|\bar{y}'|$ is essentially $q$-independent in the range $0<q<1$ where we trust our numerical method. This is what one expects from the definition of $y$ as a property of the RG fixed point, and not of a particular observable. On the other hand, the subleading exponents $\gamma_q$ are expected to depend on $q$ in a way that is qualitatively similar to the leading ones $X_q$. Indeed, this is what is seen in fig. \ref{f4}(b) where we show an entire spectrum of the subleading exponents $\bar{\gamma}'_q$ for the IQH.

Finally, we establish the validity of eqs.\ (\ref{eq:F(a)_f(alpha)}) and
(\ref{eq:lng_falpha_Q1D}). Figure \ref{fig:PDF} shows the probability
distribution functions for the random variable $\ln g$ on the Q1D
cylinder with different aspect ratios $L/M=5, 10$. The solid curves
represent ${\cal P}_{\text{Q1D}}(\ln g)$ computed from eqs.\
(\ref{eq:F(a)_f(alpha)}) and (\ref{eq:lng_falpha_Q1D}) by using
${\cal N}_\text{Q1D}$ specified in eq.\ (\ref{eq:N_Q1D}) and the
singularity spectrum $f(\alpha)$ as inputs. Here $f(\alpha)$ is
calculated from the LDOS exponents $\Delta_q$ from Ref. \cite{Obuse08}.
We see a very reasonable agreement between the curves and the symbols.

\footnotetext[6]{%
We account for a single correction term to $\langle T^q\rangle$ only,
i.e.
$N_p{=}1$ and $d_q{=}0$ in eq.\ (\ref{e3})
because our data quality does not allow to distinguish the additional
	terms $\sim r^{-\gamma_q}$.
}
\footnotetext[7]{%
We note that it is important to take $N_p{\geq}2$ in order to obtain
consistent results. With $N_p{=}1$,
$\bar{X}'_q$ and $2\Delta_q$ deviate and $|y|$ is too small: $|y|\sim 0.2$.
See also Ref.~\cite{stabilitymap}.
}

\section{Remark on experiments}
\begin{figure}[t]
\centering
\includegraphics[width=0.9\linewidth]{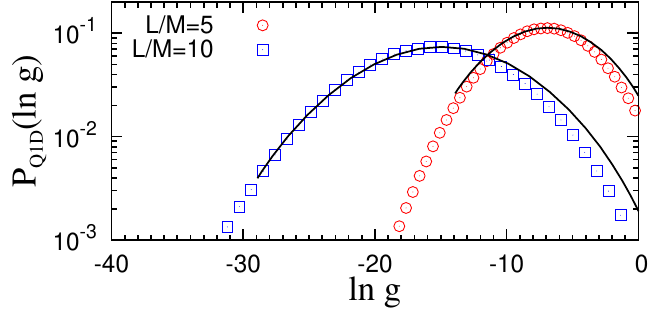}
\vspace{-3mm}
\caption{
Symbols show the probability distribution of the Q1D conductance, ${\cal P}_{\text{Q1D}}(\ln  g)$, for $M=128$ at $L/M=5$ and $10$. The curves are computed by eq.\ (\ref{eq:lng_falpha_Q1D}) with the input of $f(\alpha)$ obtained from the LDOS exponents $\Delta_q$ from Ref. \cite{Obuse08}.
}
\vspace{-5mm}
\label{fig:PDF}
\end{figure}

We need not vary the system size but need only the fixed aspect
ratio $L/M$ to obtain the solid curves in fig.\ \ref{fig:PDF}.
In experiments it is not easy to change geometric quantities like $L,M$
or $r$, used in equations above. At the same time, disorder realizations
can presumably be changed, for example, by varying a gate voltage. By
sweeping different disorder realizations, one can experimentally access
the distribution function of, say, the two-terminal conductance in Q1D,
${\cal P} (\ln g, L/M)$, in a fixed geometry. By applying eqs.\
(\ref{eq:F(a)_f(alpha)}) and (\ref{eq:lng_falpha_Q1D}) to ${\cal P} (\ln
g, L/M)$ inversely as we demonstrated above, this gives another way to access
multifractality.

\section{Conclusions}
In this Letter, we have numerically established relations between spectra of scaling exponents for disorder average moments
of PCC and of the LDOS. These results were achieved by a careful finite-size scaling analysis accounting
for subleading and irrelevant corrections, and augmented by the use of the Legendre functions as well as of the stability map.
\vspace{-3mm}
\acknowledgments
\vspace{-3mm}
FE thanks the IAS at the Hebrew University of Jerusalem for its kind
hospitality while this work was finished. HO thanks A.\ Furusaki for
helpful discussions about conformal mapping.
We also thank I. Kondov for computational support.
Numerical simulations were performed using the
resources provided by PADS at the University of Chicago (NSF grant
OCI-0821678) and the University of Chicago Research Computing Center,
and JUROPA at the Juelich Supercomputer Center (project HKA12). H.\ O.\
was supported by a Grant-in-Aid for Research Abroad and Nos.\ 25800213
and 25390113 from the
Japan Society for Promotion of Science, and I.\ A.\ G.\ was supported by NSF Grants No.\ DMR-1105509 and No.\ DMR-0820054. This work was supported, in part, by the NSF under grant DMR- 0706140 (A.W.W.L.).


\clearpage
\setcounter{equation}{0}
\setcounter{table}{0}
\setcounter{figure}{0}
\renewcommand{\theequation}{S\arabic{equation}}
\renewcommand{\thefigure}{S\arabic{figure}}
\renewcommand{\thetable}{S\arabic{table}}

\begin{widetext}
{\large {\bf Supplemental material for ``Statistics of Conductances and
 Subleading Corrections to Scaling near the Integer Quantum Hall Plateau
 Transition''}}

\vskip 5mm

In this supplemental material we present details of the numerical simulation for the two-terminal conductance in Q1D and the point contact conductance in 2D.

\end{widetext}

\clearpage

\section{Two-terminal conductance in Q1D}

\subsection{Scaling analysis for two-terminal conductances in Q1D}

As we discuss in the main text of our paper, in the quasi-one dimensional (Q1D) system of length $L$ and width $M$ the scaling function for the $q$-th moment of the two-terminal conductance $g$, is approximated as
\begin{eqnarray}
\left\langle g^q\right\rangle &=&
c^\prime_q \exp\left[ -\pi \left( X_q^\prime + \sum_{n=1}^{N_p} a_q^{\prime(n)} M^{n y^\prime}\right)
\frac{L}{M} \right] \nonumber \\
&+& d_q^\prime \exp\left[ - \pi
\left( \gamma_q^\prime + \sum_{n=1}^{N_s} b_q^{\prime(n)} M^{n y^\prime}\right)
\frac{L}{M}
\right],
\label{eq:Q1D_gq}
\end{eqnarray}
where all the parameters can be used in the fitting procedure. Similarly, for the quantity $p_q(g) = P_{-q}(2/g-1)$ which involves the Legendre function $P_{-q}(x)$, the corresponding scaling function is approximated by
\begin{eqnarray}
\left\langle p_{q}(g)\right\rangle
&\equiv&
\left\langle P_{-q}(2/g-1)\right\rangle=
\nonumber \\
&&
\bar{c}_q^\prime \exp\left[ -\pi
\left( \bar{X}_q^\prime + \sum_{n=1}^{N_p} \bar{a}_q^{\prime(n)}
 M^{n \bar{y}^\prime}\right)
\frac{L}{M}
\right] \nonumber \\
&+&
\bar{d}_q^\prime \exp\left[ - \pi
\left( \bar{\gamma}_q^\prime + \sum_{n=1}^{N_s}
 \bar{b}_q^{\prime(n)} M^{n \bar{y}^\prime}\right)
\frac{L}{M}
\right].
\label{eq:Q1D_Legendre}
\end{eqnarray}
The coefficients $c_q^\prime, d_q^\prime, \bar{c}_q^\prime,$ and $\bar{d}_q^\prime$ exhibit a weak dependence on the system width $M$ \cite{sm:Bondesan12}, and we take this fact into account by keeping terms in the Taylor expansion of these coefficients in $1/M$ up to the first order:
\begin{eqnarray}
\begin{split}
c_q^\prime &\rightarrow& c_q^\prime + \frac{c_q^{\prime(1)}}{M},\quad
d_q^\prime \rightarrow d_q^\prime + \frac{d_q^{\prime(1)}}{M}, \\
\bar{c}_q^\prime  &\rightarrow& \bar{c}_q^\prime + \frac{\bar{c}_q^{\prime(1)}}{M},\quad
\bar{d}_q^\prime \rightarrow \bar{d}_q^\prime +
 \frac{\bar{d}_q^{\prime(1)}}{M}.
\end{split}
\label{eq:c_d_M}
\end{eqnarray}

In our numerical simulations, we calculated two-terminal conductances in Q1D systems of widths $M = 32, 48, 64, 96, 128, 192, 256, 384, 512$ and maximum length $L_\text{max}=10M$. The number of samples for each $M$ is $10^6$. We obtain the fitting parameters from the scaling function (\ref{eq:Q1D_gq})--(\ref{eq:c_d_M}) by varying $M$ and $L$. We prepare four data sets with the same minimum width $M_\text{min}=32$, but different maximum widths;
$M_\text{max}=192,256,384,512$. Since the nonlinear fitting to the functions (\ref{eq:Q1D_gq})--(\ref{eq:c_d_M}) strongly depends on the initial values of the fitting parameters, we quantify the goodness of fitting by calculating the $\chi^2$ value for each fitting trial. We had more than 1000 fitting trials for each input data set starting with different initial fitting parameters chosen at random, and found the most reliable fitting which gives the minimum $\chi^2$. This complicated analysis is clearly displayed by using the so-called ``stability map'' \cite{sm:Obuse12}.

\begin{figure*}[h]
\rotatebox{90}{
\begin{minipage}{1.0\textheight}
\vspace{-0.5cm}
\flushleft
(a) $q=-0.2$
\begin{center}
\includegraphics[width=20cm]{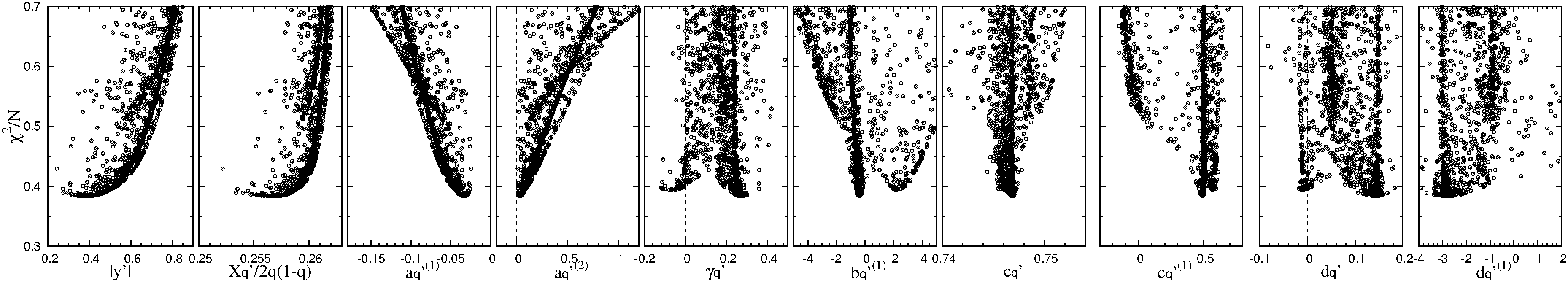}
\end{center}
\vspace{-0.5cm}
(b) $q=0.1$
\begin{center}
\includegraphics[width=20cm]{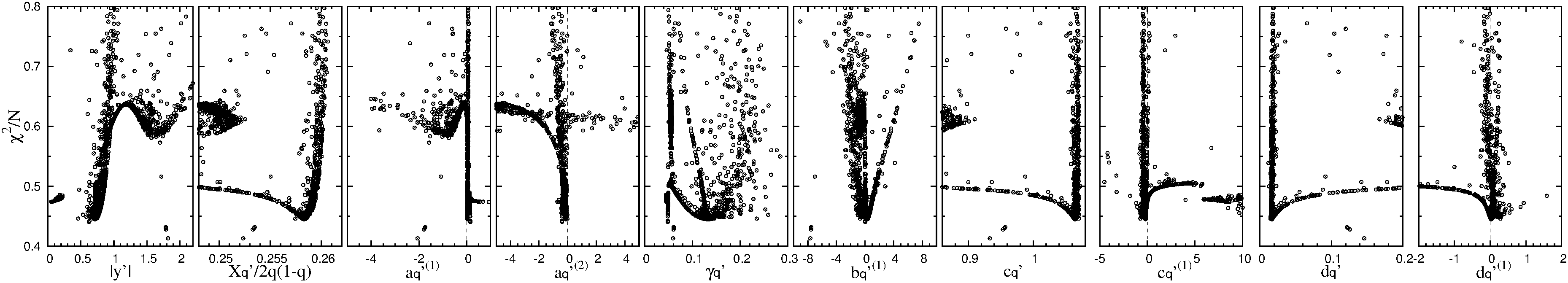}
\end{center}
\vspace{-0.5cm}
(c) $q=0.2$
\begin{center}
\includegraphics[width=20cm]{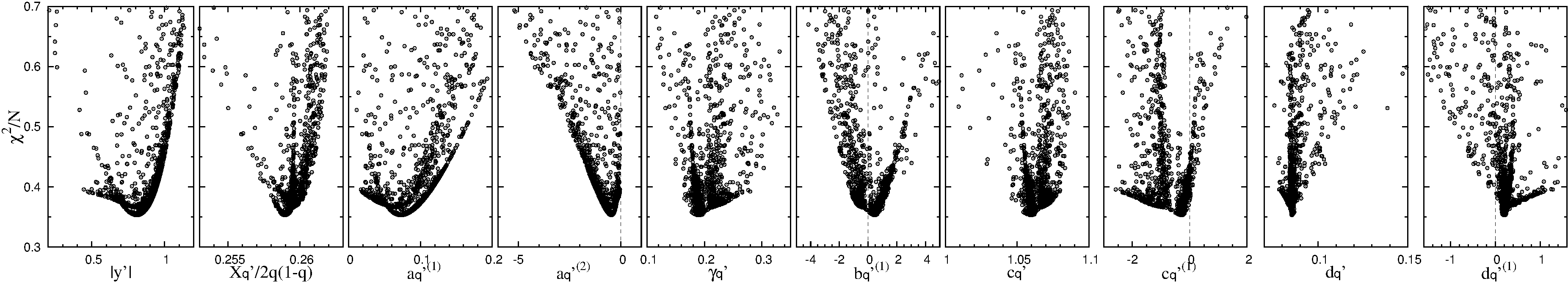}
\end{center}
\vspace{-0.5cm}
(d) $q=0.5$
\begin{center}
\includegraphics[width=20cm]{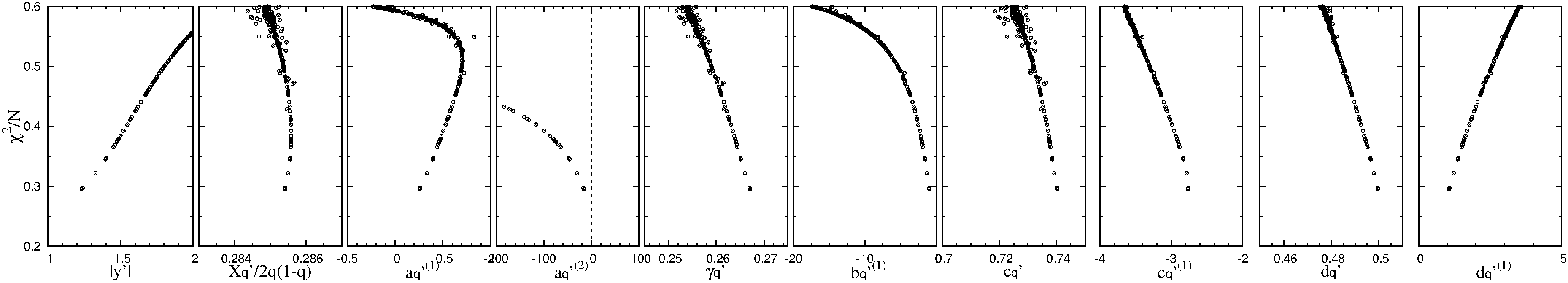}
\end{center}
\caption{
Stability maps from the scaling analysis of $\langle g^q \rangle$ in case of $M_\text{min}=32$ and $M_\text{max}=192$ for (a) $q=-0.2$, (b) $0.1$, (c) $0.2$, and (d) $0.5$. The values of fitting parameters at the global minimum are listed in Table. \ref{tab:fitting_gq_192}.
} \label{fig:stability_gq_192}
\end{minipage}
}
\end{figure*}

\begin{figure*}[h]
\rotatebox{90}{
\begin{minipage}{1.0\textheight}
\vspace{-0.5cm}
\flushleft{(a) $q=-0.2$}
\begin{center}
\includegraphics[width=19cm]{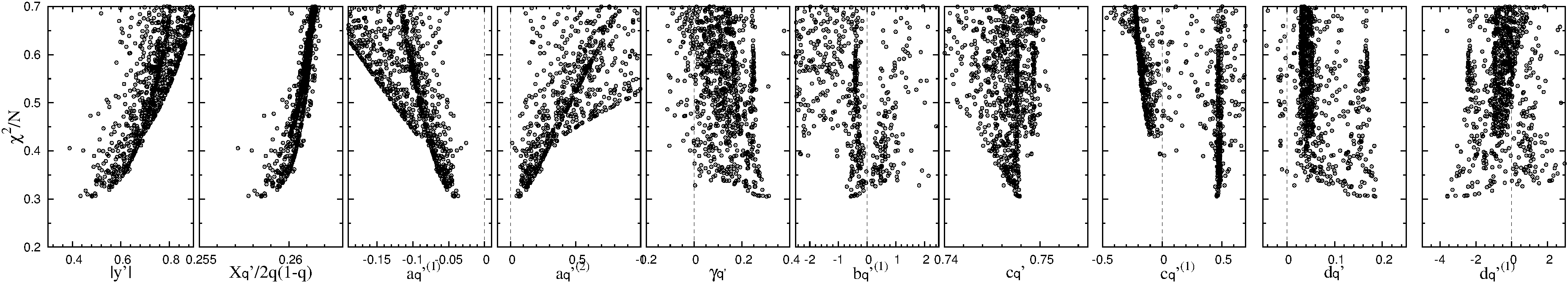}
\end{center}
\vspace{-0.5cm}
(b) $q=0.1$\\
\begin{center}
\includegraphics[width=19cm]{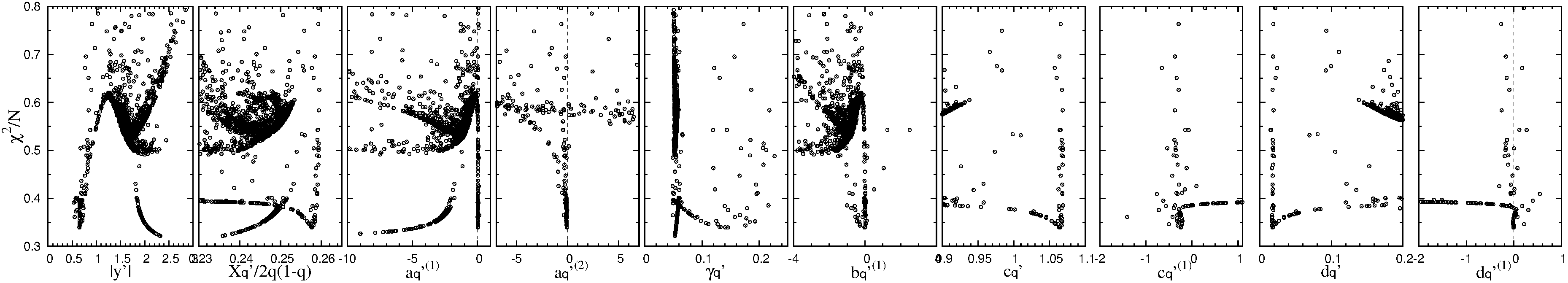}
\end{center}
\vspace{-0.5cm}
(c) $q=0.2$\\
\begin{center}
\includegraphics[width=19cm]{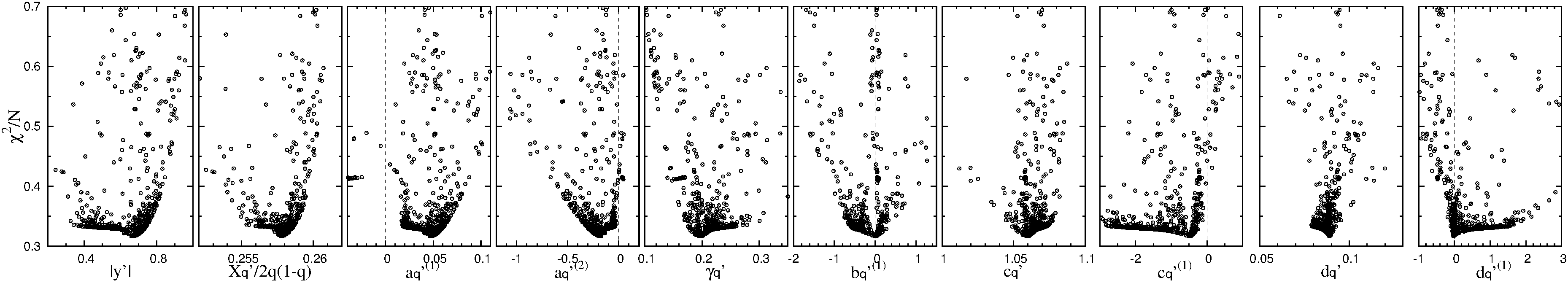}
\end{center}
\vspace{-0.5cm}
(d) $q=0.5$
\begin{center}
\includegraphics[width=19cm]{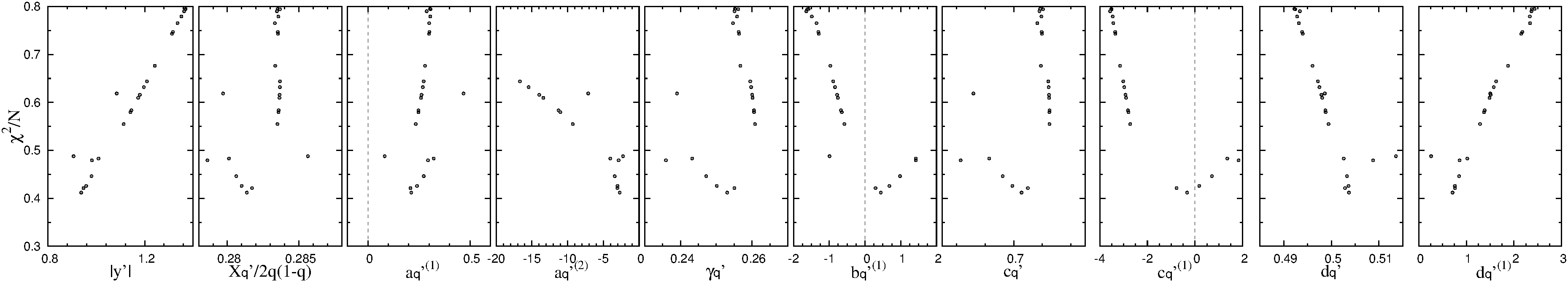}
\end{center}
\caption{
Stability maps from the scaling analysis of $\langle g^q \rangle$ in case of $M_\text{min}=32$ and $M_\text{max}=256$ for (a) $q=-0.2$, (b) $0.1$, (c) $0.2$, and (d) $0.5$. The values of fitting parameters at the global minimum are listed in Table. \ref{tab:fitting_gq_256}.
} \label{fig:stability_gq_256}
\end{minipage}
}
\end{figure*}

\begin{figure*}[h]
\rotatebox{90}{
\begin{minipage}{1.0\textheight}
\vspace{-0.5cm}
\flushleft{(a) $q=-0.2$}
\begin{center}
\includegraphics[width=19cm]{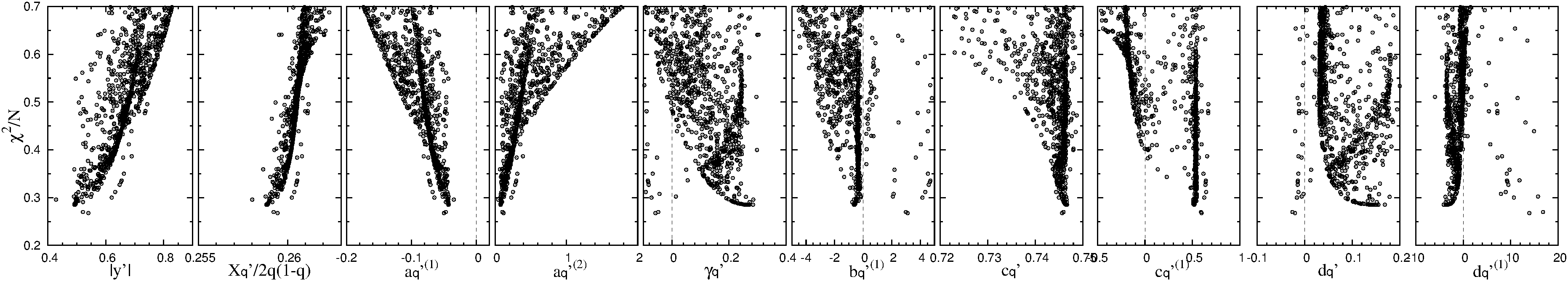}
\end{center}
\vspace{-0.5cm}
(b) $q=0.1$\\
\begin{center}
\includegraphics[width=19cm]{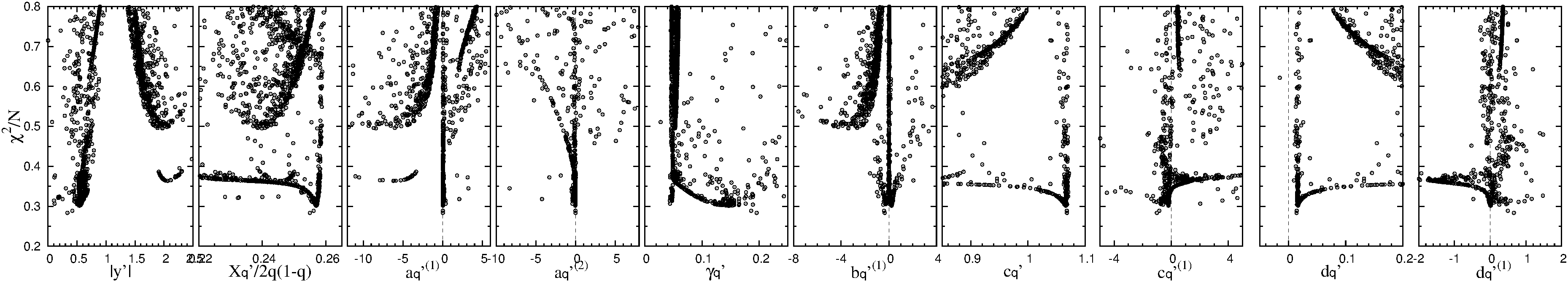}
\end{center}
\vspace{-0.5cm}
(c) $q=0.2$\\
\begin{center}
\includegraphics[width=19cm]{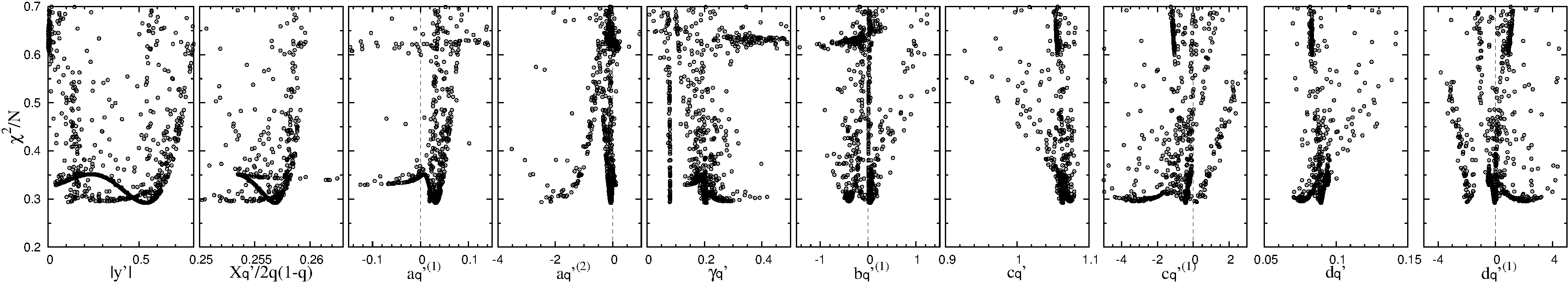}
\end{center}
\vspace{-0.5cm}
(d) $q=0.5$
\begin{center}
\includegraphics[width=19cm]{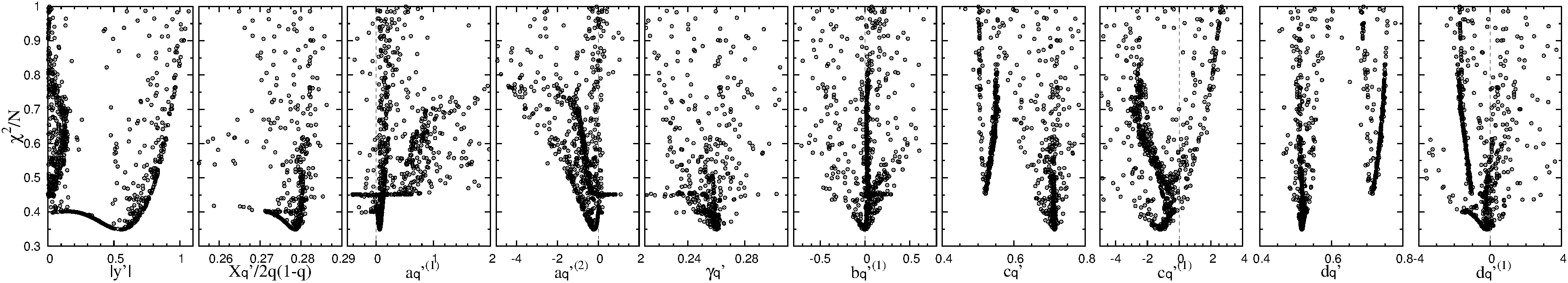}
\end{center}
\caption{
Stability maps from the scaling analysis of $\langle g^q \rangle$ in case of $M_\text{min}=32$ and $M_\text{max}=384$ for (a) $q=-0.2$, (b) $0.1$, (c) $0.2$, and (d) $0.5$. The values of fitting parameters at the global minimum are listed in Table. \ref{tab:fitting_gq_384}.
} \label{fig:stability_gq_384}
\end{minipage}
}
\end{figure*}

\begin{figure*}[h]
\rotatebox{90}{
\begin{minipage}{1.0\textheight}
\vspace{-0.5cm}
\flushleft{(a) $q=-0.2$}
\begin{center}
\includegraphics[width=19cm]{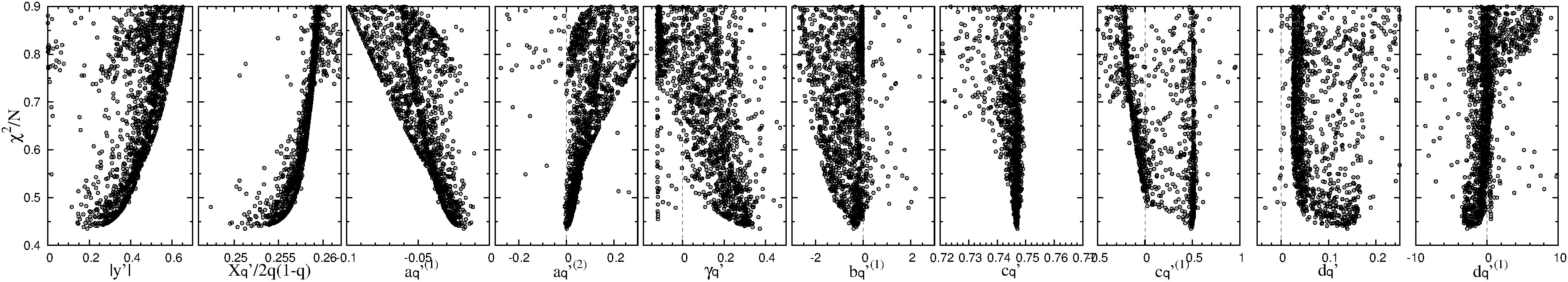}
\end{center}
\vspace{-0.5cm}
(b) $q=0.1$\\
\begin{center}
\includegraphics[width=19cm]{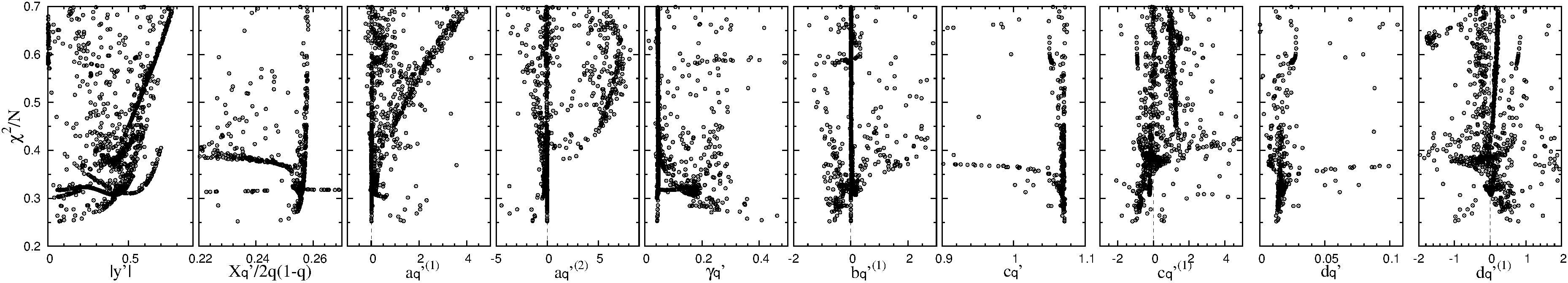}
\end{center}
\vspace{-0.5cm}
(c) $q=0.2$\\
\begin{center}
\includegraphics[width=19cm]{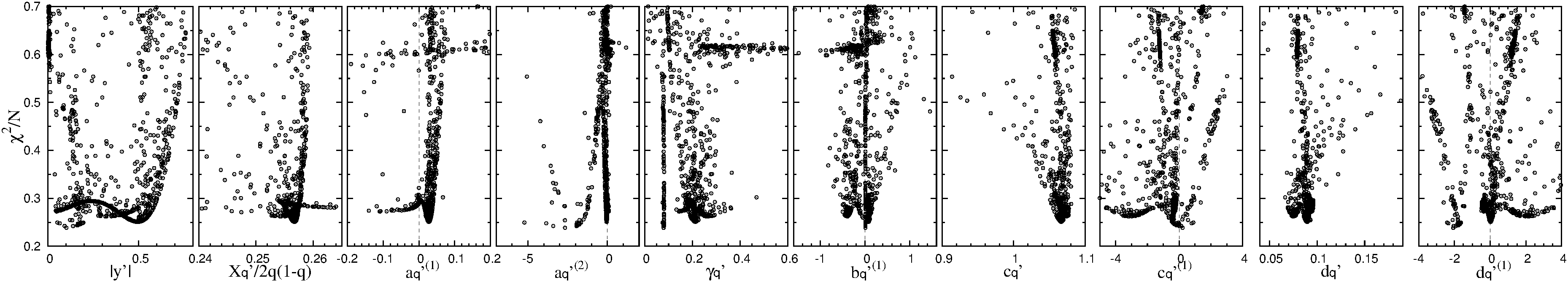}
\end{center}
\vspace{-0.5cm}
(d) $q=0.5$
\begin{center}
\includegraphics[width=19cm]{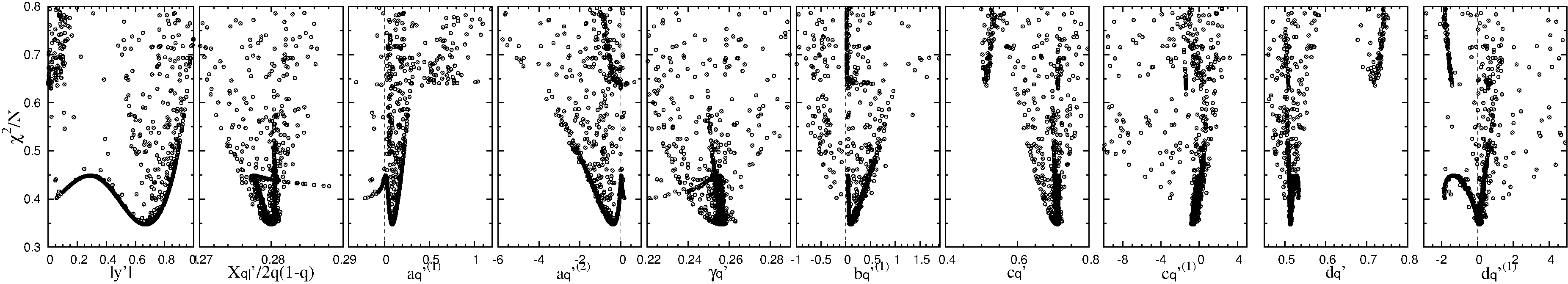}
\end{center}
\caption{
Stability maps from the scaling analysis of $\langle g^q \rangle$ in case of $M_\text{min}=32$ and $M_\text{max}=512$ for (a) $q=-0.2$, (b) $0.1$, (c) $0.2$, and (d) $0.5$. The values of fitting parameters at the global minimum are listed in Table. \ref{tab:fitting_gq_512}.
} \label{fig:stability_gq_512}
\end{minipage}
}
\end{figure*}

\begin{figure*}[h]
\rotatebox{90}{
\begin{minipage}{1.0\textheight}
\vspace{-0.5cm}
\flushleft{(a) $q=-0.2$}
\begin{center}
\includegraphics[width=19cm]{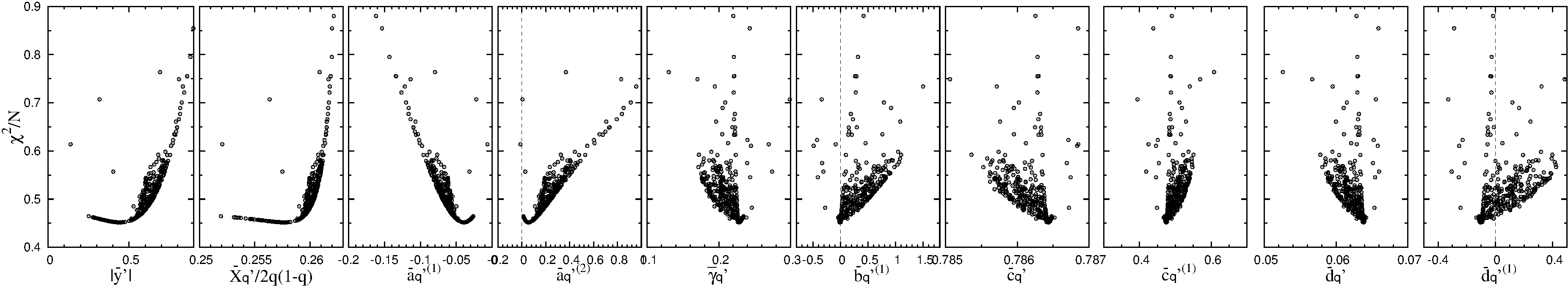}
\end{center}
\vspace{-0.5cm}
(b) $q=0.1$
\begin{center}
\includegraphics[width=19cm]{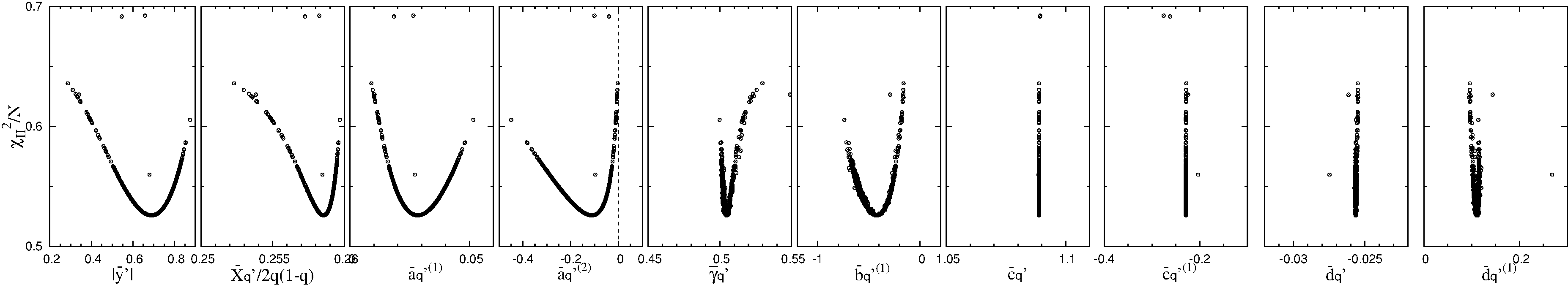}
\end{center}
\vspace{-0.5cm}
(c) $q=0.2$
\begin{center}
\includegraphics[width=19cm]{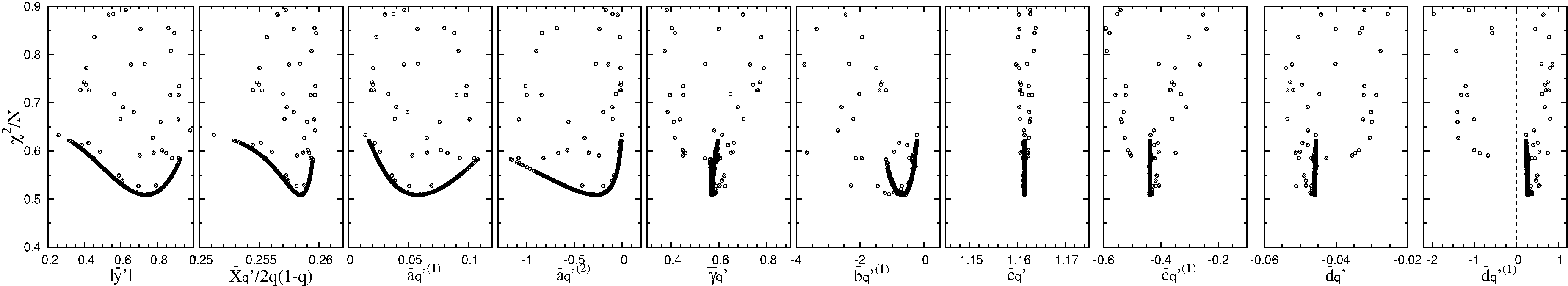}
\end{center}
\vspace{-0.5cm}
(d) $q=0.5$
\begin{center}
\includegraphics[width=19cm]{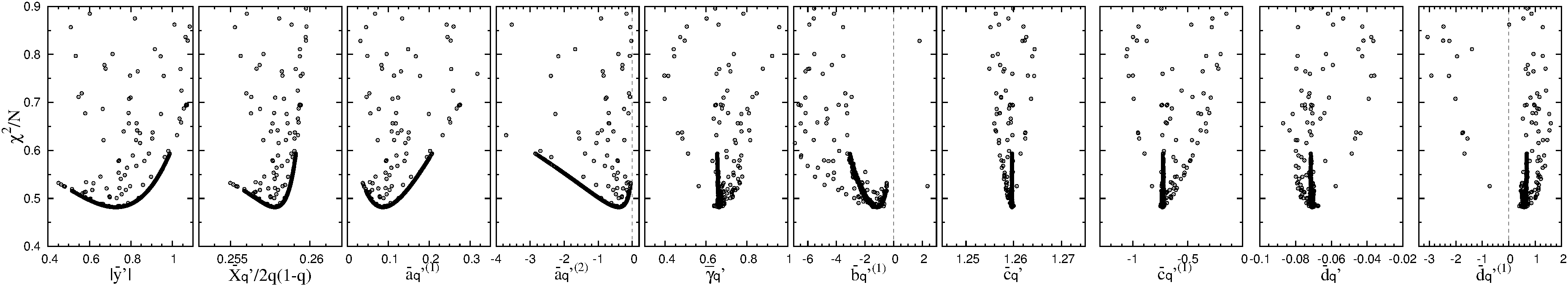}
\end{center}
\caption{
Stability maps from the scaling analysis of $\langle p_q(g) \rangle$ in case of $M_\text{min}=32$ and $M_\text{max}=192$ for (a) $q=-0.2$, (b) $0.1$, (c) $0.2$, and (d) $0.5$. The values of fitting parameters at the global minimum are listed in Table. \ref{tab:fitting_Legendre_192}.
} \label{fig:stability_Legendre_192}
\end{minipage}
}
\end{figure*}

\begin{figure*}[h]
\rotatebox{90}{
\begin{minipage}{1.0\textheight}
\vspace{-0.5cm}
\flushleft{(a) $q=-0.2$}
\begin{center}
\includegraphics[width=20cm]{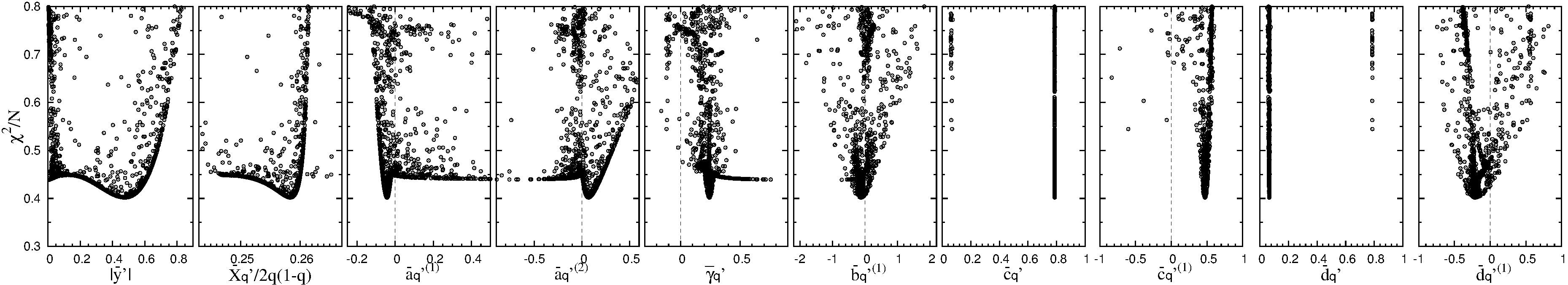}
\end{center}
\vspace{-0.5cm}
(b) $q=0.1$
\begin{center}
\includegraphics[width=20cm]{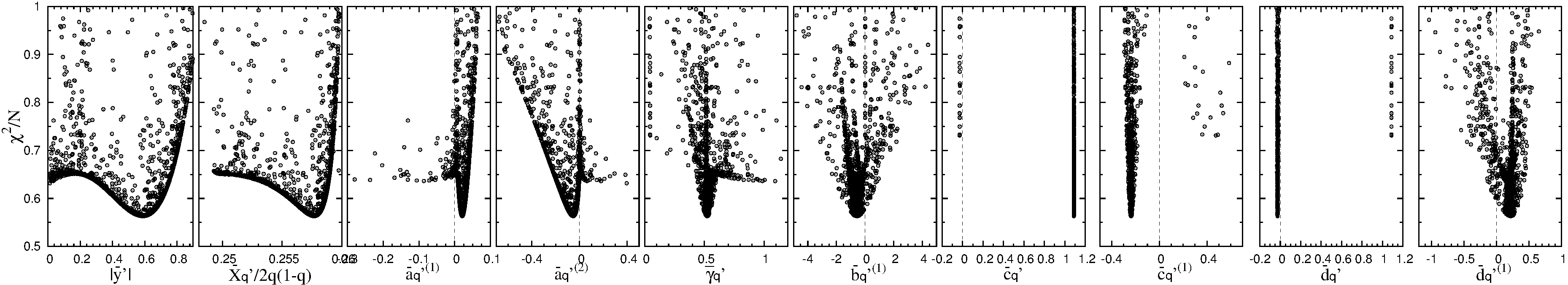}
\end{center}
\vspace{-0.5cm}
(c) $q=0.2$
\begin{center}
\includegraphics[width=20cm]{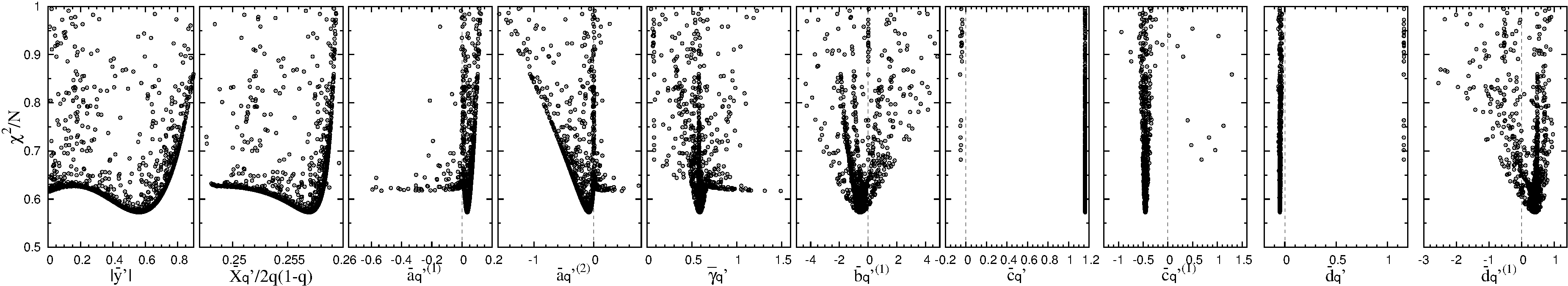}
\end{center}
\vspace{-0.5cm}
(d) $q=0.5$
\begin{center}
\includegraphics[width=20cm]{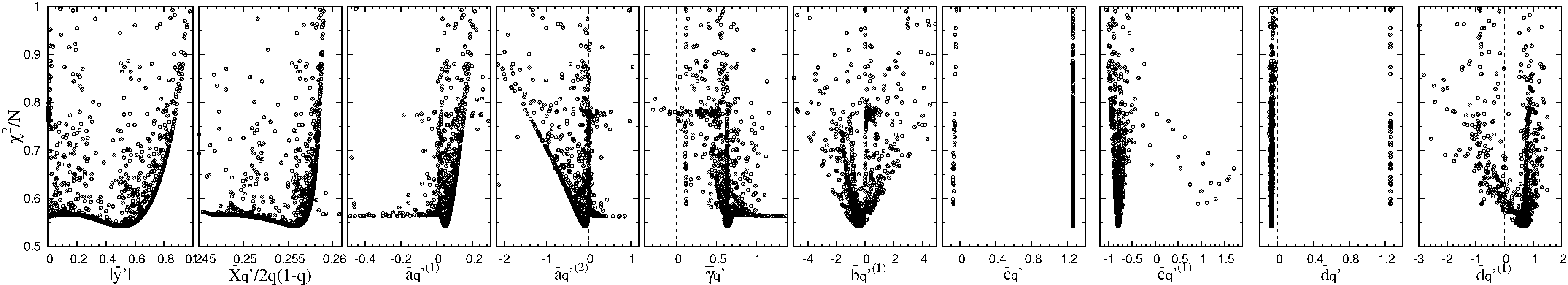}
\end{center}
\caption{
Stability maps from the scaling analysis of $\langle p_q(g) \rangle$ in case of $M_\text{min}=32$ and $M_\text{max}=256$ for (a) $q=-0.2$, (b) $0.1$, (c) $0.2$, and (d) $0.5$. The values of fitting parameters at the global minimum are listed in Table. \ref{tab:fitting_Legendre_256}.
}
\label{fig:stability_Legendre_256}
\end{minipage}
}
\end{figure*}

\begin{figure*}[h]
\rotatebox{90}{
\begin{minipage}{1.0\textheight}
\vspace{-0.5cm}
\flushleft{(a) $q=-0.2$}
\begin{center}
\includegraphics[width=19cm]{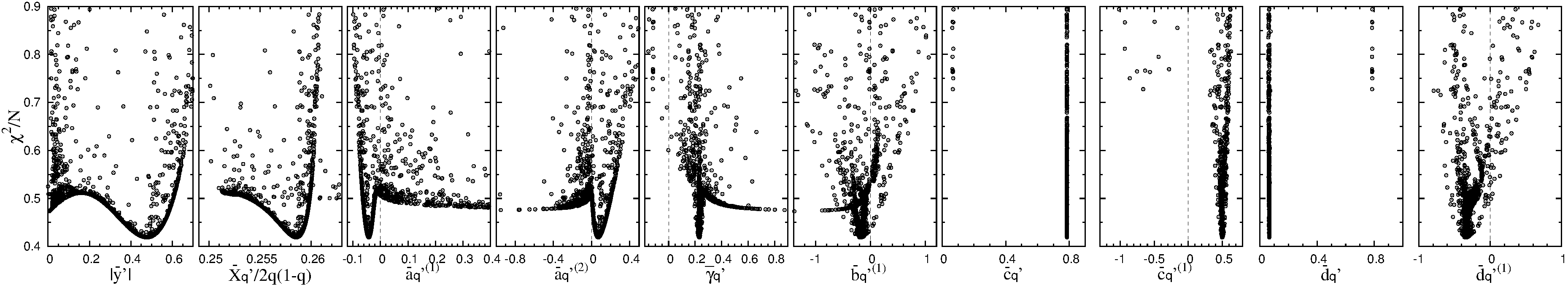}
\end{center}
\vspace{-0.5cm}
(b) $q=0.1$
\begin{center}
\includegraphics[width=19cm]{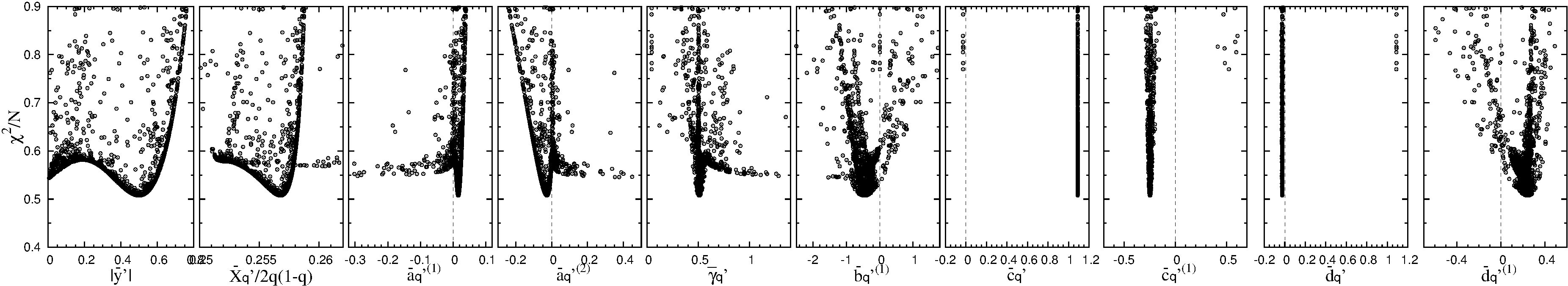}
\end{center}
\vspace{-0.5cm}
(c) $q=0.2$
\begin{center}
\includegraphics[width=19cm]{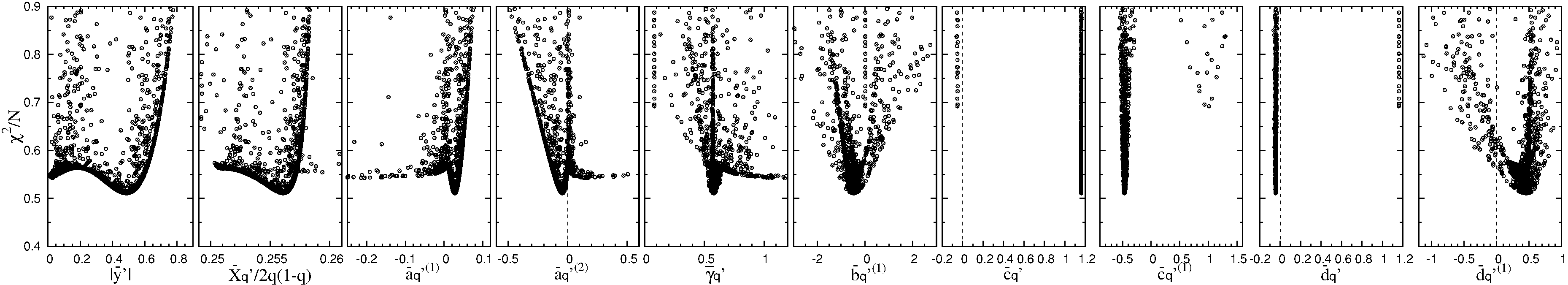}
\end{center}
\vspace{-0.5cm}
(d) $q=0.5$
\begin{center}
\includegraphics[width=19cm]{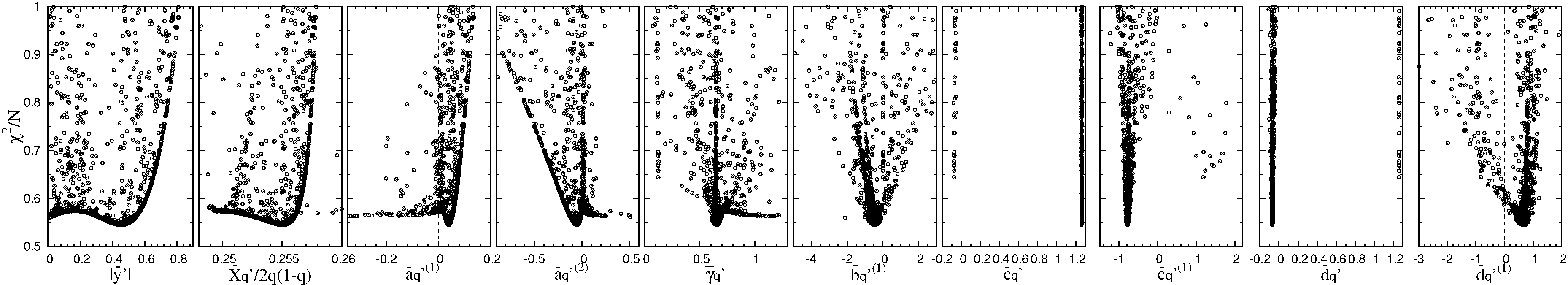}
\end{center}
\caption{
Stability maps from the scaling analysis of $\langle p_q(g) \rangle$ in case of $M_\text{min}=32$ and $M_\text{max}=384$ for (a) $q=-0.2$, (b) $0.1$, (c) $0.2$, and (d) $0.5$. The values of fitting parameters at the global minimum are listed in Table. \ref{tab:fitting_Legendre_384}.
} \label{fig:stability_Legendre_384}
\end{minipage}
}
\end{figure*}

\begin{figure*}[h]
\rotatebox{90}{
\begin{minipage}{1.0\textheight}
\vspace{-0.5cm}
\flushleft{(a) $q=-0.2$}
\begin{center}
\includegraphics[width=20cm]{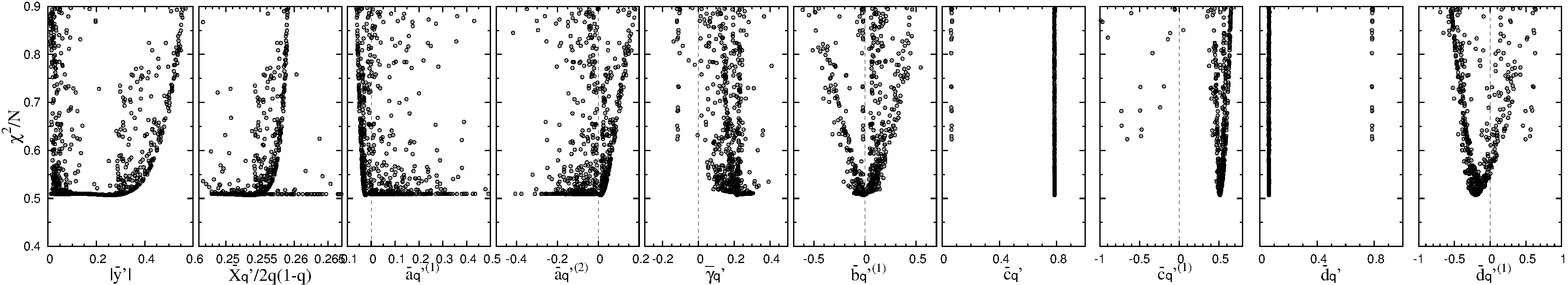}
\end{center}
\vspace{-0.5cm}
(b) $q=0.1$
\begin{center}
\includegraphics[width=20cm]{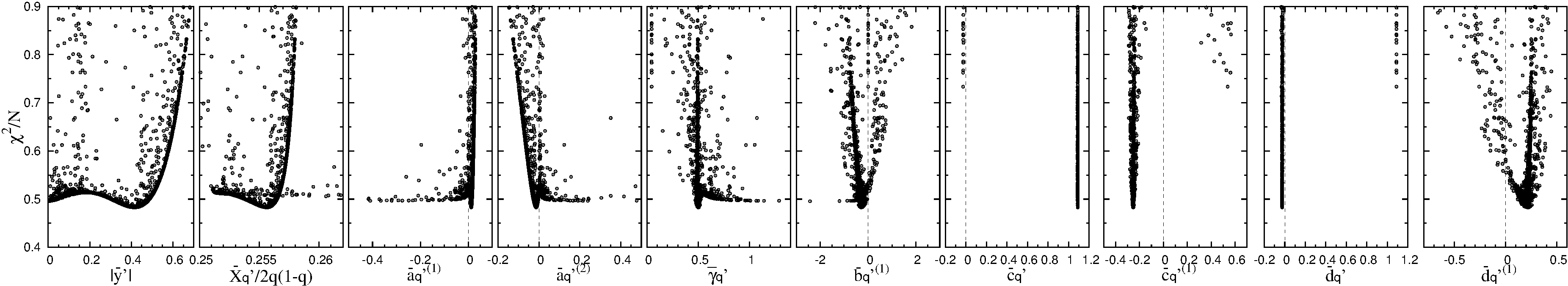}
\end{center}
\vspace{-0.5cm}
(c) $q=0.2$
\begin{center}
\includegraphics[width=20cm]{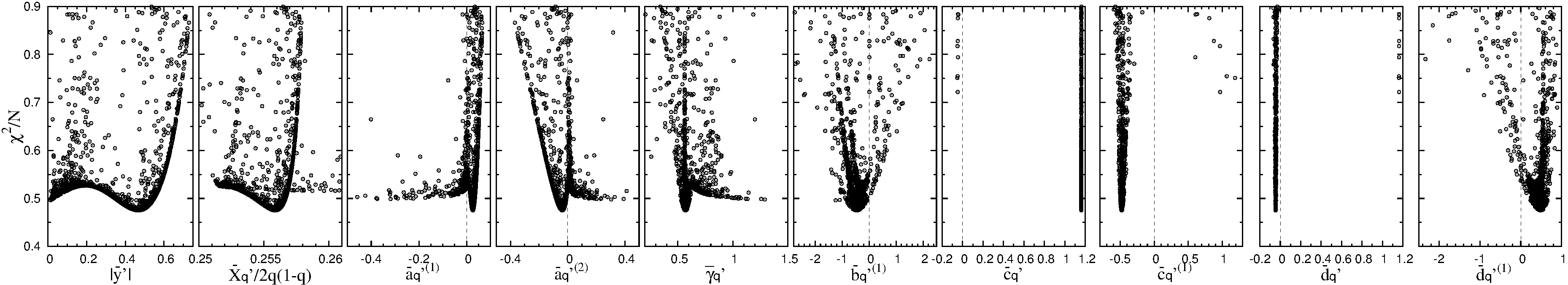}
\end{center}
\vspace{-0.5cm}
(d) $q=0.5$
\begin{center}
\includegraphics[width=20cm]{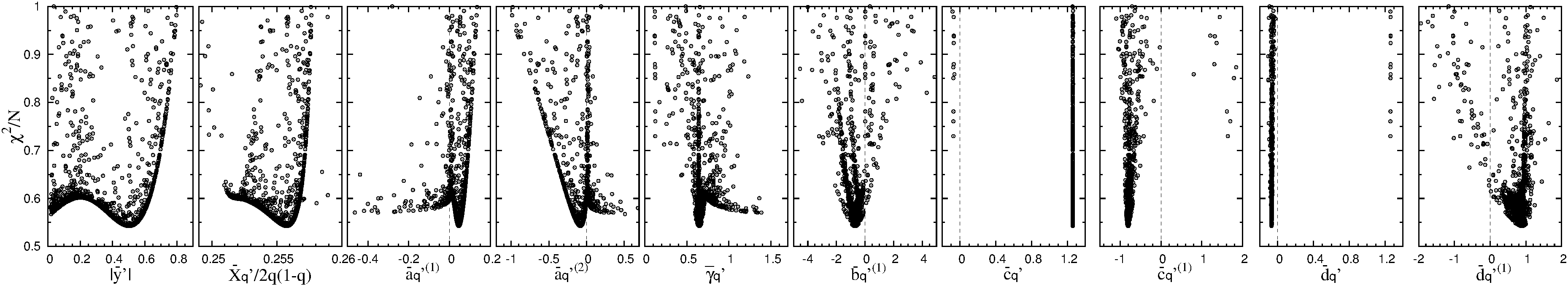}
\end{center}
\caption{
Stability maps from the scaling analysis of $\langle p_q(g) \rangle$ in case of $M_\text{min}=32$ and $M_\text{max}=512$ for (a) $q=-0.2$, (b) $0.1$, (c) $0.2$, and (d) $0.5$. The values of fitting parameters at the global minimum are listed in Table. \ref{tab:fitting_Legendre_512}.
} \label{fig:stability_Legendre_512}
\end{minipage}
}
\end{figure*}

The most reliable fitting was obtained when we chose $N_p=2$ and $N_s=1$ in Eqs.\ (\ref{eq:Q1D_gq}) and (\ref{eq:Q1D_Legendre}). The details of the most reliable fitting with different $M_\text{max}$
for $\langle g^q \rangle$  and $\langle p_q(g)\rangle$ are summarized in Tables  \ref{tab:fitting_gq_192}--\ref{tab:fitting_gq_512} and \ref{tab:fitting_Legendre_192}--\ref{tab:fitting_Legendre_512}, respectively. The corresponding stability maps are shown in Figs.\ \ref{fig:stability_gq_192}--\ref{fig:stability_gq_512} for $\langle g^q \rangle$ and Figs.\ \ref{fig:stability_Legendre_192}--\ref{fig:stability_Legendre_512} for $\langle p_q(g) \rangle$.

From the stability maps, we find that the irrelevant exponents $y^\prime$ and $\bar{y}^\prime$ are broadly
distributed in the interval $[0,1]$. For larger values of $M_\text{max}$, the $\chi^2$ exhibits two minima as a function of $y^\prime$ and $\bar{y}^\prime$. The global (deeper) minimum of the $\chi^2$ as a function of $\bar{y}^\prime$ is located at $|\bar{y}^\prime| \approx 1/2$ for all $q$ in the range of $0<q<1$. We remark that at this minimum the coefficients $a_q^{\prime(1)}$ and $a_q^{\prime(2)}$ (also $\bar{a}_q^{\prime(1)}$ and $\bar{a}_q^{\prime(2)}$) take on opposite signs (except near $q\simeq 0$ for $\langle g^q\rangle$ where the scaling analysis becomes unstable). The other minimum, with a higher value of $\chi^2$, is located in the region of very small values of $|y^\prime|$ or $|\bar{y}^\prime|$, consistent with $|y|\sim 0$, as found previously by other groups. However, at this (higher) minimum with larger $\chi^2$, the values of $a_q^{\prime(2)}$ and $\bar{a}_q^{\prime(2)}$ become almost zero. Consistently, If we set $N_p=1$ (in other words, $a_q^{\prime(2)}$ and $\bar{a}_q^{\prime(2)}$ are fixed to zero), we obtain $|y|$ close to zero, but a larger $\chi^2$. This observation is completely analogous to what we saw in the scaling analysis for the Lyapunov exponent in Ref.\
\cite{sm:Obuse12}.

We also remark that when fitting data for $\langle g^q \rangle$, the $\chi^2$ as a function $\gamma_q^\prime$ possesses two local minima, with close values of $\chi^2$, when $q < 0.5$. One minimum is located at larger $\gamma_q^\prime$, where most of the fitting results lie (a dense cloud of points). The other minimum at smaller $\gamma_q^\prime$ is comprised of a smaller number of points, see Figs.\ \ref{fig:stability_gq_192} - \ref{fig:stability_gq_512}. This makes the reliable determination of $\gamma_q^\prime$ difficult. In contrast, for $\langle p_q(g) \rangle$, the $\chi^2$ clearly shows a well defined global minimum as a function of $\bar{\gamma}_q^\prime$, as shown in Figs.\
\ref{fig:stability_Legendre_192}--\ref{fig:stability_Legendre_512}. Since the larger $\gamma_q^\prime$ is close to $\bar{\gamma}_q^\prime$, we believe that the larger $\gamma_q^\prime$ might be closer to the correct results. However, a further careful analysis is needed to firmly establish this.

\subsection{Practical error bars}

\begin{figure*}[h]
\flushleft
\vspace{-0.5cm}\hspace{1cm}(a)
\begin{center}
\includegraphics[width=15cm]{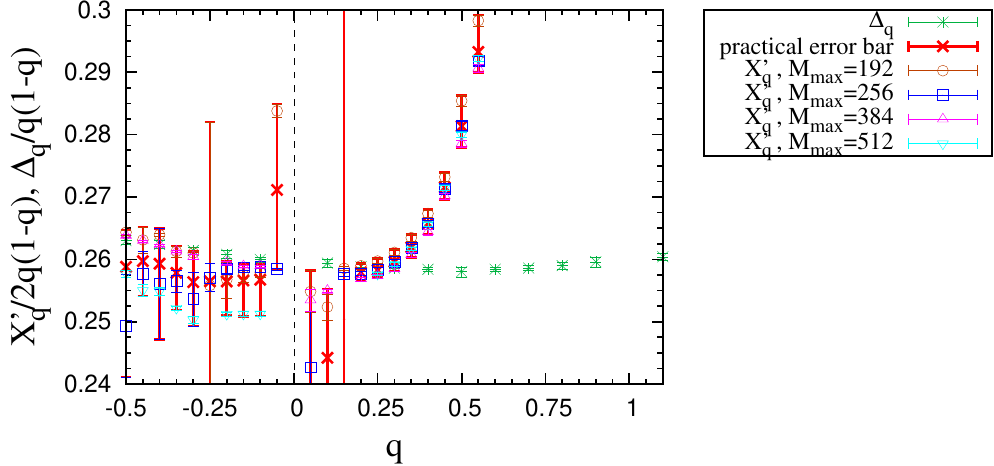}\\
\end{center}
\vspace{-0.5cm}\hspace{1cm}(b)
\begin{center}
\includegraphics[width=15cm]{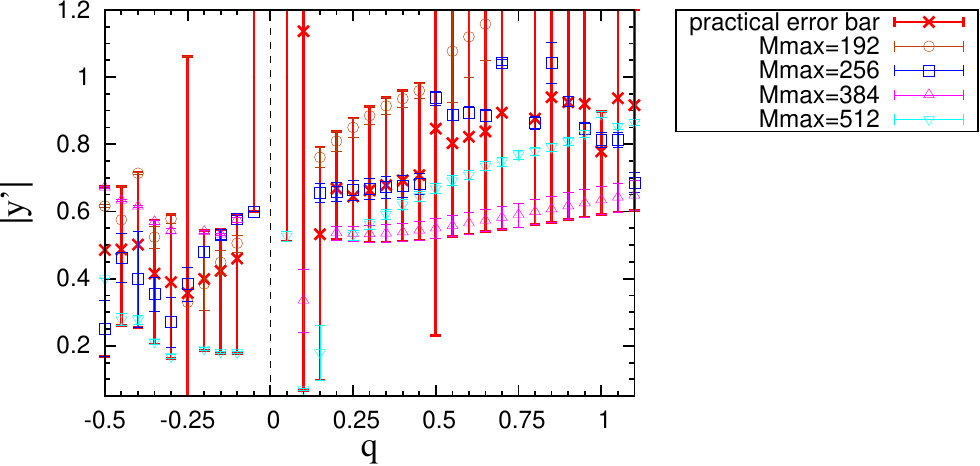}\\
\end{center}
\vspace{-0.5cm}\hspace{1cm}(c)
\begin{center}
\includegraphics[width=15cm]{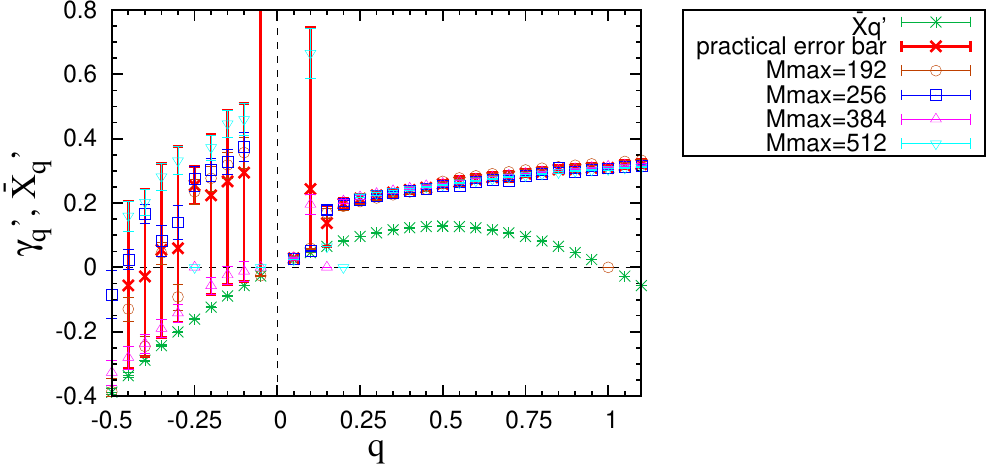}
\end{center}
\caption{
$q$ dependence of (a) $X_q^\prime$, (b) $y^\prime$, and (c) $\gamma_q^\prime$ from $\langle g^q \rangle$. The error bars by thin and thick lines represent the error bars estimated from the error-propagation theory and the practical error bars, respectively.
} \label{fig:practical_gq}
\end{figure*}

\begin{figure*}[h]
\flushleft
\vspace{-0.5cm}\hspace{1cm}(a)
\begin{center}
\includegraphics[width=15cm]{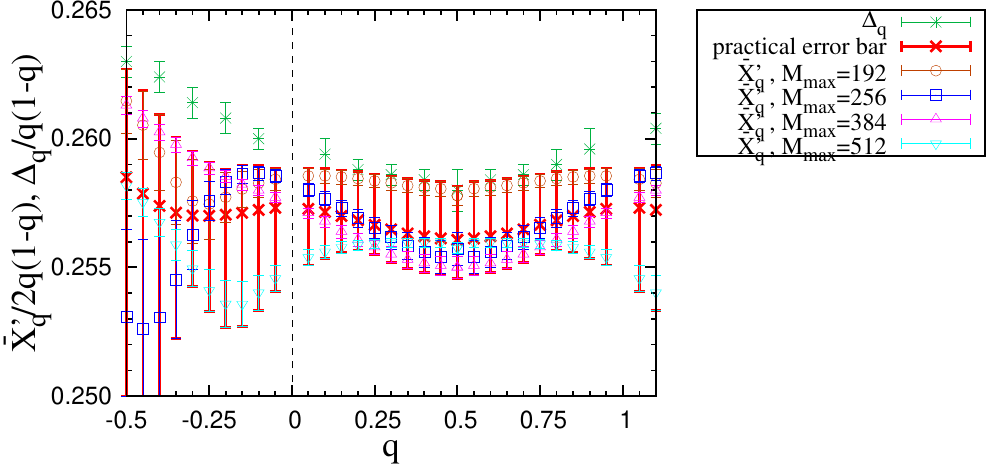}
\end{center}
\vspace{-0.5cm}\hspace{1cm}(b)
\begin{center}
\includegraphics[width=15cm]{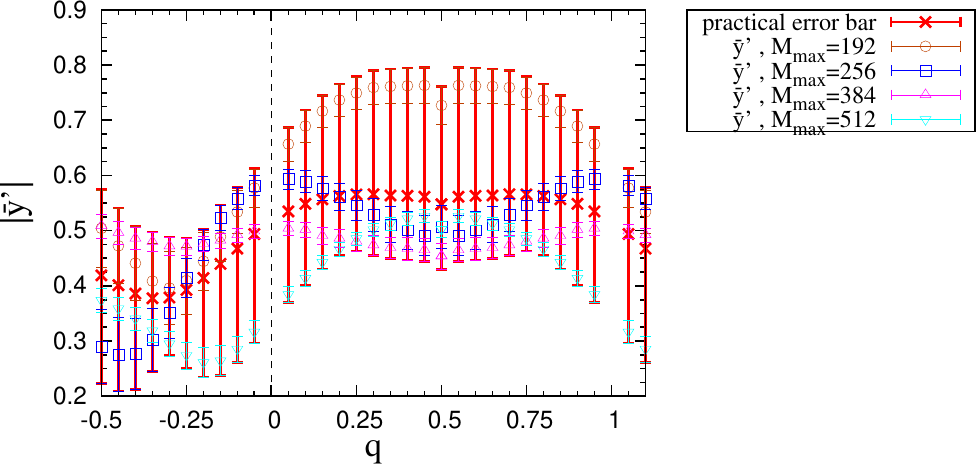}
\end{center}
\vspace{-0.5cm}\hspace{1cm}(c)
\begin{center}
\includegraphics[width=15cm]{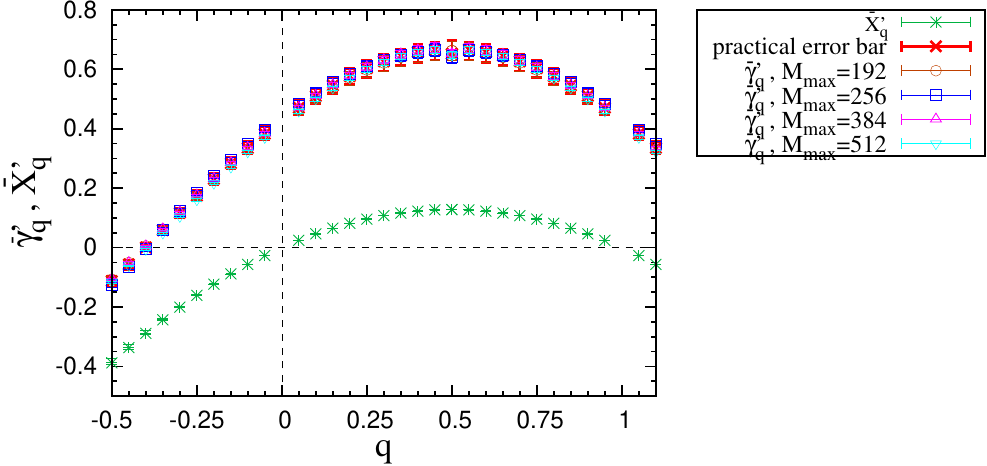}
\end{center}
\caption{
$q$ dependence of (a) $\bar{X}_q^\prime$, (b) $\bar{y}^\prime$, and (c) $\bar{\gamma}_q^\prime$ from $\langle p_q(g) \rangle$. The thin and thick lines represent the error bars estimated from the error-propagation theory and the practical error bars, respectively.
} \label{fig:practical_Legendre}
\end{figure*}

The $q$ dependence of the numerically obtained exponents $X_q^\prime$, $y^\prime$, and $\gamma_q^\prime$ is shown in Fig.\ \ref{fig:practical_gq}. Fig.\ \ref{fig:practical_Legendre} shows the $q$ dependence of the exponents $\bar{X}_q^\prime$, $\bar{y}^\prime$, and $\bar{\gamma}_q^\prime$. Both figures show that the exponents obtained from different data sets do not agree with each other even when the error bars (estimated from the error-propagation theory, thin lines) are taken into account. We believe that the insufficient truncations for the irrelevant exponent $y$ and the subleading scaling dimension $\gamma_q$ of the scaling functions give rise to this inconsistency, while increasing the truncation orders is impossible with our numerical accuracy. Therefore, as a conservative upper bound for the errors, we introduce the ``practical error bars'' that represent the union of all ``statistical error bars'' obtained from the different data sets, as shown by the thick red lines in Figs.\ \ref{fig:practical_gq} and \ref{fig:practical_Legendre}. The red cross symbol represents the average of the mean values obtained from all data sets.


\begin{largetable}
\footnotesize{
\rotatebox{90}{
\begin{minipage}[t]{1.0\textheight}
\caption{
The details of the most reliable fitting by the scaling analysis for the $q$th moment of Q1D with the minimum width $M_\text{min}=32$ and the maximum width $M_\text{max}=192$. The aspect ratio is varied from $L/M=4$ to $10$ and the number of total data point $N$ is $376$. The scaling function in Eq.\ (\ref{eq:Q1D_gq}) with $N_p=2$ and $N_s=1$ is employed. $\chi_\text{min}^2$, and $Q$ in the tables represent the minimum value of $\chi^2$, and the goodness of fit, respectively. In the table, the value with $\pm$ means the error bar for its above value.
\label{tab:fitting_gq_192}
}
\begin{tabular}{r |r r |r  r| r| r r r| r  r | r r}
\hline\hline
$q\ $ &
$X_q^\prime\quad$ & $c_q^\prime\quad\ $ &
$\gamma_q^\prime\quad$ & $d_q^\prime\quad\ $ &
$|y^\prime|\quad\ $ &
$c_q^{\prime(1)}\quad $&  $a_q^{\prime(1)}\ $ & $a_q^{\prime(2)}\ $ &
$d_q^{\prime(1)}\quad $&  $b_q^{\prime(1)}\ $  &$\chi_\text{min}^2/N$ & $Q$\\
\hline
$-0.5$ & $-0.396484$& $0.334988$& $-0.385740$& $-0.037842$& $0.615961$& $1.512426$& $-0.191301$& $0.864221$& $21.300731$& $4.329661$&
$0.49$& $1.0$\\
& $\pm0.000583$& $\pm0.004119$& $\pm0.041406$& $\pm0.021902$& $\pm0.003918$& $\pm0.151471$& $\pm0.012190$& $\pm0.923938$& $\pm6.374798$& $\pm0.376499$&  &  \\
$-0.4$ & $-0.295630$& $0.456913$& $-0.245458$& $-0.023908$& $0.713652$& $1.300010$& $-0.186081$& $1.098436$& $16.920341$& $5.527381$&
 $0.44$& $1.0$\\
& $\pm0.000281$& $\pm0.002763$& $\pm0.031209$& $\pm0.013898$& $\pm0.003596$& $\pm0.102123$& $\pm0.004999$& $\pm0.847415$& $\pm4.070400$& $\pm0.419233$&  &  \\
$-0.3$ & $-0.203606$& $0.600194$& $-0.091364$& $0.003585$& $0.578024$& $0.835473$& $-0.091235$& $0.269477$& $14.502444$& $2.760727$&
 $0.40$& $1.0$\\
& $\pm0.000173$& $\pm0.001332$& $\pm0.039858$& $\pm0.015298$& $\pm0.012599$& $\pm0.057955$& $\pm0.022121$& $\pm0.802268$& $\pm4.063576$& $\pm0.356532$&  &  \\
$-0.2$ & $-0.123226$& $0.746903$& $0.276641$& $0.146892$& $0.384617$& $0.491493$& $-0.033404$& $0.035837$& $-2.953260$& $-0.406236$&
 $0.38$& $1.0$\\
& $\pm0.001385$& $\pm0.000531$& $\pm0.038642$& $\pm0.040293$& $\pm0.078072$& $\pm0.033022$& $\pm0.050944$& $\pm0.168455$& $\pm1.094452$& $\pm0.139222$&  &  \\
$-0.1$ & $-0.056740$& $0.885483$& $0.356603$& $0.079055$& $0.505266$& $0.250012$& $-0.020543$& $0.037323$& $-1.557074$& $-0.466589$& $0.45$& $1.0$\\
& $\pm0.000159$& $\pm0.000239$& $\pm0.047604$& $\pm0.029580$& $\pm0.025619$& $\pm0.015214$& $\pm0.007365$& $\pm0.035740$& $\pm0.795048$& $\pm0.267361$&  &  \\
$0.1$ & $0.045423$& $0.933196$& $0.059392$& $0.145973$& $1.819709$& $-12.443617$& $-2.059302$& $2962.0$& $12.073200$& $-7.657070$&  $0.41$& $1.0$\\
& $\pm0.000373$& $\pm0.044217$& $\pm0.001791$& $\pm0.044286$& $\pm0.042547$& $\pm1.166978$& $\pm0.843347$& $\pm760.7$& $\pm1.164921$& $\pm1.303236$&  &  \\
$0.2$ & $0.082870$& $1.059712$& $0.189950$& $0.085366$& $0.809381$& $-0.283963$& $0.072965$& $-0.464585$& $0.203530$& $0.467173$&  $0.35$& $1.0$\\
& $\pm0.000118$& $\pm0.004121$& $\pm0.010884$& $\pm0.002654$& $\pm0.029176$& $\pm0.170854$& $\pm0.011911$& $\pm0.117745$& $\pm0.207244$& $\pm0.287221$&  &  \\
$0.3$ & $0.109664$& $0.976775$& $0.215777$& $0.211366$& $0.885299$& $-0.164929$& $0.125748$& $-1.078541$& $0.373257$& $0.594414$& $0.28$& $1.0$\\
& $\pm0.000193$& $\pm0.006798$& $\pm0.006812$& $\pm0.003975$& $\pm0.026777$& $\pm0.307484$& $\pm0.026669$& $\pm0.283988$& $\pm0.294275$& $\pm0.250716$&  &  \\
$0.4$ & $0.128321$& $0.843614$& $0.234724$& $0.363946$& $0.935891$& $0.110747$& $0.180456$& $-1.871690$& $0.535342$& $0.719350$&  $0.25$& $1.0$\\
& $\pm0.000290$& $\pm0.009104$& $\pm0.005187$& $\pm0.005022$& $\pm0.024327$& $\pm0.439973$& $\pm0.048966$& $\pm0.534315$& $\pm0.363941$& $\pm0.243008$&  &  \\
$0.5$ & $0.142707$& $0.740297$& $0.267093$& $0.499548$& $1.231197$& $-2.764682$& $0.259055$& $-16.035714$& $1.068127$& $-1.004037$&
 $0.30$& $1.0$\\
& $\pm0.000423$& $\pm0.007577$& $\pm0.003670$& $\pm0.005412$& $\pm1.000671$& $\pm0.331021$& $\pm2.237841$& $\pm88.658580$& $\pm0.407653$
& $\pm0.266620$& &  \\
$0.6$ & $0.151768$& $0.627423$& $0.284501$& $0.609876$& $1.120021$& $-3.141768$& $0.144743$& $-8.451723$& $0.700441$& $-0.872151$&  $0.32$& $1.0$\\
& $\pm0.000451$& $\pm0.007929$& $\pm0.003592$& $\pm0.006450$& $\pm0.119334$& $\pm0.433632$& $\pm0.404229$& $\pm4.747448$& $\pm0.441764$& $\pm0.236465$& &  \\
$0.7$ & $0.158097$& $0.530112$& $0.296865$& $0.683248$& $1.201198$& $-2.853750$& $0.160148$& $-14.984336$& $0.316265$& $-0.990592$& $0.39$& $1.0$\\
& $\pm0.000488$& $\pm0.007237$& $\pm0.003263$& $\pm0.007467$& $\pm0.105530$& $\pm0.379234$& $\pm0.548963$& $\pm7.321061$& $\pm0.464925$& $\pm0.280980$&  &  \\
$0.8$ & $0.162652$& $0.453488$& $0.307941$& $0.726259$& $1.264022$& $-2.528339$& $0.173028$& $-23.274193$& $-0.035229$& $-1.083277$& $0.46$& $1.0$\\
& $\pm0.000522$& $\pm0.006546$& $\pm0.003120$& $\pm0.008520$& $\pm0.102332$& $\pm0.334664$& $\pm0.729360$& $\pm10.609417$& $\pm0.490302$& $\pm0.329744$& &  \\
$0.9$ & $0.166012$& $0.392766$& $0.317282$& $0.745728$& $1.347204$& $-2.177898$& $0.212830$& $-40.816932$& $-0.248847$& $-1.204268$& $0.52$& $1.0$\\
& $\pm0.000542$& $\pm0.005839$& $\pm0.003020$& $\pm0.009373$& $\pm0.117205$& $\pm0.283894$& $\pm1.096200$& $\pm21.180793$& $\pm0.506087$& $\pm0.399400$&  &  \\
$1.1$ & $0.170613$& $0.306800$& $0.332747$& $0.742630$& $1.466717$& $-1.683694$& $0.288416$& $-91.947556$& $-0.508610$& $-1.409160$
&  $0.62$& $1.0$\\
& $\pm0.000582$& $\pm0.004818$& $\pm0.003029$& $\pm0.010748$& $\pm0.144936$& $\pm0.222427$& $\pm2.015521$& $\pm58.559267$& $\pm0.525351$
& $\pm0.553678$& &  \\
\hline\hline
\end{tabular}
\end{minipage}
}
}
\end{largetable}



\begin{table*}[p]
\footnotesize{
\rotatebox{90}{
\begin{minipage}[t]{1.0\textheight}
\caption{
The details of the most reliable fitting by the scaling analysis for the $q$th moment of Q1D two-terminal conductances, $\langle g^q \rangle$, with  the minimum width $M_\text{min}=32$ and the maximum width
$M_\text{max}=256$. The aspect ratio is varied from $L/M=4$ to $10$ and the number of total data point $N$ is $486$. The scaling function in Eq.\ (\ref{eq:Q1D_gq}) with $N_p=2$ and $N_s=1$ is employed. $\chi_\text{min}^2$, and $Q$ in the tables represent the minimum value of $\chi^2$, and the goodness of fit, respectively. In the table, the value with $\pm$ means the error bar for its above value.
\label{tab:fitting_gq_256}
}
\begin{tabular}{r |r r |r  r| r| r r r| r  r | r r}
\hline\hline
$q\ $ &
$X_q^\prime\quad$ & $c_q^\prime\quad\ $ &
$\gamma_q^\prime\quad$ & $d_q^\prime\quad\ $ &
$|y^\prime|\quad\ $ &
$c_q^{\prime(1)}\quad $&  $a_q^{\prime(1)}\ $ & $a_q^{\prime(2)}\ $ &
$d_q^{\prime(1)}\quad $&  $b_q^{\prime(1)}\ $  &$\chi_\text{min}^2/N$ & $Q$\\
\hline
$-0.5$ & $-0.374025$& $0.346534$& $-0.084504$& $0.185819$& $0.251868$& $0.940797$& $-0.141092$& $0.144466$&
$-1.251424$& $-0.025914$&  $0.58$& $1.0$\\
& $\pm0.012324$& $\pm0.002122$& $\pm0.074455$& $\pm0.071099$& $\pm0.082999$& $\pm0.118917$& $\pm0.008548$& $\pm0.
058412$& $\pm2.040957$& $\pm0.209068$&  &  \\
$-0.4$ & $-0.286756$& $0.473392$& $0.167188$& $0.102006$& $0.398169$& $-0.627689$& $-0.116813$& $0.160025$&
$-1.503558$& $-1.679846$&  $0.47$& $1.0$\\
& $\pm0.009950$& $\pm0.001203$& $\pm0.028322$& $\pm0.008591$& $\pm0.142818$& $\pm0.149515$& $\pm0.374925$& $\pm1.
175179$& $\pm0.288082$& $\pm0.124436$&  &  \\
$-0.3$ & $-0.197819$& $0.603422$& $0.140468$& $0.132614$& $0.270860$& $0.608887$& $-0.046220$& $0.030099$& $
-0.668488$& $-0.150138$& $0.35$& $1.0$\\
& $\pm0.003366$& $\pm0.000786$& $\pm0.053621$& $\pm0.034540$& $\pm0.074159$& $\pm0.057062$& $\pm0.021813$& $\pm0.
045291$& $\pm1.077597$& $\pm0.169112$& &  \\
$-0.2$ & $-0.124062$& $0.747853$& $0.303349$& $0.185041$& $0.480888$& $0.458064$& $-0.042083$& $0.069109$& $
-3.587427$& $-0.563118$& $0.31$& $1.0$\\
& $\pm0.000297$& $\pm0.000364$& $\pm0.036161$& $\pm0.053405$& $\pm0.016811$& $\pm0.029080$& $\pm0.015642$& $\pm0.
069764$& $\pm1.403837$& $\pm0.199787$& &  \\
$-0.1$ & $-0.056940$& $0.885701$& $0.374825$& $0.089072$& $0.578410$& $0.241196$& $-0.025016$& $0.062067$& $
-1.754863$& $-0.662210$& $0.33$& $1.0$\\
& $\pm0.000063$& $\pm0.000175$& $\pm0.044380$& $\pm0.033261$& $\pm0.011101$& $\pm0.013827$& $\pm0.004651$& $\pm0.
028487$& $\pm0.882509$& $\pm0.348013$&  &  \\
$0.1$ & $0.042460$& $0.557804$& $0.052706$& $0.521044$& $2.320982$& $-11.656885$& $-13.155058$& $167457.1$& $11.299898$& $-17.740150$& $0.32$& $1.0$\\
& $\pm0.001092$& $\pm0.118796$& $\pm0.001165$& $\pm0.118722$& $\pm0.137773$& $\pm0.528684$& $\pm25.287906$& $\pm1
34345.3$& $\pm0.531595$& $\pm5.152539$&  &  \\
$0.2$ & $0.082486$& $1.060131$& $0.198163$& $0.088975$& $0.659897$& $-0.478111$& $0.046565$& $-0.171579$& $-
0.027949$& $0.010017$&  $0.32$& $1.0$\\
& $\pm0.000112$& $\pm0.003438$& $\pm0.009352$& $\pm0.002178$& $\pm0.028616$& $\pm0.177592$& $\pm0.004462$& $\pm0.
039908$& $\pm0.180031$& $\pm0.128064$&  &  \\
$0.3$ & $0.109019$& $0.978669$& $0.221489$& $0.216708$& $0.668204$& $-0.697550$& $0.063620$& $-0.259034$& $0
.064887$& $0.040643$& $0.30$& $1.0$\\
& $\pm0.000186$& $\pm0.005436$& $\pm0.005629$& $\pm0.003258$& $\pm0.031414$& $\pm0.271629$& $\pm0.007652$& $\pm0.
068076$& $\pm0.266990$& $\pm0.078697$&  &  \\
$0.4$ & $0.127528$& $0.848067$& $0.239141$& $0.369736$& $0.676025$& $-0.839083$& $0.076953$& $-0.349509$& $0.237335$& $0.068236$& $0.31$& $1.0$\\
& $\pm0.000271$& $\pm0.006943$& $\pm0.004114$& $\pm0.004101$& $\pm0.033118$& $\pm0.336161$& $\pm0.012272$& $\pm0.102073$& $\pm0.330397$& $\pm0.058631$&  &  \\
$0.5$ & $0.140682$& $0.705505$& $0.253032$& $0.503682$& $0.937536$& $-0.332060$& $0.212552$& $-2.718920$& $0.710788$& $0.442181$& $0.41$& $1.0$\\
& $\pm0.000358$& $\pm0.009290$& $\pm0.004080$& $\pm0.004469$& $\pm0.020115$& $\pm0.599821$& $\pm0.045467$& $\pm0.590415$& $\pm0.358266$& $\pm0.222669$&  &  \\
$0.6$ & $0.148957$& $0.582303$& $0.265807$& $0.605345$& $0.895239$& $-0.001142$& $0.201420$& $-2.061302$& $0.980186$& $0.492228$&  $0.42$& $1.0$\\
& $\pm0.000400$& $\pm0.008183$& $\pm0.003344$& $\pm0.004775$& $\pm0.018798$& $\pm0.468364$& $\pm0.057284$& $\pm0.529787$& $\pm0.404138$& $\pm0.143559$&  &  \\
$0.7$ & $0.154006$& $0.465265$& $0.269879$& $0.654282$& $1.044011$& $1.877978$& $0.475836$& $-6.498344$& $2.597988$& $1.893654$& $0.56$& $1.0$\\
& $\pm0.000477$& $\pm0.008247$& $\pm0.003174$& $\pm0.005845$& $\pm0.008468$& $\pm0.450299$& $\pm0.116308$& $\pm1.420084$& $\pm0.597372$& $\pm0.271606$&  &  \\
$0.8$ & $0.159207$& $0.417027$& $0.290285$& $0.701083$& $0.863284$& $-0.091367$& $0.191149$& $-1.834479$& $1.618341$& $0.459509$& $0.51$& $1.0$\\
& $\pm0.000507$& $\pm0.006730$& $\pm0.003201$& $\pm0.006286$& $\pm0.019653$& $\pm0.402536$& $\pm0.081644$& $\pm0.574854$& $\pm0.517534$& $\pm0.124122$& &  \\
$0.9$ & $0.162161$& $0.355298$& $0.297012$& $0.704691$& $0.927948$& $0.415482$& $0.270416$& $-2.934130$& $2.611038$& $0.829648$& $0.57$& $1.0$\\
& $\pm0.000567$& $\pm0.006615$& $\pm0.003434$& $\pm0.007738$& $\pm0.014348$& $\pm0.417159$& $\pm0.122227$& $\pm0.901206$& $\pm0.682387$& $\pm0.186203$& &  \\
$1.0$ & $0.164898$& $0.318221$& $0.308348$& $0.711567$& $0.813504$& $-0.210609$& $0.160795$& $-1.358406$& $1.988185$& $0.392484$& $0.59$& $1.0$\\
& $\pm0.000609$& $\pm0.005372$& $\pm0.003249$& $\pm0.007860$& $\pm0.021413$& $\pm0.319641$& $\pm0.095576$& $\pm0.508608$& $\pm0.601572$& $\pm0.102036$&  &  \\
$1.1$ & $0.166763$& $0.286850$& $0.317352$& $0.707403$& $0.685913$& $-0.555660$& $0.102332$& $-0.638286$& $1.478133$& $0.184979$& $0.62$& $1.0$\\
& $\pm0.000786$& $\pm0.004490$& $\pm0.003123$& $\pm0.008028$& $\pm0.030186$& $\pm0.240422$& $\pm0.068681$& $\pm0.275390$& $\pm0.523968$& $\pm0.052628$&  &  \\
\hline\hline
\end{tabular}
\end{minipage}
}
}
\end{table*}

\begin{table*}[p]
\footnotesize{
\rotatebox{90}{
\begin{minipage}[t]{1.0\textheight}
\caption{
The details of the most reliable fitting by the scaling analysis for the $q$th moment of Q1D two-terminal conductances, $\langle g^q \rangle$, with  the minimum width $M_\text{min}=32$ and the maximum width
$M_\text{max}=384$. The aspect ratio is varied from $L/M=4$ to $10$ and the number of total data point $N$ is $717$. The scaling function in Eq.\ (\ref{eq:Q1D_gq}) with $N_p=2$ and $N_s=1$ is employed. $\chi_\text{min}^2$, and $Q$ in the tables represent the minimum value of $\chi^2$, and the goodness of fit, respectively. In the table, the value with $\pm$ means the error bar for its above value.
\label{tab:fitting_gq_384}
}
\begin{tabular}{c |r r |r  r| r| r r r| r  r | r r}
\hline\hline
$q\ $ &
$X_q^\prime\quad$ & $c_q^\prime\quad\ $ &
$\gamma_q^\prime\quad$ & $d_q^\prime\quad\ $ &
$|y^\prime|\quad\ $ &
$c_q^{\prime(1)}\quad $&  $a_q^{\prime(1)}\ $ & $a_q^{\prime(2)}\ $ &
$d_q^{\prime(1)}\quad $&  $b_q^{\prime(1)}\ $  &$\chi_\text{min}^2/N$ & $Q$\\
\hline
$-0.5$ & $-0.395720$& $0.340006$& $-0.326910$& $-0.017778$& $0.671876$& $1.317255$& $-0.249383$& $1.356729$& $20.077285$& $4.696611$&  $0.66$& $1.0$\\
& $\pm0.000352$& $\pm0.003285$& $\pm0.038849$& $\pm0.013783$& $\pm0.002961$& $\pm0.119577$& $\pm0.005936$& $\pm0.849714$& $\pm6.070049$& $\pm0.328406$&  &  \\
$-0.4$ & $-0.293761$& $0.461188$& $-0.236389$& $-0.021383$& $0.614671$& $1.138517$& $-0.156072$& $0.617612$& $19.336989$& $3.921922$& $0.52$& $1.0$\\
& $\pm0.000139$& $\pm0.001987$& $\pm0.030214$& $\pm0.010638$& $\pm0.003517$& $\pm0.074826$& $\pm0.011188$& $\pm0.569776$& $\pm4.772048$& $\pm0.217845$&  &  \\
$-0.3$ & $-0.203141$& $0.600304$& $-0.142294$& $-0.025818$& $0.541516$& $0.867346$& $-0.084613$& $0.220073$& $19.421414$& $3.105257$&  $0.35$& $1.0$\\
& $\pm0.000297$& $\pm0.001009$& $\pm0.027045$& $\pm0.010032$& $\pm0.007270$& $\pm0.041900$& $\pm0.030561$& $\pm0.657579$& $\pm4.667253$& $\pm0.156559$&  &  \\
$-0.2$ & $-0.124533$& $0.745625$& $-0.057750$& $-0.020423$& $0.539831$& $0.595920$& $-0.050023$& $0.118521$& $13.909845$& $3.112354$&  $0.27$& $1.0$\\
& $\pm0.000111$& $\pm0.000569$& $\pm0.026918$& $\pm0.007383$& $\pm0.003667$& $\pm0.024887$& $\pm0.009911$& $\pm0.197698$& $\pm3.364895$& $\pm0.162738$&  &  \\
$-0.1$ & $-0.056965$& $0.884568$& $-0.011977$& $-0.008031$& $0.575543$& $0.304396$& $-0.025089$& $0.066775$& $4.821933$& $3.660181$&  $0.27$& $1.0$\\
& $\pm0.000042$& $\pm0.000329$& $\pm0.030842$& $\pm0.003290$& $\pm0.008134$& $\pm0.013567$& $\pm0.006657$& $\pm0.166176$& $\pm1.362451$& $\pm0.216194$&  &  \\
$0.1$ & $0.045894$& $1.067966$& $0.195670$& $0.014069$& $0.334925$& $-1.007886$& $0.008435$& $-0.007873$& $0.663214$& $-0.351494$&  $0.28$& $1.0$\\
& $\pm0.000053$& $\pm0.001569$& $\pm0.030526$& $\pm0.000715$& $\pm0.093046$& $\pm0.152661$& $\pm0.003236$& $\pm0.018802$& $\pm0.180493$& $\pm0.088396$&  &  \\
$0.2$ & $0.082196$& $1.063537$& $0.205673$& $0.089457$& $0.535820$& $-0.431712$& $0.031856$& $-0.065195$& $-0.051170$& $0.042202$&  $0.29$& $1.0$\\
& $\pm0.000091$& $\pm0.002328$& $\pm0.007443$& $\pm0.002182$& $\pm0.020155$& $\pm0.105426$& $\pm0.002575$& $\pm0.013886$& $\pm0.164434$& $\pm0.053395$&  &  \\
$0.3$ & $0.108576$& $0.984334$& $0.228695$& $0.218769$& $0.533450$& $-0.717027$& $0.043372$& $-0.099664$& $-0.116867$& $0.004686$&  $0.29$& $1.0$\\
& $\pm0.000154$& $\pm0.003887$& $\pm0.004536$& $\pm0.002963$& $\pm0.023946$& $\pm0.186956$& $\pm0.003585$& $\pm0.022972$& $\pm0.252757$& $\pm0.031933$&  &  \\
$0.4$ & $0.126913$& $0.854166$& $0.245322$& $0.374784$& $0.538326$& $-0.992159$& $0.054072$& $-0.146638$& $-0.136713$& $-0.002037$&  $0.31$& $1.0$\\
& $\pm0.000225$& $\pm0.005171$& $\pm0.003329$& $\pm0.003617$& $\pm0.027028$& $\pm0.251016$& $\pm0.004509$& $\pm0.035294$& $\pm0.323453$& $\pm0.023722$&  &  \\
$0.5$ & $0.139277$& $0.714367$& $0.260212$& $0.517803$& $0.549803$& $-1.150657$& $0.064446$& $-0.206920$& $-0.213162$& $-0.001927$& $0.35$& $1.0$\\
& $\pm0.000292$& $\pm0.005661$& $\pm0.002749$& $\pm0.004112$& $\pm0.029594$& $\pm0.278989$& $\pm0.005549$& $\pm0.051709$& $\pm0.359718$& $\pm0.020714$&  &  \\
$0.6$ & $0.147631$& $0.592085$& $0.273844$& $0.624854$& $0.564980$& $-1.178008$& $0.074186$& $-0.276204$& $-0.373175$& $-0.000696$& $0.41$& $1.0$\\
& $\pm0.000347$& $\pm0.005516$& $\pm0.002456$& $\pm0.004569$& $\pm0.032071$& $\pm0.279778$& $\pm0.006796$& $\pm0.072206$& $\pm0.370595$& $\pm0.020158$& &  \\
$0.7$ & $0.153418$& $0.494535$& $0.286028$& $0.693488$& $0.581560$& $-1.122735$& $0.083005$& $-0.349747$& $-0.573348$& $0.000425$& $0.48$& $1.0$\\
& $\pm0.000393$& $\pm0.005098$& $\pm0.002315$& $\pm0.005111$& $\pm0.034642$& $\pm0.268121$& $\pm0.008255$& $\pm0.096497$& $\pm0.371057$& $\pm0.020956$&  &  \\
$0.8$ & $0.157563$& $0.419028$& $0.296664$& $0.730811$& $0.598397$& $-1.030758$& $0.090876$& $-0.424499$& $-0.768317$& $0.001410$& $0.55$& $1.0$\\
& $\pm0.000429$& $\pm0.004631$& $\pm0.002260$& $\pm0.005699$& $\pm0.037302$& $\pm0.252956$& $\pm0.009943$& $\pm0.124317$& $\pm0.369912$& $\pm0.022619$&  &  \\
$0.9$ & $0.160643$& $0.360612$& $0.305810$& $0.745425$& $0.616105$& $-0.929237$& $0.098217$& $-0.502770$& $-0.931498$& $0.002489$& $0.62$& $1.0$\\
& $\pm0.000459$& $\pm0.004206$& $\pm0.002257$& $\pm0.006256$& $\pm0.039903$& $\pm0.238435$& $\pm0.011945$& $\pm0.156345$& $\pm0.370466$& $\pm0.025042$&  &  \\
$1.0$ & $0.162995$& $0.314874$& $0.313608$& $0.744568$& $0.633286$& $-0.829949$& $0.104929$& $-0.580950$& $-1.054404$& $0.004000$& $0.68$& $1.0$\\
& $\pm0.000485$& $\pm0.003845$& $\pm0.002287$& $\pm0.006743$& $\pm0.042303$& $\pm0.225571$& $\pm0.014395$& $\pm0.192205$& $\pm0.373426$& $\pm0.028053$&  &  \\
$1.1$ & $0.164827$& $0.278484$& $0.320239$& $0.733618$& $0.647788$& $-0.738353$& $0.110543$& $-0.649788$& $-1.138348$& $0.006048$&  $0.74$& $1.0$\\
& $\pm0.000509$& $\pm0.003544$& $\pm0.002338$& $\pm0.007147$& $\pm0.044373$& $\pm0.214324$& $\pm0.017403$& $\pm0.230573$& $\pm0.378406$& $\pm0.031454$& &  \\
\hline\hline
\end{tabular}
\end{minipage}
}
}
\end{table*}

\begin{table*}[p]
\footnotesize{
\rotatebox{90}{
\begin{minipage}[t]{1.0\textheight}
\caption{
The details of the most reliable fitting by the scaling analysis for the $q$th moment of Q1D two-terminal conductances, $\langle g^q \rangle$, with  the minimum width $M_\text{min}=32$ and the maximum width
$M_\text{max}=512$. The aspect ratio is varied from $L/M=4$ to $10$ and the number of total data point $N$ is $1025$. The scaling function in Eq.\ (\ref{eq:Q1D_gq}) with $N_p=2$ and $N_s=1$ is employed. $\chi_\text{min}^2$, and $Q$ in the tables represent the minimum value of $\chi^2$, and the goodness of fit, respectively. In the table, the value with $\pm$ means the error bar for its above value.
\label{tab:fitting_gq_512}
}
\begin{tabular}{c |r r |r  r| r| r r r| r  r| r r}
\hline\hline
$q\ $ &
$X_q^\prime\quad$ & $c_q^\prime\quad\ $ &
$\gamma_q^\prime\quad$ & $d_q^\prime\quad\ $ &
$|y^\prime|\quad\ $ &
$c_q^{\prime(1)}\quad $&  $a_q^{\prime(1)}\ $ & $a_q^{\prime(2)}\ $ &
$d_q^{\prime(1)}\quad $&  $b_q^{\prime(1)}\ $  &$\chi_\text{min}^2/N$ & $Q$\\
\hline
$-0.5$ & $-0.386598$& $0.355199$& $1156121.4$& $-73655.7$& $0.400457$& $1.233223$& $-0.145448$& $0.284554$& $2863053.8$& $1768961.8$&  $1.1$& $0.1$\\
& $\pm0.000851$& $\pm0.000426$& $\pm0.000000$& $\pm0.000000$& $\pm0.000000$& $\pm0.040532$& $\pm0.143532$& $\pm0.636583$& $\pm0.000000$& $\pm0.000
000$& &  \\
$-0.4$ & $-0.285471$& $0.468068$& $0.202874$& $0.284177$& $0.278106$& $0.748748$& $-0.068749$& $0.051729$& $-5.903556$& $-0.554412$&  $0.52$& $1.0$\\
& $\pm0.000753$& $\pm0.000608$& $\pm0.041075$& $\pm0.087671$& $\pm0.013802$& $\pm0.073958$& $\pm0.047575$& $\pm0.111431$& $\pm2.348020$& $\pm0.115
770$& &  \\
$-0.3$ & $-0.195259$& $0.603252$& $0.331848$& $0.120448$& $0.168540$& $0.555443$& $-0.029432$& $0.000781$& $-2.151591$& $-0.587345$&  $0.46$& $1.0$\\
& $\pm0.000432$& $\pm0.000461$& $\pm0.043747$& $\pm0.026305$& $\pm0.006185$& $\pm0.070884$& $\pm0.012892$& $\pm0.018963$& $\pm0.723725$& $\pm0.095
318$&  &  \\
$-0.2$ & $-0.120591$& $0.747082$& $0.372864$& $0.139630$& $0.190979$& $0.484278$& $-0.018086$& $0.001351$& $-2.783243$& $-0.405702$&  $0.43$& $1.0$\\
& $\pm0.000140$& $\pm0.000251$& $\pm0.039187$& $\pm0.033539$& $\pm0.004331$& $\pm0.028305$& $\pm0.005386$& $\pm0.008701$& $\pm0.889503$& $\pm0.090
513$&  &  \\
$-0.1$ & $-0.055282$& $0.885258$& $0.460095$& $0.054447$& $0.180447$& $0.251179$& $-0.006403$& $-0.001728$& $-1.159937$& $-0.452971$&  $0.44$& $1.0$\\
& $\pm0.000086$& $\pm0.000125$& $\pm0.048905$& $\pm0.015723$& $\pm0.003900$& $\pm0.014625$& $\pm0.002073$& $\pm0.003151$& $\pm0.434422$& $\pm0.107
653$& &  \\
$0.1$ & $0.042051$& $1.070980$& $0.665873$& $0.012572$& $0.070900$& $-1.173374$& $0.008771$& $-0.001957$& $0.786909$& $-0.741842$& $0.25$& $1.0$\\
& $\pm0.002294$& $\pm0.000516$& $\pm0.078593$& $\pm0.000620$& $\pm0.003522$& $\pm0.094736$& $\pm0.007228$& $\pm0.005814$& $\pm0.120696$& $\pm0.098
851$&  &  \\
$0.2$ & $-0.492833$& $0.090772$& $0.078921$& $1.073297$& $0.101720$& $0.216614$& $2.772094$& $-2.668405$& $-2.065717$& $0.008485$& $0.24$& $1.0$\\
& $\pm0.921458$& $\pm0.002326$& $\pm0.000135$& $\pm0.001364$& $\pm0.051380$& $\pm0.175778$& $\pm2.507368$& $\pm1.483259$& $\pm0.112880$& $\pm0.000
251$&  &  \\
$0.3$ & $0.108860$& $0.987807$& $0.229516$& $0.217078$& $0.562664$& $-0.581562$& $0.046330$& $-0.113620$& $0.027827$& $0.066343$& $0.25$& $1.0$\\
& $\pm0.000109$& $\pm0.003174$& $\pm0.003909$& $\pm0.002857$& $\pm0.014958$& $\pm0.147459$& $\pm0.003595$& $\pm0.020374$& $\pm0.231225$& $\pm0.028948$&  &  \\
$0.4$ & $0.127492$& $0.855787$& $0.243509$& $0.371636$& $0.620295$& $-0.673724$& $0.066371$& $-0.227364$& $0.068935$& $0.078907$&  $0.29$& $1.0$\\
& $\pm0.000158$& $\pm0.004451$& $\pm0.002851$& $\pm0.003238$& $\pm0.014752$& $\pm0.210683$& $\pm0.005992$& $\pm0.039849$& $\pm0.292004$& $\pm0.027045$&  &  \\
$0.5$ & $0.140026$& $0.712788$& $0.256927$& $0.513739$& $0.667955$& $-0.656912$& $0.086441$& $-0.383218$& $0.089494$& $0.101106$&  $0.35$& $1.0$\\
& $\pm0.000207$& $\pm0.005014$& $\pm0.002348$& $\pm0.003637$& $\pm0.014518$& $\pm0.248694$& $\pm0.009667$& $\pm0.068776$& $\pm0.326587$& $\pm0.028483$&  &  \\
$0.6$ & $0.148437$& $0.587813$& $0.269673$& $0.619456$& $0.708522$& $-0.562059$& $0.106394$& $-0.575608$& $0.084932$& $0.128993$& $0.41$& $1.0$\\
& $\pm0.000249$& $\pm0.004954$& $\pm0.002097$& $\pm0.004045$& $\pm0.014092$& $\pm0.263111$& $\pm0.014764$& $\pm0.107626$& $\pm0.340412$& $\pm0.032351$&  &  \\
$0.7$ & $0.154193$& $0.488094$& $0.281016$& $0.685866$& $0.748709$& $-0.414572$& $0.128962$& $-0.826268$& $0.101280$& $0.168673$& $0.46$& $1.0$\\
& $\pm0.000283$& $\pm0.004631$& $\pm0.001982$& $\pm0.004505$& $\pm0.013204$& $\pm0.266119$& $\pm0.021752$& $\pm0.161469$& $\pm0.347094$& $\pm0.039636$&  &  \\
$0.8$ & $0.158237$& $0.411238$& $0.290897$& $0.720467$& $0.779904$& $-0.269262$& $0.149988$& $-1.075971$& $0.150156$& $0.212284$&  $0.52$& $1.0$\\
& $\pm0.000313$& $\pm0.004239$& $\pm0.001943$& $\pm0.005010$& $\pm0.012416$& $\pm0.261034$& $\pm0.029543$& $\pm0.219484$& $\pm0.357585$& $\pm0.047417$&  &  \\
$0.9$ & $0.161171$& $0.351922$& $0.299270$& $0.732002$& $0.809492$& $-0.119697$& $0.173328$& $-1.357778$& $0.259230$& $0.269275$& $0.56$& $1.0$\\
& $\pm0.000339$& $\pm0.003886$& $\pm0.001955$& $\pm0.005490$& $\pm0.011384$& $\pm0.255154$& $\pm0.038760$& $\pm0.290155$& $\pm0.377828$& $\pm0.058121$&  &  \\
$1.0$ & $0.163260$& $0.303332$& $0.304956$& $0.723150$& $0.889849$& $0.287054$& $0.258561$& $-2.509821$& $0.905139$& $0.524088$&  $0.61$& $1.0$\\
& $\pm0.000375$& $\pm0.003755$& $\pm0.002080$& $\pm0.005950$& $\pm0.008173$& $\pm0.273767$& $\pm0.059153$& $\pm0.498663$& $\pm0.460971$& $\pm0.101022$& &  \\
$1.1$ & $0.164987$& $0.268512$& $0.311882$& $0.712968$& $0.864948$& $0.184440$& $0.231403$& $-2.034678$& $0.728268$& $0.446166$& $0.64$& $1.0$\\
& $\pm0.000391$& $\pm0.003371$& $\pm0.002084$& $\pm0.006322$& $\pm0.009014$& $\pm0.245992$& $\pm0.060894$& $\pm0.470888$& $\pm0.458256$& $\pm0.089392$&  &  \\
\hline\hline
\end{tabular}
\end{minipage}
}
}
\end{table*}

\begin{table*}[p]
\footnotesize{
\rotatebox{90}{
\begin{minipage}[t]{1.0\textheight}
\caption{
The details of the most reliable fitting by the scaling analysis for the Legendre function as a function of  Q1D  two-terminal conductances, $\langle p_q(g) \rangle$, with  the minimum width $M_\text{min}=32$ and the maximum width $M_\text{max}=192$. The aspect ratio is varied from $L/M=2$ to $10$ and the number of total data points $N$ is $420$ except at $q=0.5$. At $q=0.5$, the maximum aspect ratio is limited to $8$ due to our numerical accuracy and the number of total data points $N$ is $324$. The scaling function in Eq.\ (\ref{eq:Q1D_Legendre}) with $N_p=2$ and $N_s=1$ is employed. $\chi_\text{min}^2$, and $Q$ in the tables represent the minimum value of $\chi^2$, and the goodness of fit, respectively. In the table, the value with $\pm$ means the error bar for its above value. Note that the Legendre function satisfies $p_{q}(x)=p_{1-q}(x)$.
\label{tab:fitting_Legendre_192}
}
\begin{tabular}{r |r r |r  r| r| r r r| r  r| r r}
\hline\hline
$q\ $ &
$\bar{X}_q^\prime\quad$ & $\bar{c}_q^\prime\quad$ &
$\bar{\gamma}_q^\prime\quad$ & $\bar{d}_q^\prime\quad$ &
$|\bar{y}^\prime|\quad\ $ &
$\bar{c}_q^{\prime(1)}\quad$&  $\bar{a}_q^{\prime(1)}\ $ &
$\bar{a}_q^{\prime(2)}\ $ &
$\bar{d}_q^{\prime(1)}\ $&  $\bar{b}_q^{\prime(1)}\ $ & $\chi_\text{min}^2/N$ & $Q$\\
\hline
$0.5$ & $0.128892$& $1.259646$& $0.663351$& $-0.070738$& $0.727002$& $-0.730887$& $0.086980$& $-0.400856$& $0.578356$& $-1.153155$&  $0.48$& $1.0$\\
& $\pm0.000186$& $\pm0.000539$& $\pm0.032673$& $\pm0.004694$& $\pm0.034635$& $\pm0.039244$& $\pm0.019364$& $\pm0.162621$& $\pm0.283376$& $\pm0.743605$&  &  \\
$0.6$ & $0.123889$& $1.248526$& $0.653068$& $-0.069492$& $0.762378$& $-0.727757$& $0.092810$& $-0.505492$& $0.552401$& $-0.928184$&   $0.47$& $1.0$\\
& $\pm0.000137$& $\pm0.000404$& $\pm0.029383$& $\pm0.004502$& $\pm0.032117$& $\pm0.029553$& $\pm0.016732$& $\pm0.167836$& $\pm0.291098$& $\pm0.802835$&  &  \\
$0.7$ & $0.108477$& $1.215405$& $0.621069$& $-0.060637$& $0.759376$& $-0.612117$& $0.080862$& $-0.433034$& $0.434811$& $-0.858006$&  $0.49$& $1.0$\\
& $\pm0.000115$& $\pm0.000334$& $\pm0.026634$& $\pm0.003541$& $\pm0.030471$& $\pm0.024617$& $\pm0.013396$& $\pm0.130034$& $\pm0.234375$& $\pm0.712931$&  &  \\
$0.8$ & $0.082704$& $1.161538$& $0.570880$& $-0.045972$& $0.736683$& $-0.438758$& $0.057818$& $-0.280092$& $0.270943$& $-0.676493$&  $0.51$& $1.0$\\
& $\pm0.000085$& $\pm0.000233$& $\pm0.022989$& $\pm0.002286$& $\pm0.028905$& $\pm0.017225$& $\pm0.008676$& $\pm0.078295$& $\pm0.156146$& $\pm0.571802$&  &  \\
$0.9$ & $0.046539$& $1.088872$& $0.504700$& $-0.025636$& $0.689558$& $-0.228973$& $0.028392$& $-0.112529$& $0.110458$& $-0.420938$&  $0.53$& $1.0$\\
& $\pm0.000051$& $\pm0.000118$& $\pm0.019248$& $\pm0.001030$& $\pm0.028965$& $\pm0.008590$& $\pm0.003829$& $\pm0.029896$& $\pm0.072334$& $\pm0.397680$&  &  \\
$1.1$ & $-0.056819$& $0.898015$& $0.330462$& $0.030151$& $0.533879$& $0.236937$& $-0.022552$& $0.046539$& $-0.053905$& $-0.054616$&  $0.48$& $1.0$\\
& $\pm0.000116$& $\pm0.000119$& $\pm0.013362$& $\pm0.000698$& $\pm0.038792$& $\pm0.008180$& $\pm0.002781$& $\pm0.014215$& $\pm0.050159$& $\pm0.143303$&  &  \\
$1.2$ & $-0.123708$& $0.786433$& $0.228334$& $0.063849$& $0.444370$& $0.474406$& $-0.039654$& $0.057091$& $-0.103179$& $-0.010408$&  $0.45$& $1.0$\\
& $\pm0.000472$& $\pm0.000255$& $\pm0.011620$& $\pm0.001093$& $\pm0.052059$& $\pm0.016664$& $\pm0.005330$& $\pm0.021706$& $\pm0.077748$& $\pm0.084169$&  &  \\
$1.3$ & $-0.200909$& $0.669308$& $0.118980$& $0.099579$& $0.395918$& $0.707446$& $-0.060041$& $0.076333$& $-0.179595$& $0.001151$&  $0.45$& $1.0$\\
& $\pm0.001277$& $\pm0.000451$& $\pm0.010354$& $\pm0.001318$& $\pm0.066295$& $\pm0.027767$& $\pm0.008268$& $\pm0.033156$& $\pm0.094707$& $\pm0.060417$&  &  \\
$1.4$ & $-0.290603$& $0.552025$& $0.006154$& $0.134874$& $0.440857$& $0.920555$& $-0.099891$& $0.172678$& $-0.248832$& $0.008996$&  $0.48$& $1.0$\\
& $\pm0.001644$& $\pm0.000779$& $\pm0.009057$& $\pm0.001529$& $\pm0.067659$& $\pm0.045873$& $\pm0.013283$& $\pm0.070931$& $\pm0.115358$& $\pm0.063575$&  &  \\
$1.5$ & $-0.392190$& $0.441646$& $-0.106291$& $0.165908$& $0.504153$& $1.090853$& $-0.160472$& $0.396645$& $-0.283190$& $0.017538$&  $0.53$& $1.0$\\
& $\pm0.001864$& $\pm0.001255$& $\pm0.008823$& $\pm0.001810$& $\pm0.069851$& $\pm0.072198$& $\pm0.021969$& $\pm0.162391$& $\pm0.143105$& $\pm0.080184$&  &  \\
\hline\hline
\end{tabular}
\end{minipage}
}
}
\end{table*}


\begin{table*}[p]
\footnotesize{
\rotatebox{90}{
\begin{minipage}[t]{1.0\textheight}
\caption{
The details of the most reliable fitting by the scaling analysis for the Legendre function as a function of  Q1D  two-terminal conductances, $\langle p_q(g) \rangle$, with  the minimum width $M_\text{min}=32$ and the maximum width $M_\text{max}=256$. The aspect ratio is varied from $L/M=2$ to $10$ and the number of total data points $N$ is $651$ except at $q=0.5$. At $q=0.5$, the maximum aspect ratio is limited to $8$ due to our numerical accuracy and the number of total data points $N$ is $501$. The scaling function in Eq.\ (\ref{eq:Q1D_Legendre}) with $N_p=2$ and $N_s=1$ is employed. $\chi_\text{min}^2$, and $Q$ in the tables represent the minimum value of $\chi^2$, and the goodness of fit, respectively. In the table, the value with $\pm$ means the error bar for its above value. Note that the Legendre function satisfies $p_{q}(x)=p_{1-q}(x)$.
\label{tab:fitting_Legendre_256}
}
\begin{tabular}{c |r r |r  r| r| r r r| r  r | r r}
\hline\hline
$q\ $ &
$\bar{X}_q^\prime\quad$ & $\bar{c}_q^\prime\quad$ &
$\bar{\gamma}_q^\prime\quad$ & $\bar{d}_q^\prime\quad$ &
$|\bar{y}^\prime|\quad\ $ &
$\bar{c}_q^{\prime(1)}\quad$&  $\bar{a}_q^{\prime(1)}\ $ &
$\bar{a}_q^{\prime(2)}\ $ &
$\bar{d}_q^{\prime(1)}\ $&  $\bar{b}_q^{\prime(1)}\ $ & $\chi_\text{min}^2/N$ & $Q$\\
\hline
$0.5$ & $0.127866$& $1.261090$& $0.643272$& $-0.070508$& $0.507135$& $-0.799714$& $0.045831$& $-0.084169$& $0.662712$& $-0.416566$&  $0.54$& $1.0$\\
& $\pm0.000315$& $\pm0.000376$& $\pm0.021914$& $\pm0.002875$& $\pm0.038514$& $\pm0.026853$& $\pm0.009781$& $\pm0.041462$& $\pm0.165803$& $\pm0.198128$&  &  \\
$0.6$ & $0.122687$& $1.249100$& $0.659716$& $-0.071088$& $0.500801$& $-0.758539$& $0.043466$& $-0.077862$& $0.642750$& $-0.384260$& $0.55$& $1.0$\\
& $\pm0.000281$& $\pm0.000271$& $\pm0.021146$& $\pm0.002907$& $\pm0.034920$& $\pm0.020608$& $\pm0.008335$& $\pm0.034498$& $\pm0.172550$& $\pm0.193570$&  &  \\
$0.7$ & $0.107596$& $1.215897$& $0.632244$& $-0.062683$& $0.528901$& $-0.637870$& $0.041063$& $-0.082668$& $0.561028$& $-0.456715$&  $0.56$& $1.0$\\
& $\pm0.000195$& $\pm0.000222$& $\pm0.018603$& $\pm0.002293$& $\pm0.029497$& $\pm0.017332$& $\pm0.007099$& $\pm0.031778$& $\pm0.139394$& $\pm0.194168$&  &  \\
$0.8$ & $0.082213$& $1.161896$& $0.586949$& $-0.048142$& $0.561493$& $-0.456236$& $0.034334$& $-0.079043$& $0.417977$& $-0.540379$& $0.57$& $1.0$\\
& $\pm0.000112$& $\pm0.000155$& $\pm0.015510$& $\pm0.001488$& $\pm0.023762$& $\pm0.012458$& $\pm0.005101$& $\pm0.025003$& $\pm0.094039$& $\pm0.189623$&  &  \\
$0.9$ & $0.046383$& $1.089063$& $0.522848$& $-0.027068$& $0.588911$& $-0.237169$& $0.021005$& $-0.054337$& $0.215229$& $-0.561435$&  $0.56$& $1.0$\\
& $\pm0.000049$& $\pm0.000080$& $\pm0.012617$& $\pm0.000682$& $\pm0.019689$& $\pm0.006553$& $\pm0.002605$& $\pm0.013829$& $\pm0.045778$& $\pm0.178173$& &  \\
$1.1$ & $-0.056907$& $0.897894$& $0.348975$& $0.031561$& $0.558288$& $0.239629$& $-0.023878$& $0.054711$& $-0.162930$& $-0.309616$&  $0.47$& $1.0$\\
& $\pm0.000062$& $\pm0.000086$& $\pm0.008584$& $\pm0.000495$& $\pm0.020100$& $\pm0.006973$& $\pm0.002508$& $\pm0.012125$& $\pm0.037352$& $\pm0.110300$&  &  \\
$1.2$ & $-0.123995$& $0.786445$& $0.242881$& $0.065509$& $0.474472$& $0.471068$& $-0.042281$& $0.068967$& $-0.215574$& $-0.127185$& $0.40$& $1.0$\\
& $\pm0.000231$& $\pm0.000185$& $\pm0.007642$& $\pm0.000794$& $\pm0.027543$& $\pm0.014267$& $\pm0.004695$& $\pm0.018227$& $\pm0.062479$& $\pm0.068129$&  &  \\
$1.3$ & $-0.199888$& $0.669903$& $0.123813$& $0.099677$& $0.351452$& $0.690151$& $-0.055131$& $0.056758$& $-0.139264$& $-0.000915$& $0.39$& $1.0$\\
& $\pm0.001040$& $\pm0.000323$& $\pm0.007751$& $\pm0.000949$& $\pm0.046133$& $\pm0.022413$& $\pm0.005466$& $\pm0.018588$& $\pm0.078571$& $\pm0.039917$&  &  \\
$1.4$ & $-0.283435$& $0.553779$& $-0.003166$& $0.132114$& $0.276741$& $0.878793$& $-0.081561$& $0.071765$& $0.015882$& $0.062143$&  $0.45$& $1.0$\\
& $\pm0.003525$& $\pm0.000545$& $\pm0.008384$& $\pm0.001070$& $\pm0.064327$& $\pm0.033813$& $\pm0.002545$& $\pm0.018261$& $\pm0.096293$& $\pm0.031161$&  &  \\
$1.5$ & $-0.379616$& $0.445143$& $-0.123804$& $0.159553$& $0.289822$& $1.014620$& $-0.127824$& $0.136705$& $0.230040$& $0.108348$& $0.54$& $1.0$\\
& $\pm0.005078$& $\pm0.000866$& $\pm0.008410$& $\pm0.001250$& $\pm0.067184$& $\pm0.049782$& $\pm0.008976$& $\pm0.025515$& $\pm0.122484$& $\pm0.032876$& &  \\
\hline\hline
\end{tabular}
\end{minipage}
}
}
\end{table*}

\begin{table*}[p]
\footnotesize{
\rotatebox{90}{
\begin{minipage}[t]{1.0\textheight}
\caption{
The details of the most reliable fitting by the scaling analysis for the Legendre function as a function of  Q1D  two-terminal conductances, $\langle p_q(g) \rangle$, with  the minimum width $M_\text{min}=32$ and the maximum width $M_\text{max}=384$. The aspect ratio is varied from $L/M=2$ to $10$ and the number of total data points $N$ is $997$ except at $q=0.5$. At $q=0.5$, the maximum aspect ratio is limited to $8$ due to our numerical accuracy and the number of total data points $N$ is $784$. The scaling function in Eq.\ (\ref{eq:Q1D_Legendre}) with $N_p=2$ and $N_s=1$ is employed. $\chi_\text{min}^2$, and $Q$ in the tables represent the minimum value of $\chi^2$, and the goodness of fit, respectively. In the table, the value with $\pm$ means the error bar for its above value. Note that the Legendre function satisfies $p_{q}(x)=p_{1-q}(x)$.
\label{tab:fitting_Legendre_384}
}
\begin{tabular}{c |r r |r  r| r| r r r| r  r | r r}
\hline\hline
$q\ $ &
$\bar{X}_q^\prime\quad$ & $\bar{c}_q^\prime\quad$ &
$\bar{\gamma}_q^\prime\quad$ & $\bar{d}_q^\prime\quad$ &
$|\bar{y}^\prime|\quad\ $ &
$\bar{c}_q^{\prime(1)}\quad$&  $\bar{a}_q^{\prime(1)}\ $ &
$\bar{a}_q^{\prime(2)}\ $ &
$\bar{d}_q^{\prime(1)}\ $&  $\bar{b}_q^{\prime(1)}\ $ & $\chi_\text{min}^2/N$ & $Q$\\
\hline
$0.5$ & $0.127501$& $1.261022$& $0.652190$& $-0.069552$& $0.453816$& $-0.787101$& $0.039208$& $-0.054526$& $0.667292$& $-0.445857$&  $0.54$& $1.0$\\
& $\pm0.000214$& $\pm0.000273$& $\pm0.016101$& $\pm0.002174$& $\pm0.023808$& $\pm0.023206$& $\pm0.006251$& $\pm0.021709$& $\pm0.126287$& $\pm0.116736$&  &  \\
$0.6$ & $0.122479$& $1.249190$& $0.664432$& $-0.069886$& $0.467256$& $-0.753561$& $0.039347$& $-0.059339$& $0.693239$& $-0.461968$&  $0.53$& $1.00$\\
& $\pm0.000172$& $\pm0.000203$& $\pm0.014975$& $\pm0.002151$& $\pm0.020249$& $\pm0.018122$& $\pm0.005422$& $\pm0.019691$& $\pm0.125898$& $\pm0.118097$&  &  \\
$0.7$ & $0.107314$& $1.216132$& $0.631616$& $-0.061303$& $0.473782$& $-0.640731$& $0.034896$& $-0.053675$& $0.606057$& $-0.472567$&   $0.52$& $1.0$\\
& $\pm0.000132$& $\pm0.000167$& $\pm0.013134$& $\pm0.001651$& $\pm0.018269$& $\pm0.015055$& $\pm0.004447$& $\pm0.016539$& $\pm0.098988$& $\pm0.109018$&  &  \\
$0.8$ & $0.081944$& $1.162250$& $0.579014$& $-0.046800$& $0.485959$& $-0.466603$& $0.027406$& $-0.044194$& $0.453100$& $-0.474084$&  $0.51$& $1.0$\\
& $\pm0.000084$& $\pm0.000117$& $\pm0.010826$& $\pm0.001033$& $\pm0.015765$& $\pm0.010624$& $\pm0.003041$& $\pm0.011692$& $\pm0.064637$& $\pm0.096816$&  &  \\
$0.9$ & $0.046222$& $1.089350$& $0.511341$& $-0.026322$& $0.501638$& $-0.246559$& $0.016146$& $-0.027994$& $0.244495$& $-0.465251$&  $0.51$& $1.0$\\
& $\pm0.000038$& $\pm0.000060$& $\pm0.008625$& $\pm0.000459$& $\pm0.013319$& $\pm0.005500$& $\pm0.001506$& $\pm0.006074$& $\pm0.030382$& $\pm0.084861$&  &  \\
$1.1$ & $-0.056757$& $0.897441$& $0.336311$& $0.031047$& $0.492275$& $0.254076$& $-0.019632$& $0.033602$& $-0.212346$& $-0.286249$&   $0.47$& $1.0$\\
& $\pm0.000043$& $\pm0.000065$& $\pm0.005644$& $\pm0.000326$& $\pm0.012582$& $\pm0.005806$& $\pm0.001454$& $\pm0.005727$& $\pm0.025268$& $\pm0.056218$&  &  \\
$1.2$ & $-0.124094$& $0.785431$& $0.232415$& $0.065074$& $0.476356$& $0.497634$& $-0.042224$& $0.070737$& $-0.346189$& $-0.188538$&  $0.42$& $1.0$\\
& $\pm0.000109$& $\pm0.000143$& $\pm0.004752$& $\pm0.000531$& $\pm0.014107$& $\pm0.012403$& $\pm0.003024$& $\pm0.011741$& $\pm0.044895$& $\pm0.044628$&  &  \\
$1.3$ & $-0.202217$& $0.668175$& $0.119088$& $0.100056$& $0.471735$& $0.721573$& $-0.071246$& $0.127772$& $-0.381763$& $-0.104912$& $0.44$& $1.0$\\
& $\pm0.000212$& $\pm0.000262$& $\pm0.004180$& $\pm0.000651$& $\pm0.016643$& $\pm0.021100$& $\pm0.004919$& $\pm0.020713$& $\pm0.061094$& $\pm0.038431$&  &  \\
$1.4$ & $-0.291488$& $0.551240$& $0.005018$& $0.134044$& $0.486075$& $0.897656$& $-0.112490$& $0.235457$& $-0.355429$& $-0.067673$& $0.51$& $1.0$\\
& $\pm0.000349$& $\pm0.000460$& $\pm0.003985$& $\pm0.000743$& $\pm0.019264$& $\pm0.034172$& $\pm0.007913$& $\pm0.038894$& $\pm0.077932$& $\pm0.038444$&  &  \\
$1.5$ & $-0.391936$& $0.441601$& $-0.106193$& $0.163659$& $0.508576$& $1.000598$& $-0.166680$& $0.411058$& $-0.303176$& $-0.072968$&  $0.60$& $1.0$\\
& $\pm0.000524$& $\pm0.000744$& $\pm0.004192$& $\pm0.000870$& $\pm0.021719$& $\pm0.052708$& $\pm0.013288$& $\pm0.073914$& $\pm0.098547$& $\pm0.043762$&  & \\
\hline\hline
\end{tabular}
\end{minipage}
}
}
\end{table*}

\begin{table*}[p]
\footnotesize{
\rotatebox{90}{
\begin{minipage}[t]{1.0\textheight}
\caption{
The details of the most reliable fitting by the scaling analysis for the Legendre function as a function of  Q1D  two-terminal conductances, $\langle p_q(g) \rangle$, with  the minimum width $M_\text{min}=32$ and the maximum width $M_\text{max}=512$. The aspect ratio is varied from $L/M=2$ to $10$ and the number of total data points $N$ is $1458$ except at $q=0.5$. At $q=0.5$, the maximum aspect ratio is limited to $8$ due to our numerical accuracy and the number of total data points $N$ is $1164$. The scaling function in Eq.\ (\ref{eq:Q1D_Legendre}) with $N_p=2$ and $N_s=1$ is employed. $\chi_\text{min}^2$, and $Q$ in the tables represent the minimum value of $\chi^2$, and the goodness of fit, respectively. In the table, the value with $\pm$ means the error bar for its above value. Note that the Legendre function satisfies $p_{q}(x)=p_{1-q}(x)$.
\label{tab:fitting_Legendre_512}
}
\begin{tabular}{c |r r |r  r| r| r r r| r  r | r r}
\hline\hline
$q\ $ &
$\bar{X}_q^\prime\quad$ & $\bar{c}_q^\prime\quad$ &
$\bar{\gamma}_q^\prime\quad$ & $\bar{d}_q^\prime\quad$ &
$|\bar{y}^\prime|\quad\ $ &
$\bar{c}_q^{\prime(1)}\quad$&  $\bar{a}_q^{\prime(1)}\ $ &
$\bar{a}_q^{\prime(2)}\ $ &
$\bar{d}_q^{\prime(1)}\ $&  $\bar{b}_q^{\prime(1)}\ $ & $\chi_\text{min}^2/N$ & $Q$\\
\hline
$0.5$ & $0.127880$& $1.261637$& $0.648772$& $-0.070035$& $0.502680$& $-0.805567$& $0.044965$& $-0.080693$& $0.887891$& $-0.695250$& $0.54$& $1.0$\\
& $\pm0.000105$& $\pm0.000216$& $\pm0.011418$& $\pm0.001699$& $\pm0.014023$& $\pm0.022389$& $\pm0.005253$& $\pm0.022452$& $\pm0.099765$& $\pm0.101576$&  &  \\
$0.6$ & $0.122863$& $1.249876$& $0.657074$& $-0.069912$& $0.521865$& $-0.778881$& $0.045937$& $-0.091743$& $0.948794$& $-0.749678$& $0.53$& $1.0$\\
& $\pm0.000081$& $\pm0.000164$& $\pm0.010421$& $\pm0.001651$& $\pm0.011684$& $\pm0.017488$& $\pm0.004616$& $\pm0.021417$& $\pm0.095329$& $\pm0.103671$&  &  \\
$0.7$ & $0.107501$& $1.216620$& $0.624096$& $-0.061013$& $0.502006$& $-0.658657$& $0.037769$& $-0.067574$& $0.754075$& $-0.629617$&  $0.50$& $1.0$\\
& $\pm0.000074$& $\pm0.000134$& $\pm0.009406$& $\pm0.001272$& $\pm0.011748$& $\pm0.014212$& $\pm0.003632$& $\pm0.015436$& $\pm0.077517$& $\pm0.088716$&  &  \\
$0.8$ & $0.081881$& $1.162523$& $0.570967$& $-0.046199$& $0.466725$& $-0.477109$& $0.025794$& $-0.037911$& $0.488745$& $-0.459206$&  $0.48$& $1.0$\\
& $\pm0.000063$& $\pm0.000093$& $\pm0.008100$& $\pm0.000794$& $\pm0.012180$& $\pm0.009688$& $\pm0.002352$& $\pm0.008630$& $\pm0.052529$& $\pm0.068309$&  &  \\
$0.9$ & $0.046009$& $1.089469$& $0.500696$& $-0.025704$& $0.413905$& $-0.252312$& $0.012375$& $-0.013410$& $0.218066$& $-0.281125$&  $0.48$& $1.0$\\
& $\pm0.000045$& $\pm0.000048$& $\pm0.006871$& $\pm0.000348$& $\pm0.013681$& $\pm0.004784$& $\pm0.001107$& $\pm0.003307$& $\pm0.025304$& $\pm0.047794$& &  \\
$1.1$ & $-0.055885$& $0.897339$& $0.322820$& $0.030164$& $0.285426$& $0.261068$& $-0.010860$& $0.005095$& $-0.141016$& $-0.065599$&  $0.54$& $1.0$\\
& $\pm0.000153$& $\pm0.000051$& $\pm0.005354$& $\pm0.000241$& $\pm0.024099$& $\pm0.004703$& $\pm0.001203$& $\pm0.002347$& $\pm0.020916$& $\pm0.021922$&  &  \\
$1.2$ & $-0.121720$& $0.785270$& $0.217176$& $0.063496$& $0.261409$& $0.511558$& $-0.024145$& $0.011502$& $-0.205564$& $-0.018865$&  $0.51$& $1.0$\\
& $\pm0.000439$& $\pm0.000113$& $\pm0.004667$& $\pm0.000394$& $\pm0.026830$& $\pm0.009878$& $\pm0.002189$& $\pm0.004404$& $\pm0.037777$& $\pm0.017265$&  &  \\
$1.3$ & $-0.198862$& $0.668117$& $0.105317$& $0.098168$& $0.295404$& $0.736462$& $-0.047657$& $0.036608$& $-0.201395$& $0.012571$&  $0.48$& $1.0$\\
& $\pm0.000542$& $\pm0.000206$& $\pm0.003772$& $\pm0.000486$& $\pm0.022178$& $\pm0.016758$& $\pm0.002509$& $\pm0.006996$& $\pm0.053106$& $\pm0.016293$& &  \\
$1.4$ & $-0.287574$& $0.551334$& $-0.005989$& $0.131921$& $0.340645$& $0.914536$& $-0.082930$& $0.093218$& $-0.152016$& $0.027427$& $0.50$& $1.0$\\
& $\pm0.000618$& $\pm0.000357$& $\pm0.003337$& $\pm0.000557$& $\pm0.020483$& $\pm0.026899$& $\pm0.003590$& $\pm0.013745$& $\pm0.069479$& $\pm0.017388$&  &  \\
$1.5$ & $-0.387254$& $0.441683$& $-0.115036$& $0.161359$& $0.373929$& $1.023250$& $-0.127047$& $0.181595$& $-0.098250$& $0.017967$&  $0.55$& $1.0$\\
& $\pm0.000791$& $\pm0.000575$& $\pm0.003372$& $\pm0.000651$& $\pm0.021038$& $\pm0.040883$& $\pm0.006114$& $\pm0.027293$& $\pm0.088935$& $\pm0.019809$&  &  \\
\hline\hline
\end{tabular}
\end{minipage}
}
}
\end{table*}

\clearpage

\clearpage

\section{Point-contact conductance in 2D}

\subsection{ Boundary condition effects}

At an Anderson transition in two dimensions (2D), the $q$-th moment of the two point-contact conductance (PCC) between two points separated by distance $r$ in the plane is expected to behave as $\langle T(r)^q\rangle \propto r^{-X_q}$, plus power-law corrections due to irrelevant exponents and subleading scaling dimensions. However, since we calculate the PCC in a system with periodic boundary conditions (the geometry of a torus), and the
expected simple power law cannot appear for large $r$. To avoid this effect caused by the periodic boundary conditions, we have focused on relatively short distances. To determine the appropriate range of $r$ that we use in our scaling analysis, we compared the PCC for two different system sizes.

\begin{figure*}[t]
\includegraphics[width=15cm]{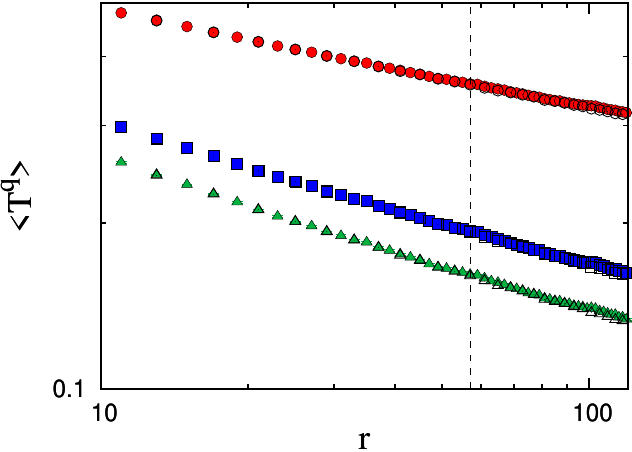}
\caption{
$q$th moment of PCC in 2D, $\langle T^q \rangle$, for the system $L=480$ (filled  dots) and $L=1024$ (open dots) for $q=0.5$ (circles), $1$ (rectangles), and $1.2$ (triangles). The vertical dashed lines represent $r=57$.
} \label{fig:PCC_Ldep}
\end{figure*}

Figure \ref{fig:PCC_Ldep} shows the $r$ dependence of the moments $\langle T^q\rangle$ in 2D with the different system size; $L=480$ (filled dots) and $1024$ (open dots). Since $\langle T^q \rangle$ for both systems [practically coincide for $r < 57$, we can safely use the numerical data for $r<57$ in the fitting.

\subsection{Scaling analysis for PCC in 2D}

Since the $T^q$ is not a pure scaling operator, as we explained in the main paper, the appropriate scaling function in 2D is a combination of the leading power laws with exponent $X_q$, and correction terms with  the irrelevant exponent $y$ and the subleading dimension $\gamma_q$:
\begin{align}
\langle T^{q} \rangle &= c_q r^{-X_q} \Big(1+\sum_{n=1}^{N_p} a_q^{(n)}r^{n y}\Big) \nonumber \\
& \quad + d_q r^{-\gamma_q} \Big(1+\sum_{n=1}^{N_s} b_q^{(n)}r^{n y}\Big).
\label{eq:2D_Tq}
\end{align}

The use of the Legendre functions allows to omit the corrections with subleading scaling dimensions for the quantity $p_q(T) \equiv P_{-q}(2/T-1)$:
\begin{eqnarray}
\langle p_q(T) \rangle &=& \bar{c}_q r^{-\bar{X}_q}
\Big(1+\sum_{n=1}^{N_p} \bar{a}_q^{(n)}r^{-n y}\Big).
\label{eq:2D_Legendre}
\end{eqnarray}

We fit the numerical data in the range $r < 57$ (chosen as explained in the previous section) to these power laws to extract exponents and coefficients. Because of the difficulty of fitting the data to multiple power laws with limited data points,  we can introduce only a single correction term to obtain the reliable fitting. Therefore, we truncate Eq.\ (\ref{eq:2D_Legendre}) at $N_p=1$, and Eq.\ (\ref{eq:2D_Tq}) at $N_p=1, d_q=0$. This simplification makes the simultaneous determination of $y$ and $\gamma_q$ in Eq.\ (\ref{eq:2D_Tq}) impossible. 

\begin{figure*}[h]
\flushleft
\vspace{-0.5cm}\hspace{1.5cm}(a) $r_\text{max}=41$
\begin{center}
\includegraphics[width=13cm]{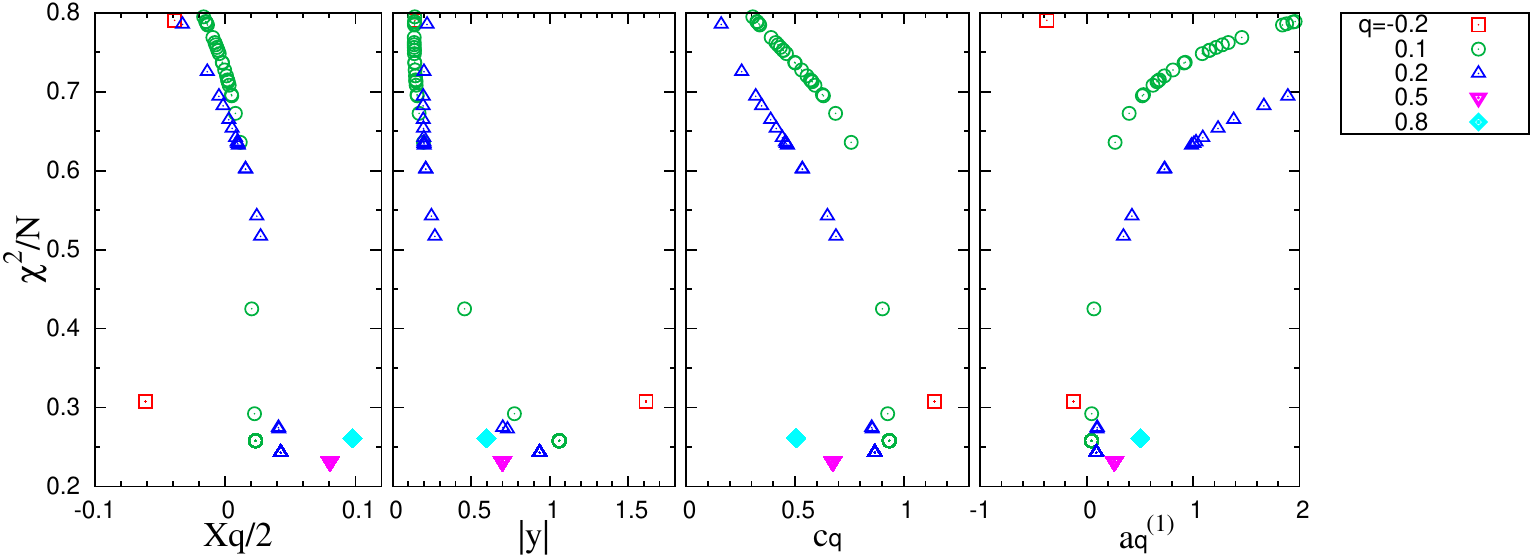}
\end{center}
\vspace{-0.5cm}\hspace{1.5cm}(b) $r_\text{max}=49$
\begin{center}
\includegraphics[width=13cm]{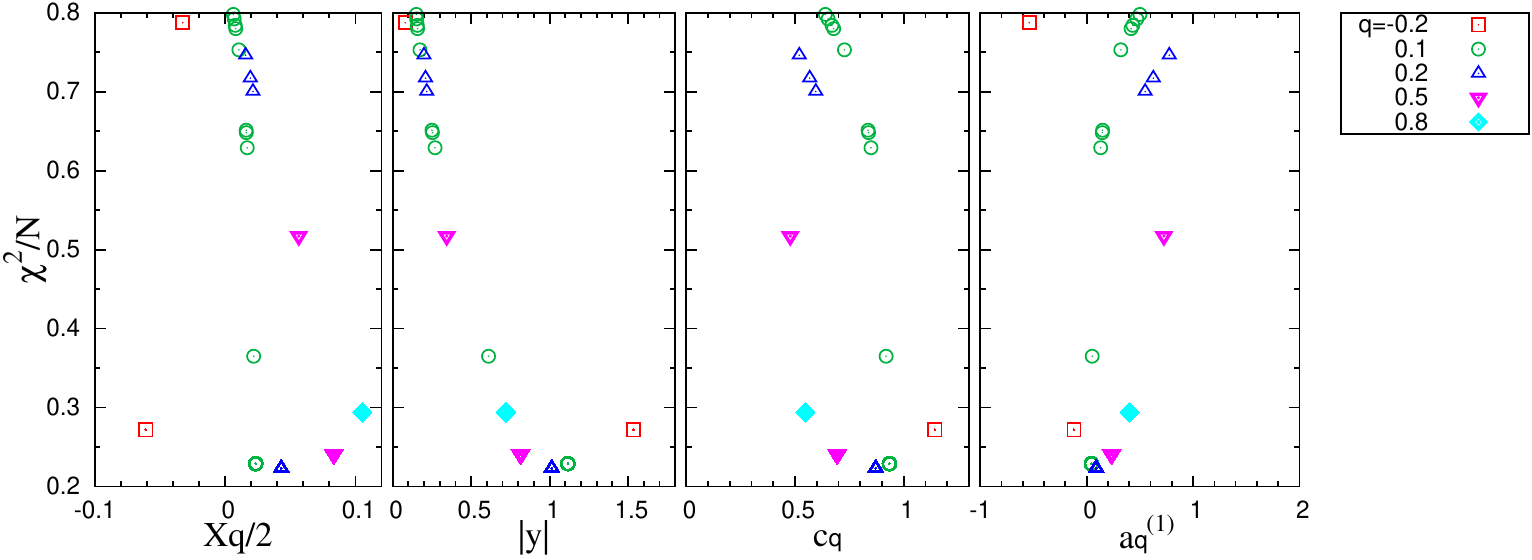}
\end{center}
\vspace{-0.5cm}\hspace{1.5cm}(c) $r_\text{max}=57$
\begin{center}
\includegraphics[width=13cm]{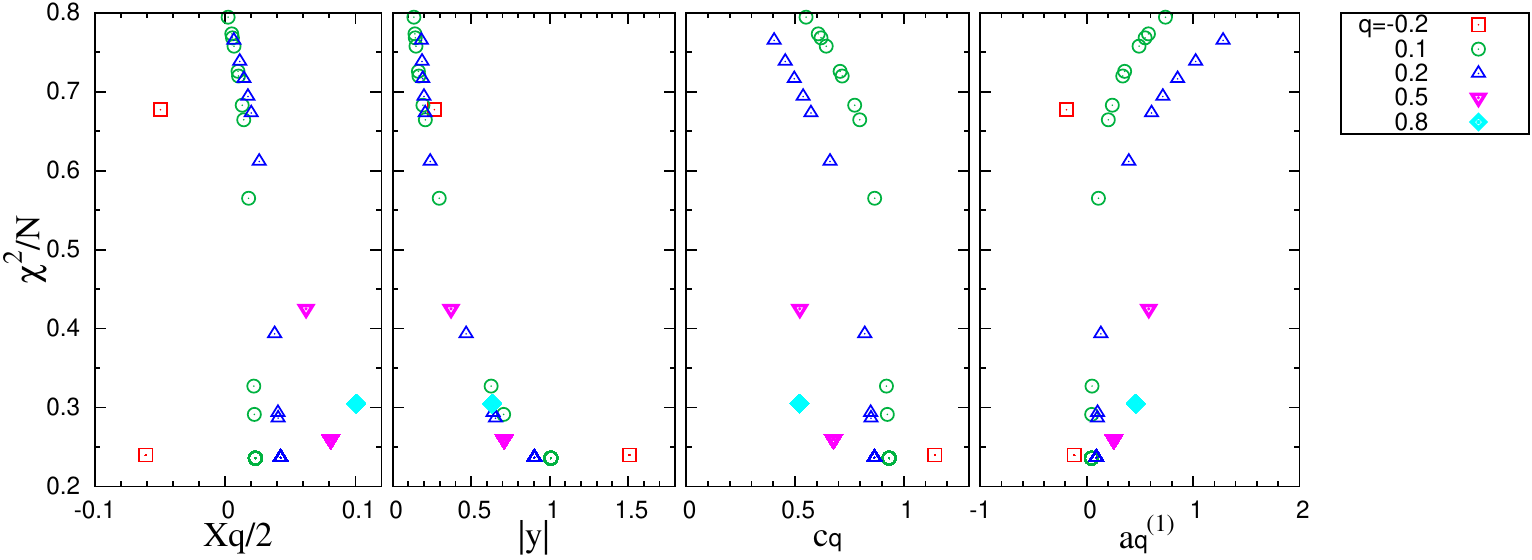}
\end{center}
\caption{
Stability maps from the scaling analysis of $\langle T^q \rangle$ for various $q$ in case of (a) the maximum distance $r_\text{max}=41$, (b) $r_\text{max}=49$, and (c) $r_\text{max}=57$.  The value of $q$ is listed in the legend at the right. The values of fitting parameters at the global minimum are listed in Table. \ref{tab:2D_Tq}.
}
\label{fig:2D_Tq_stability}
\end{figure*}

\begin{figure*}[h]
\flushleft
\vspace{-0.5cm}\hspace{1.5cm}(a) $r_\text{max}=41$
\begin{center}
\includegraphics[width=13cm]{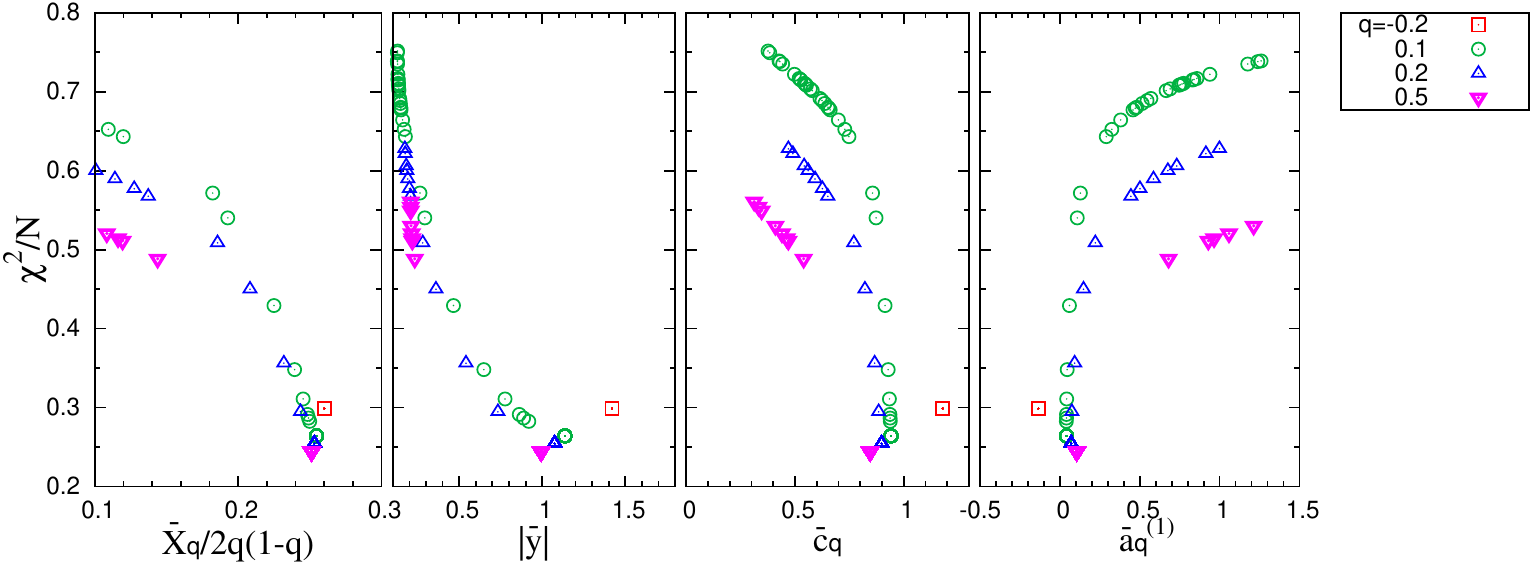}
\end{center}
\vspace{-0.5cm}\hspace{1.5cm}(b) $r_\text{max}=49$
\begin{center}
\includegraphics[width=13cm]{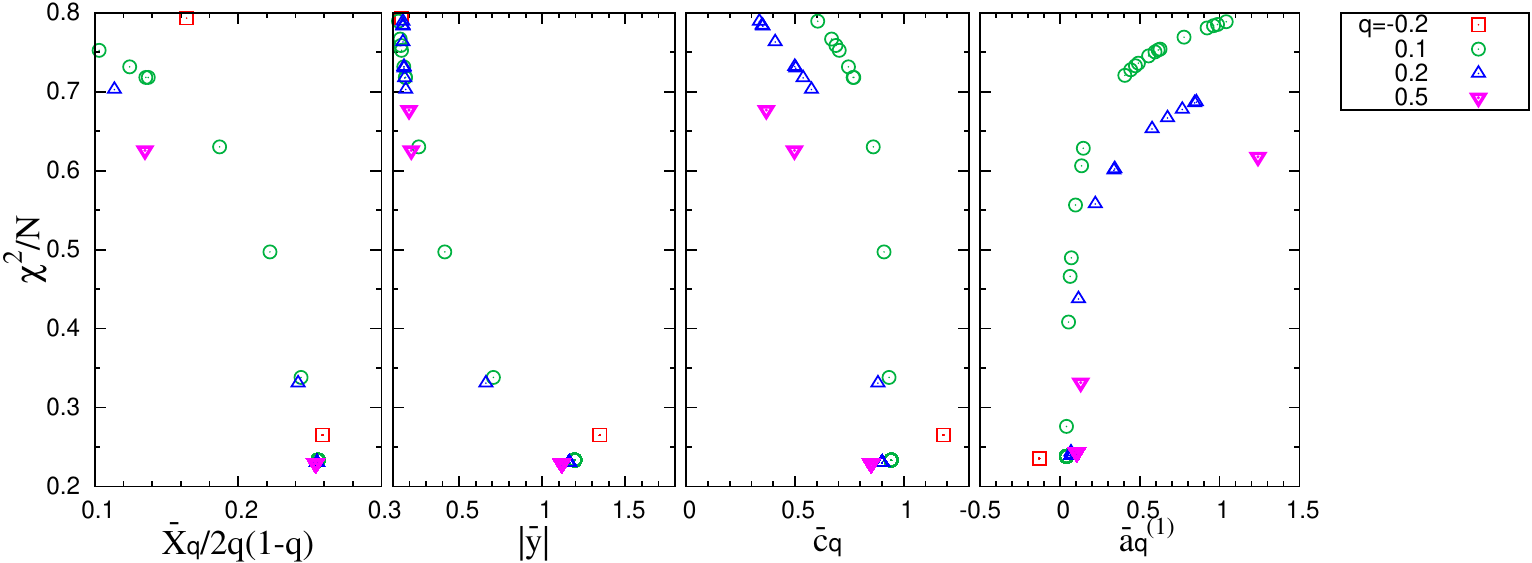}
\end{center}
\vspace{-0.5cm}\hspace{1.5cm}(c) $r_\text{max}=57$
\begin{center}
\includegraphics[width=13cm]{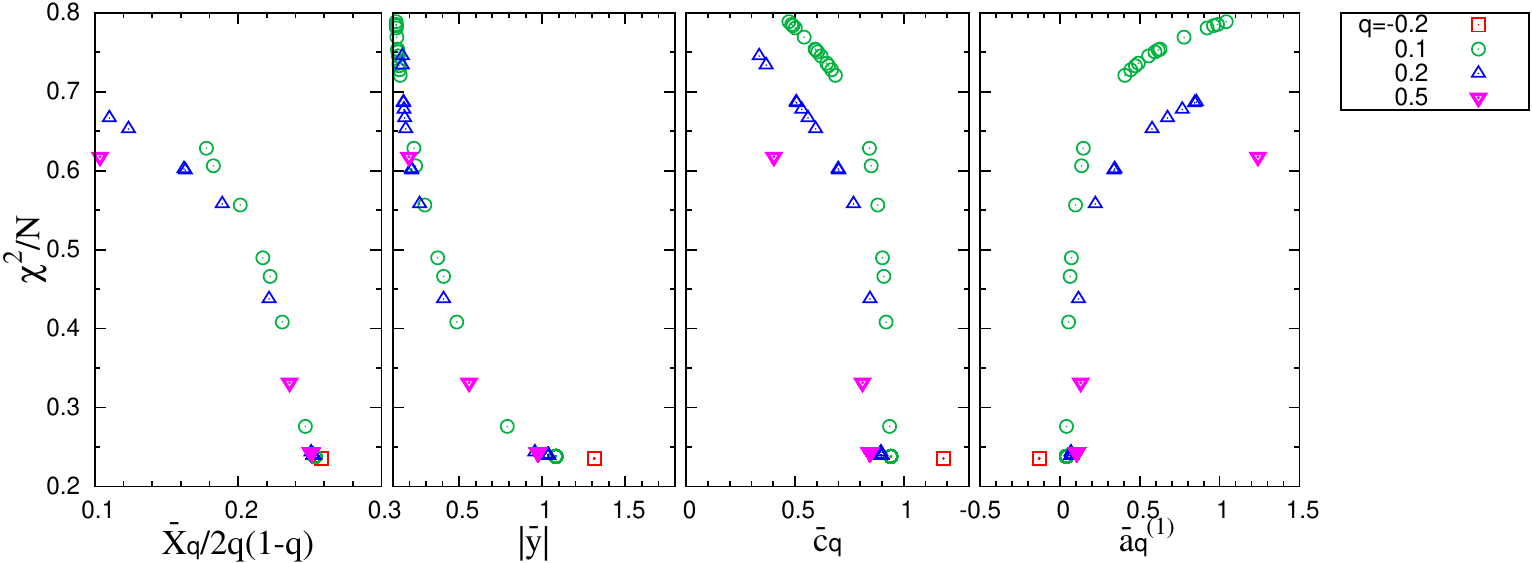}
\end{center}
\caption{
Stability maps from the scaling analysis of $\langle p_q(T) \rangle$ for various $q$ in; case of (a) the maximum distance $r_\text{max}=41$, (b) $r_\text{max}=49$, and (c) $r_\text{max}=57$.  The value of $q$ is listed in the legend at the right. The values of fitting parameters at the global minimum are listed in Table. \ref{tab:2D_Legendre}.
} \label{fig:2D_Legendre_stability}
\end{figure*}

The results of the fitting are given in Tables \ref{tab:2D_Tq} and \ref{tab:2D_Legendre}. We also show the corresponding stability maps \cite{sm:Obuse12} in Figs.\ \ref{fig:2D_Tq_stability} and \ref{fig:2D_Legendre_stability}.

\begin{figure*}[h]
\flushleft
\vspace{-0.5cm}\hspace{3cm}(a)
\begin{center}
\includegraphics[width=12cm]{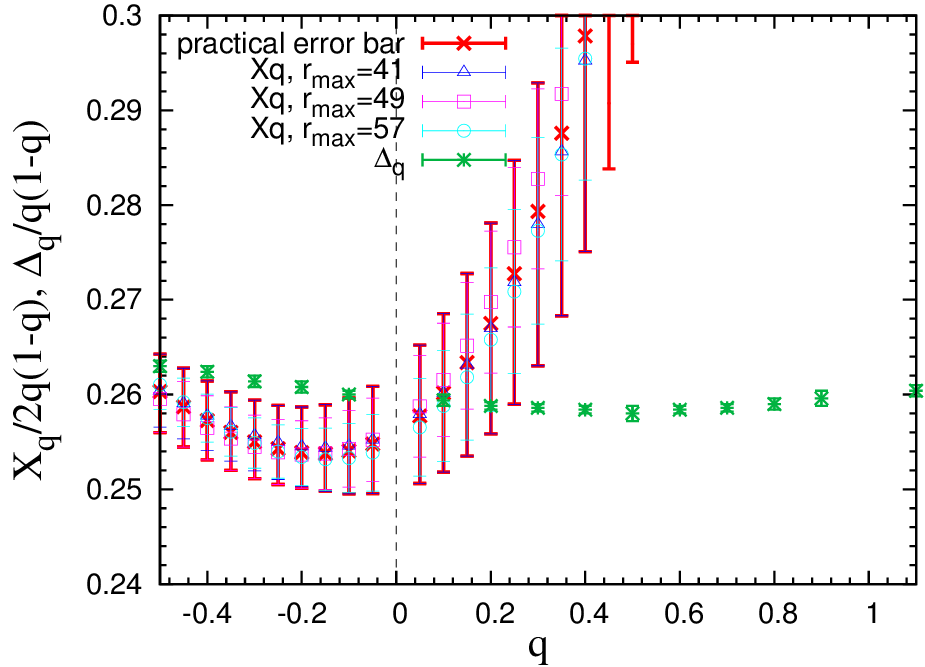}
\end{center}
\vspace{-0.5cm}\hspace{3cm}(b)
\begin{center}
\includegraphics[width=12cm]{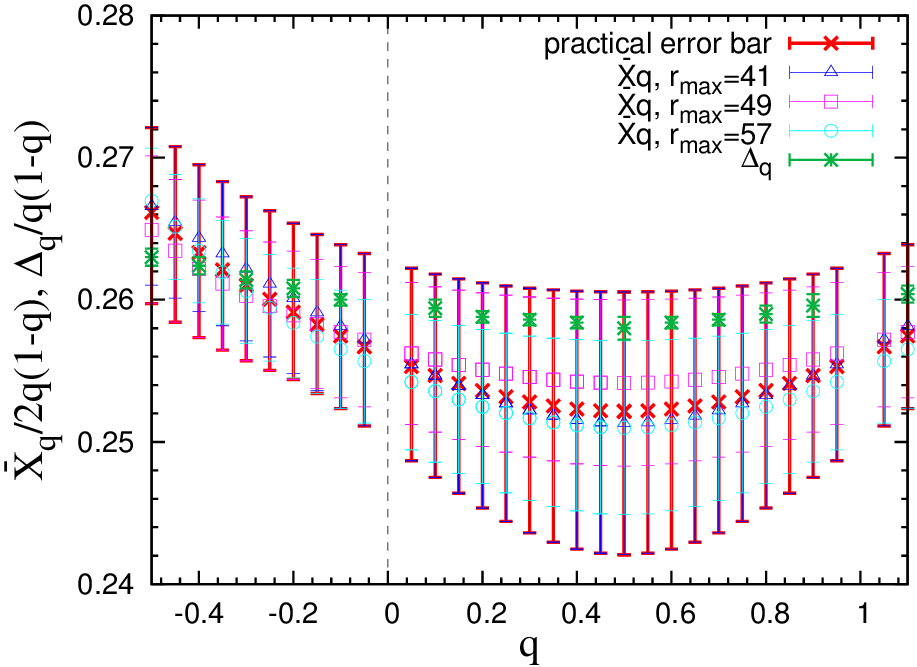}
\end{center}
\caption{
The $q$ dependence of (a) $X_q$ from $\langle T^q \rangle$ and (b) $\bar{X}_q$ from $\langle p_q(T) \rangle$. The thin and thick lines represent the error bars estimated from the error-propagation theory and the practical error bars, respectively.
} \label{fig:2D_practical-err}
\end{figure*}

Finally, we estimate the practical error bars for $X_q$ and $\bar{X}_q$ in the same way as in the Q1D case. These are shown in Fig.\ \ref{fig:2D_practical-err}.


\begin{table*}
\rotatebox{90}{
\begin{minipage}{1.0\textheight}
\footnotesize{
\vspace{-0.3cm}
\caption{
The details of the most reliable fitting by the scaling analysis for $q$th moment of 2D PCC, $\langle T^q \rangle$, with  the minimum $r$, $r_\text{min}=3$ and the maximum $r$, (a) $r_\text{max}=41$, (b) $r_\text{max}=49$, and (c) $r_\text{max}=57$. The scaling function in Eq.\ (\ref{eq:2D_Tq}) with $N_p=1$ and $d_q=0$ is employed. $N$, $\chi_\text{min}^2$, and $Q$ in the tables represent the number of data point used in the fitting, the minimum $\chi^2$, and the goodness of fit, respectively.
\label{tab:2D_Tq}
}
\vspace{-0.3cm}
\flushleft{(a)}
\begin{tabular}{c c c|r|c c |r r r r r }
\hline\hline
$r_\text{min}$ &  $r_\text{max}$ & $N$ & $q\ $ & $\chi^2_\text{min}/N$ & $Q$
 &$X_q/2\quad\quad\ $ & $X_q/2q(1-q)\quad $ & $c_q\quad\quad\quad$ & $|y|\quad\quad\quad$ & $a_q^{(1)}\quad\quad\quad$ \\
\hline
$3$ & $41$ & $20$ & $-0.5$ & $0.491$ &  $0.876$ &  $-0.39062 \pm
			 0.00579$ &  $0.26041 \pm  0.00386$ &   $\ \
				 1.4544 \pm 0.0280$ &  $\ \ 2.6689 \pm 1.5472$ &  $-1.0132 \pm 1.5362$ \\
$3$ & $41$ & $20$ & $-0.4$ & $0.380$ &  $0.960$ &  $-0.28871 \pm  0.00410$ &  $0.25777 \pm  0.00366$ &   $1.3234 \pm 0.0187$ &  $2.1724 \pm 0.9425$ &  $-0.4464 \pm 0.3766$ \\
$3$ & $41$ & $20$ & $-0.3$ & $0.324$ &  $0.982$ &  $-0.19943 \pm  0.00292$ &  $0.25568 \pm  0.00374$ &   $1.2240 \pm 0.0128$ &  $1.8622 \pm 0.6894$ &  $-0.2359 \pm 0.1312$ \\
$3$ & $41$ & $20$ & $-0.2$ & $0.307$ &  $0.986$ &  $-0.12216 \pm  0.00201$ &  $0.25450 \pm  0.00419$ &   $1.1417 \pm 0.0086$ &  $1.6139 \pm 0.5522$ &  $-0.1241 \pm 0.0481$ \\
$3$ & $41$ & $20$ & $-0.1$ & $0.295$ &  $0.989$ &  $-0.05601 \pm  0.00111$ &  $0.25458 \pm  0.00505$ &   $1.0686 \pm 0.0046$ &  $1.3995 \pm 0.4645$ &  $-0.0528 \pm 0.0140$ \\
$3$ & $41$ & $20$ & $0.1$ & $0.258$ &  $0.995$ &  $0.04683 \pm  0.00151$ &  $0.26017 \pm  0.00837$ &   $0.9334 \pm 0.0061$ &  $1.0612 \pm 0.3575$ &  $0.0457 \pm 0.0041$ \\
$3$ & $41$ & $20$ & $0.2$ & $0.243$ &  $0.996$ &  $0.08543 \pm  0.00356$ &  $0.26697 \pm  0.01113$ &   $0.8675 \pm 0.0142$ &  $0.9359 \pm 0.3223$ &  $0.0913 \pm 0.0066$ \\
$3$ & $41$ & $20$ & $0.3$ & $0.234$ &  $0.997$ &  $0.11675 \pm  0.00627$ &  $0.27797 \pm  0.01493$ &   $0.8020 \pm 0.0243$ &  $0.8364 \pm 0.2939$ &  $0.1408 \pm 0.0179$ \\
$3$ & $41$ & $20$ & $0.4$ & $0.230$ &  $0.997$ &  $0.14171 \pm  0.00966$ &  $0.29523 \pm  0.02013$ &   $0.7372 \pm 0.0361$ &  $0.7589 \pm 0.2699$ &  $0.1971 \pm 0.0383$ \\
$3$ & $41$ & $20$ & $0.5$ & $0.232$ &  $0.997$ &  $0.16124 \pm  0.01370$ &  $0.32248 \pm  0.02740$ &   $0.6740 \pm 0.0489$ &  $0.6996 \pm 0.2489$ &  $0.2615 \pm 0.0680$ \\
$3$ & $41$ & $20$ & $0.6$ & $0.238$ &  $0.997$ &  $0.17622 \pm  0.01827$ &  $0.36713 \pm  0.03806$ &   $0.6136 \pm 0.0615$ &  $0.6549 \pm 0.2301$ &  $0.3344 \pm 0.1073$ \\
$3$ & $41$ & $20$ & $0.7$ & $0.248$ &  $0.996$ &  $0.18747 \pm  0.02321$ &  $0.44636 \pm  0.05527$ &   $0.5568 \pm 0.0731$ &  $0.6217 \pm 0.2129$ &  $0.4157 \pm 0.1562$ \\
$3$ & $41$ & $20$ & $0.8$ & $0.261$ &  $0.995$ &  $0.19572 \pm  0.02838$ &  $0.61161 \pm  0.08868$ &   $0.5045 \pm 0.0828$ &  $0.5973 \pm 0.1973$ &  $0.5045 \pm 0.2142$ \\
$3$ & $41$ & $20$ & $0.9$ & $0.275$ &  $0.993$ &  $0.20157 \pm  0.03363$ &  $1.11984 \pm  0.18681$ &   $0.4569 \pm 0.0905$ &  $0.5797 \pm 0.1829$ &  $0.5997 \pm 0.2805$ \\
$3$ & $41$ & $20$ & $1.1$ & $0.308$ &  $0.986$ &  $0.20806 \pm  0.04399$ &  $-0.94573 \pm  0.19995$ &   $0.3760 \pm 0.0997$ &  $0.5582 \pm 0.1580$ &  $0.8047 \pm 0.4348$ \\
\hline\hline
\end{tabular}
\vspace{-0.2cm}
\flushleft{(b)}
\begin{tabular}{c c c|r|c c | r r r r r }
\hline\hline
$r_\text{min}$ &  $r_\text{max}$ & $N$ & $q\ $ & $\chi^2_\text{min}/N$ & $Q$
 &$X_q/2\quad\quad\ $ & $X_q/2q(1-q)\quad $ & $c_q\quad\quad\quad$ & $|y|\quad\quad\quad$ & $a_q^{(1)}\quad\quad\quad$ \\
\hline
$3$ & $49$ & $24$ & $-0.5$ & $0.427$ &  $0.964$ &  $-0.38930 \pm  0.00531$ &  $0.25954 \pm  0.00354$ &   $\ \ 1.4605 \pm 0.0270$ &  $\ \ 2.4458 \pm 1.2788$ &  $-0.8301 \pm 1.0249$ \\
$3$ & $49$ & $24$ & $-0.4$ & $0.335$ &  $0.992$ &  $-0.28727 \pm  0.00379$ &  $0.25649 \pm  0.00338$ &   $1.3297 \pm 0.0184$ &  $1.9559 \pm 0.7597$ &  $-0.3775 \pm 0.2484$ \\
$3$ & $49$ & $24$ & $-0.3$ & $0.290$ &  $0.997$ &  $-0.19849 \pm  0.00261$ &  $0.25448 \pm  0.00335$ &   $1.2279 \pm 0.0121$ &  $1.7097 \pm 0.5636$ &  $-0.2132 \pm 0.0939$ \\
$3$ & $49$ & $24$ & $-0.2$ & $0.272$ &  $0.998$ &  $-0.12177 \pm  0.00171$ &  $0.25368 \pm  0.00356$ &   $1.1433 \pm 0.0076$ &  $1.5349 \pm 0.4624$ &  $-0.1188 \pm 0.0383$ \\
$3$ & $49$ & $24$ & $-0.1$ & $0.256$ &  $0.999$ &  $-0.05594 \pm  0.00089$ &  $0.25428 \pm  0.00405$ &   $1.0688 \pm 0.0038$ &  $1.3792 \pm 0.3958$ &  $-0.0524 \pm 0.0124$ \\
$3$ & $49$ & $24$ & $0.1$ & $0.229$ &  $0.999$ &  $0.04708 \pm  0.00108$ &  $0.26154 \pm  0.00597$ &   $0.9344 \pm 0.0044$ &  $1.1158 \pm 0.3096$ &  $0.0460 \pm 0.0050$ \\
$3$ & $49$ & $24$ & $0.2$ & $0.223$ &  $1.000$ &  $0.08632 \pm  0.00240$ &  $0.26976 \pm  0.00749$ &   $0.8709 \pm 0.0094$ &  $1.0139 \pm 0.2804$ &  $0.0903 \pm 0.0062$ \\
$3$ & $49$ & $24$ & $0.3$ & $0.223$ &  $1.000$ &  $0.11876 \pm  0.00399$ &  $0.28277 \pm  0.00950$ &   $0.8094 \pm 0.0152$ &  $0.9315 \pm 0.2572$ &  $0.1359 \pm 0.0082$ \\
$3$ & $49$ & $24$ & $0.4$ & $0.230$ &  $0.999$ &  $0.14538 \pm  0.00583$ &  $0.30288 \pm  0.01215$ &   $0.7502 \pm 0.0212$ &  $0.8664 \pm 0.2383$ &  $0.1842 \pm 0.0152$ \\
$3$ & $49$ & $24$ & $0.5$ & $0.241$ &  $0.999$ &  $0.16711 \pm  0.00788$ &  $0.33422 \pm  0.01576$ &   $0.6940 \pm 0.0273$ &  $0.8155 \pm 0.2227$ &  $0.2356 \pm 0.0266$ \\
$3$ & $49$ & $24$ & $0.6$ & $0.256$ &  $0.999$ &  $0.18478 \pm  0.01008$ &  $0.38496 \pm  0.02101$ &   $0.6411 \pm 0.0331$ &  $0.7762 \pm 0.2095$ &  $0.2900 \pm 0.0413$ \\
$3$ & $49$ & $24$ & $0.7$ & $0.274$ &  $0.998$ &  $0.19912 \pm  0.01238$ &  $0.47410 \pm  0.02947$ &   $0.5920 \pm 0.0383$ &  $0.7458 \pm 0.1983$ &  $0.3469 \pm 0.0589$ \\
$3$ & $49$ & $24$ & $0.8$ & $0.294$ &  $0.997$ &  $0.21075 \pm  0.01472$ &  $0.65859 \pm  0.04600$ &   $0.5469 \pm 0.0427$ &  $0.7224 \pm 0.1887$ &  $0.4056 \pm 0.0788$ \\
$3$ & $49$ & $24$ & $0.9$ & $0.314$ &  $0.994$ &  $0.22017 \pm  0.01707$ &  $1.22319 \pm  0.09481$ &   $0.5058 \pm 0.0464$ &  $0.7042 \pm 0.1804$ &  $0.4655 \pm 0.1008$ \\
$3$ & $49$ & $24$ & $1.1$ & $0.357$ &  $0.987$ &  $0.23400 \pm  0.02170$ &  $-1.06363 \pm  0.09864$ &   $0.4346 \pm 0.0518$ &  $0.6788 \pm 0.1669$ &  $0.5864 \pm 0.1496$ \\
\hline\hline
\end{tabular}
\vspace{-0.2cm}
\flushleft{(c)}
\begin{tabular}{c c c|r|c c | r r r r r  }
\hline\hline
$r_\text{min}$ &  $r_\text{max}$ & $N$ & $q\ $ & $\chi^2_\text{min}/N$ & $Q$
 &$X_q/2\quad\quad\ $ & $X_q/2q(1-q)\quad $ & $c_q\quad\quad\quad$ & $|y|\quad\quad\quad$ & $a_q^{(1)}\quad\quad\quad$ \\
\hline
$3$ & $57$ & $28$ & $-0.5$ & $0.428$ &  $0.980$ &  $-0.39155 \pm  0.00387$ &  $0.26103 \pm  0.00258$ &   $\ \ 1.4501 \pm 0.0195$ &  $\ \ 2.8735 \pm 1.5219$ &  $-1.2255 \pm 1.8972$ \\
$3$ & $57$ & $28$ & $-0.4$ & $0.307$ &  $0.998$ &  $-0.28842 \pm  0.00284$ &  $0.25752 \pm  0.00254$ &   $1.3246 \pm 0.0138$ &  $2.1404 \pm 0.7722$ &  $-0.4364 \pm 0.3115$ \\
$3$ & $57$ & $28$ & $-0.3$ & $0.250$ &  $1.000$ &  $-0.19881 \pm  0.00208$ &  $0.25488 \pm  0.00267$ &   $1.2265 \pm 0.0098$ &  $1.7633 \pm 0.5260$ &  $-0.2210 \pm 0.0959$ \\
$3$ & $57$ & $28$ & $-0.2$ & $0.240$ &  $1.000$ &  $-0.12164 \pm  0.00145$ &  $0.25341 \pm  0.00302$ &   $1.1439 \pm 0.0066$ &  $1.5080 \pm 0.4096$ &  $-0.1170 \pm 0.0342$ \\
$3$ & $57$ & $28$ & $-0.1$ & $0.239$ &  $1.000$ &  $-0.05573 \pm  0.00080$ &  $0.25331 \pm  0.00364$ &   $1.0697 \pm 0.0036$ &  $1.3068 \pm 0.3400$ &  $-0.0508 \pm 0.0100$ \\
$3$ & $57$ & $28$ & $0.1$ & $0.236$ &  $1.000$ &  $0.04658 \pm  0.00105$ &  $0.25880 \pm  0.00584$ &   $0.9324 \pm 0.0045$ &  $1.0083 \pm 0.2580$ &  $0.0456 \pm 0.0033$ \\
$3$ & $57$ & $28$ & $0.2$ & $0.237$ &  $1.000$ &  $0.08506 \pm  0.00242$ &  $0.26580 \pm  0.00757$ &   $0.8660 \pm 0.0102$ &  $0.9022 \pm 0.2316$ &  $0.0917 \pm 0.0051$ \\
$3$ & $57$ & $28$ & $0.3$ & $0.241$ &  $1.000$ &  $0.11646 \pm  0.00413$ &  $0.27728 \pm  0.00984$ &   $0.8007 \pm 0.0168$ &  $0.8199 \pm 0.2107$ &  $0.1414 \pm 0.0116$ \\
$3$ & $57$ & $28$ & $0.4$ & $0.249$ &  $1.000$ &  $0.14182 \pm  0.00616$ &  $0.29546 \pm  0.01282$ &   $0.7374 \pm 0.0239$ &  $0.7573 \pm 0.1936$ &  $0.1963 \pm 0.0232$ \\
$3$ & $57$ & $28$ & $0.5$ & $0.260$ &  $1.000$ &  $0.16214 \pm  0.00842$ &  $0.32428 \pm  0.01684$ &   $0.6769 \pm 0.0310$ &  $0.7106 \pm 0.1793$ &  $0.2567 \pm 0.0391$ \\
$3$ & $57$ & $28$ & $0.6$ & $0.273$ &  $0.999$ &  $0.17833 \pm  0.01083$ &  $0.37152 \pm  0.02257$ &   $0.6202 \pm 0.0375$ &  $0.6763 \pm 0.1672$ &  $0.3221 \pm 0.0589$ \\
$3$ & $57$ & $28$ & $0.7$ & $0.289$ &  $0.999$ &  $0.19120 \pm  0.01331$ &  $0.45524 \pm  0.03168$ &   $0.5679 \pm 0.0430$ &  $0.6515 \pm 0.1567$ &  $0.3914 \pm 0.0818$ \\
$3$ & $57$ & $28$ & $0.8$ & $0.305$ &  $0.998$ &  $0.20144 \pm  0.01577$ &  $0.62949 \pm  0.04929$ &   $0.5204 \pm 0.0474$ &  $0.6338 \pm 0.1477$ &  $0.4634 \pm 0.1072$ \\
$3$ & $57$ & $28$ & $0.9$ & $0.322$ &  $0.998$ &  $0.20959 \pm  0.01818$ &  $1.16441 \pm  0.10100$ &   $0.4776 \pm 0.0508$ &  $0.6213 \pm 0.1400$ &  $0.5369 \pm 0.1345$ \\
$3$ & $57$ & $28$ & $1.1$ & $0.356$ &  $0.995$ &  $0.22136 \pm  0.02271$ &  $-1.00620 \pm  0.10324$ &   $0.4054 \pm 0.0546$ &  $0.6067 \pm 0.1276$ &  $0.6841 \pm 0.1925$ \\
\hline\hline
\end{tabular}
}
\end{minipage}
}
\end{table*}

\begin{table*}
\rotatebox{90}{
\begin{minipage}{1.0\textheight}
\footnotesize{
\caption{
The details of the most reliable fitting by the scaling analysis for the Legendre function as a function  of 2D PCC, $\langle p_{q}(T) \rangle$, with  the minimum $r$, $r_\text{min}=3$ and the maximum $r$, (a) $r_\text{max}=41$, (b) $r_\text{max}=49$, and (c) $r_\text{max}=57$. The scaling function in Eq.\ (\ref{eq:2D_Legendre}) with $N_p=1$ and $N_s=0$ is employed. $N$, $\chi_\text{min}^2$, and $Q$ in the tables represent the number of data point used in the fitting, the minimum $\chi^2$, and the goodness of fit, respectively. Note that the Legendre function satisfies $p_{q}(x)=p_{1-q}(x)$.
\label{tab:2D_Legendre}
}
\flushleft{(a)}
\begin{tabular}{c c c|r|c c | r r r   r r }
\hline\hline
$r_\text{min}$ &  $r_\text{max}$ & $N$ & $q\ $ & $\chi^2_\text{min}/N$ & $Q$
 &$\bar{X}_q/2\quad\quad\ $ & $\bar{X}_q/2q(1-q)\quad$ & $\bar{c}_q\quad\quad\quad$ & $|\bar{y}|\quad\quad\quad$ & $\bar{a}_q^{(1)}\quad\quad\quad$ \\
\hline
$3$ & $41$ & $20$ & $0.5$ & $0.244$ &  $0.996$ &  $0.06283 \pm  0.00231$
			 &  $0.25131 \pm  0.00924$ &   $0.8454 \pm
				 0.0174$ &  $\ \ 0.9932 \pm 0.4077$ & $0.1050 \pm 0.0086$ \\
$3$ & $41$ & $20$ & $0.6$ & $0.245$ &  $0.996$ &  $0.06037 \pm  0.00218$ &  $0.25154 \pm  0.00907$ &   $0.8512 \pm 0.0164$ &  $1.0024 \pm 0.4070$ & $0.1012 \pm 0.0083$ \\
$3$ & $41$ & $20$ & $0.7$ & $0.248$ &  $0.996$ &  $0.05296 \pm  0.00181$ &  $0.25221 \pm  0.00860$ &   $0.8686 \pm 0.0138$ &  $1.0298 \pm 0.4051$ & $0.0897 \pm 0.0078$ \\
$3$ & $41$ & $20$ & $0.8$ & $0.254$ &  $0.995$ &  $0.04052 \pm  0.00127$ &  $0.25327 \pm  0.00792$ &   $0.8984 \pm 0.0098$ &  $1.0749 \pm 0.4028$ & $0.0699 \pm 0.0072$ \\
$3$ & $41$ & $20$ & $0.9$ & $0.264$ &  $0.994$ &  $0.02292 \pm  0.00064$ &  $0.25466 \pm  0.00715$ &   $0.9415 \pm 0.0051$ &  $1.1370 \pm 0.4014$ & $0.0408 \pm 0.0053$ \\
$3$ & $41$ & $20$ & $1.1$ & $0.289$ &  $0.990$ &  $-0.02839 \pm  0.00063$ &  $0.25812 \pm  0.00576$ &   $1.0775 \pm 0.0055$ &  $1.3091 \pm 0.4111$ & $-0.0566 \pm 0.0117$ \\
$3$ & $41$ & $20$ & $1.2$ & $0.299$ &  $0.988$ &  $-0.06242 \pm  0.00127$ &  $0.26009 \pm  0.00529$ &   $1.1797 \pm 0.0116$ &  $1.4193 \pm 0.4325$ & $-0.1360 \pm 0.0351$ \\
$3$ & $41$ & $20$ & $1.3$ & $0.314$ &  $0.985$ &  $-0.10225 \pm  0.00198$ &  $0.26217 \pm  0.00507$ &   $1.3160 \pm 0.0197$ &  $1.5523 \pm 0.4816$ & $-0.2506 \pm 0.0827$ \\
$3$ & $41$ & $20$ & $1.4$ & $0.373$ &  $0.963$ &  $-0.14803 \pm  0.00289$ &  $0.26434 \pm  0.00516$ &   $1.5023 \pm 0.0318$ &  $1.7327 \pm 0.5962$ & $-0.4309 \pm 0.1998$ \\
$3$ & $41$ & $20$ & $1.5$ & $0.489$ &  $0.878$ &  $-0.19993 \pm  0.00415$ &  $0.26657 \pm  0.00554$ &   $1.7653 \pm 0.0517$ &  $2.0378 \pm 0.8839$ & $-0.7879 \pm 0.6150$ \\
\hline\hline
\end{tabular}
\flushleft{(b)}
\begin{tabular}{c c c|r|c c | r r r  r r}
\hline\hline
$r_\text{min}$ &  $r_\text{max}$ & $N$ & $q\ $ & $\chi^2_\text{min}/N$ & $Q$
 &$\bar{X}_q/2\quad\quad\ $ & $\bar{X}_q/2q(1-q)\quad$ & $\bar{c}_q\quad\quad\quad$ & $|\bar{y}|\quad\quad\quad$ & $\bar{a}_q^{(1)}\quad\quad\quad$ \\
\hline
$3$ & $49$ & $24$ & $0.5$ & $0.230$ &  $0.999$ &  $0.06354 \pm  0.00146$
			 &  $0.25415 \pm  0.00585$ &   $0.8504 \pm
				 0.0108$ &  $\ \ 1.1179 \pm 0.3654$ & $0.1050 \pm 0.0128$ \\
$3$ & $49$ & $24$ & $0.6$ & $0.229$ &  $0.999$ &  $0.06102 \pm  0.00139$ &  $0.25425 \pm  0.00579$ &   $0.8558 \pm 0.0103$ &  $1.1229 \pm 0.3641$ & $0.1013 \pm 0.0126$ \\
$3$ & $49$ & $24$ & $0.7$ & $0.229$ &  $0.999$ &  $0.05346 \pm  0.00118$ &  $0.25457 \pm  0.00564$ &   $0.8722 \pm 0.0089$ &  $1.1377 \pm 0.3605$ & $0.0901 \pm 0.0117$ \\
$3$ & $49$ & $24$ & $0.8$ & $0.230$ &  $0.999$ &  $0.04081 \pm  0.00086$ &  $0.25509 \pm  0.00540$ &   $0.9005 \pm 0.0067$ &  $1.1623 \pm 0.3555$ & $0.0706 \pm 0.0098$ \\
$3$ & $49$ & $24$ & $0.9$ & $0.233$ &  $0.999$ &  $0.02302 \pm  0.00046$ &  $0.25580 \pm  0.00512$ &   $0.9422 \pm 0.0037$ &  $1.1960 \pm 0.3503$ & $0.0413 \pm 0.0062$ \\
$3$ & $49$ & $24$ & $1.1$ & $0.252$ &  $0.999$ &  $-0.02835 \pm  0.00051$ &  $0.25773 \pm  0.00462$ &   $1.0778 \pm 0.0045$ &  $1.2881 \pm 0.3484$ & $-0.0563 \pm 0.0104$ \\
$3$ & $49$ & $24$ & $1.2$ & $0.265$ &  $0.998$ &  $-0.06214 \pm  0.00108$ &  $0.25891 \pm  0.00451$ &   $1.1822 \pm 0.0104$ &  $1.3463 \pm 0.3598$ & $-0.1318 \pm 0.0278$ \\
$3$ & $49$ & $24$ & $1.3$ & $0.284$ &  $0.997$ &  $-0.10151 \pm  0.00178$ &  $0.26029 \pm  0.00457$ &   $1.3231 \pm 0.0188$ &  $1.4225 \pm 0.3938$ & $-0.2342 \pm 0.0599$ \\
$3$ & $49$ & $24$ & $1.4$ & $0.333$ &  $0.992$ &  $-0.14682 \pm  0.00271$ &  $0.26218 \pm  0.00484$ &   $1.5152 \pm 0.0319$ &  $1.5573 \pm 0.4849$ & $-0.3856 \pm 0.1372$ \\
$3$ & $49$ & $24$ & $1.5$ & $0.428$ &  $0.963$ &  $-0.19869 \pm  0.00390$ &  $0.26492 \pm  0.00520$ &   $1.7802 \pm 0.0514$ &  $1.8545 \pm 0.7350$ & $-0.6836 \pm 0.4295$\\
\hline\hline
\end{tabular}
\flushleft{(c)}
\begin{tabular}{c c c|r|c c | r r r  r r}
\hline\hline
$r_\text{min}$ &  $r_\text{max}$ & $N$ & $q\ $ & $\chi^2_\text{min}/N$ & $Q$
 &$\bar{X}_q/2\quad\quad\ $ & $\bar{X}_q/2q(1-q)\quad$ & $\bar{c}_q\quad\quad\quad$ & $|\bar{y}|\quad\quad\quad$ & $\bar{a}_q^{(1)}\quad\quad\quad$ \\
\hline
$3$ & $57$ & $28$ & $0.5$ & $0.244$ &  $1.000$ &  $0.06275 \pm  0.00153$
			 &  $0.25099 \pm  0.00611$ &   $0.8447 \pm
				 0.0121$ &  $\ \ 0.9748 \pm 0.2986$ & $0.1048 \pm 0.0076$ \\
$3$ & $57$ & $28$ & $0.6$ & $0.243$ &  $1.000$ &  $0.06028 \pm  0.00145$ &  $0.25116 \pm  0.00603$ &   $0.8504 \pm 0.0115$ &  $0.9816 \pm 0.2979$ & $0.1010 \pm 0.0074$ \\
$3$ & $57$ & $28$ & $0.7$ & $0.241$ &  $1.000$ &  $0.05285 \pm  0.00121$ &  $0.25167 \pm  0.00578$ &   $0.8677 \pm 0.0097$ &  $1.0018 \pm 0.2961$ & $0.0894 \pm 0.0070$ \\
$3$ & $57$ & $28$ & $0.8$ & $0.239$ &  $1.000$ &  $0.04040 \pm  0.00087$ &  $0.25248 \pm  0.00542$ &   $0.8974 \pm 0.0071$ &  $1.0353 \pm 0.2938$ & $0.0696 \pm 0.0061$ \\
$3$ & $57$ & $28$ & $0.9$ & $0.238$ &  $1.000$ &  $0.02282 \pm  0.00045$ &  $0.25357 \pm  0.00499$ &   $0.9407 \pm 0.0038$ &  $1.0820 \pm 0.2921$ & $0.0404 \pm 0.0042$ \\
$3$ & $57$ & $28$ & $1.1$ & $0.237$ &  $1.000$ &  $-0.02822 \pm  0.00046$ &  $0.25651 \pm  0.00417$ &   $1.0790 \pm 0.0043$ &  $1.2171 \pm 0.2983$ & $-0.0549 \pm 0.0083$ \\
$3$ & $57$ & $28$ & $1.2$ & $0.235$ &  $1.000$ &  $-0.06201 \pm  0.00093$ &  $0.25838 \pm  0.00386$ &   $1.1833 \pm 0.0092$ &  $1.3128 \pm 0.3149$ & $-0.1300 \pm 0.0244$ \\
$3$ & $57$ & $28$ & $1.3$ & $0.245$ &  $1.000$ &  $-0.10164 \pm  0.00144$ &  $0.26062 \pm  0.00368$ &   $1.3218 \pm 0.0154$ &  $1.4456 \pm 0.3563$ & $-0.2370 \pm 0.0585$ \\
$3$ & $57$ & $28$ & $1.4$ & $0.301$ &  $0.999$ &  $-0.14753 \pm  0.00205$ &  $0.26345 \pm  0.00365$ &   $1.5074 \pm 0.0240$ &  $1.6639 \pm 0.4617$ & $-0.4128 \pm 0.1530$ \\
$3$ & $57$ & $28$ & $1.5$ & $0.423$ &  $0.982$ &  $-0.20023 \pm  0.00278$ &  $0.26697 \pm  0.00371$ &   $1.7615 \pm 0.0362$ &  $2.1017 \pm 0.7676$ & $-0.8317 \pm 0.5931$ \\
\hline\hline
\end{tabular}
}
\end{minipage}
}
\end{table*}

\end{document}